\documentclass[12pt,reqno]{amsart}
\allowdisplaybreaks[4]
\usepackage{graphicx} %We can use any other package if it is necessary
\usepackage{amssymb}
\usepackage{color}
\usepackage{amsmath}
\usepackage{cite}
\usepackage{subfigure}
\usepackage{graphicx}
\usepackage{epstopdf}
\usepackage{bibentry}
\usepackage{subeqnarray}
\usepackage{cases}
\newtheorem{theorem}{Theorem}
\newtheorem{Proof}{Proof of Theorem }
\newtheorem{lemma}{Lemma}[section]

\begin{document}
\title{Semi-rational solutions of the third-type Davey-Stewartson equation}
\author{Jiguang Rao$^1$, Kuppuswamy Porsezian $^2$, Jingsong He$^1$}
\thanks{$^*$ Corresponding author: hejingsong@nbu.edu.cn, jshe@ustc.edu.cn}
\dedicatory {$^1$ Department of Mathematics, Ningbo University,
Ningbo, Zhejiang 315211, P.\ R.\ China\\
 $^2$ Department of Physics, Pondicherry University, Puducherry 605014,
India}
\begin{abstract}
General dark solitons and mixed solutions consisting of dark solitons and breathers for the third-type Davey-Stewartson (DS-III) equation are derived by employing the bilinear method. By introducing the two differential operators, semi-rational solutions  consisting of rogue waves, breathers and solitons are generated. These semi-rational solutions are given in terms of determinants whose matrix elements have simple algebraic expressions.  Under suitable parametric conditions, we derive general rogue wave solutions expressed in terms of rational functions.  It is shown that the fundamental (simplest) rogue waves are line rogue waves. It is also shown that the multi-rogue waves describe interactions of several fundamental rogue waves, which would generate interesting curvy wave patterns. The higher order rogue waves originate from a localized lump and retreat back to it. Several types of hybrid solutions composed of rogue waves, breathers and solitons have also been illustrated.  Specifically, these semi-rational solutions have a new phenomenon: lumps form on dark solitons and    gradual separation from the dark solitons is observed.
\end{abstract}

 \maketitle \vspace{-0.9cm}
 \noindent{{\bf Keywords}: Davey-Stewartson III equation, Bilinear transformation method, \\
 Rogue waves, Semi-rational solutions.}
% 02.30.Ik	Integrable systems
% 02.30.Jr	Partial differential equations
% 42.65.Tg	Optical solitons; nonlinear guided waves (for solitons in fibers, see 42.81.Dp)
% 05.45.Yv	Soliton

%%%%%%%%%%%%%%%%%%%%%%%%%%%%%%%%%%%%%%%%and the Kadomtsev-Petviashvil (KP) -hierarchy reduction method%%%%%%%%

%\begin{General Abstract}
{\bf
Like one dimensional nonlinear systems, the higher dimensional nonlinear systems are also widely used to explain the dynamics of many  higher dimensional physical systems. When compared with one dimensional nonlinear systems, the higher dimensional nonlinear systems are less studied mainly because of the non availability of analytical methods like in one dimension. For instance, during and after 1980s, several higher dimensional systems like the Davey-Stewartson (DS) equation, the Kadomtsev-Petviashvil(KP) equation, Ishimori equation, Three wave equation and so on have been introduced and investigated for Line solitons, Dromions, Lump solutions, vortex solutions etc.  For the past one decade or so, the concept of rogue waves have been well reported in many one dimensional nonlinear physical  systems but, again, not well studied in higher dimensional systems. The main focus of this paper is to address this problem and investigate rogue wave solutions of one such a higher dimensional system.  In this paper, we consider one such a equation in higher dimension, namely the DSIII equation   and derive general $N$th-order dark solitons by reducing the Gram-type solutions of the KP hierarchy. Considering suitable parametric conditions, a new kind of mixed solutions that include breathers, line breathers and dark solitons are given in terms of Gram determinants. Further, general form of semi-rational solutions containing rogue waves, breathers and dark solitons are generated and  show that the fundamental rogue waves are actually  line rogue waves, which arise from the constant background and retreat back to the constant background again.  We also show that the multi-rogue waves demonstrate the interaction of several fundamental line rogue waves, and occurrence of interesting curvy wave patterns due to  interaction.  The multi-semi-rational solutions possess interesting dynamics, such as more lumps form on  multi-dark solitons, such as interaction of lumps, line breathers and dark solitons.  A new kind of line rogue waves which are localized both in space and time are  illustrated. The higher order semi-rational solutions illustrating the dynamics that more gradual separation of lumps from the one soliton have also been shown, these dynamics correspond to the numerical study of blow up solutions in KP equation.
}
%\end{General Abstract}
%%%%%%%%%%%%%%%%%%%%%%%%%%%%%%%%%%%%%%%%%%%%%%%%

\section{Introduction}
For the past one decade or so, the investigation of rogue waves has attracted a lot of attention because of their potential applications in many areas of science and technology. Though initially this identification was done purely from mathematical point of view, over the years, it has been experimentally observed in Ocean technology, Ultra broadband supercontinuum generation, water tank,  optical fiber etc. Originally, rogue waves or freak waves are coined for vivid description of the mysterious and monstrous ocean waves. Usually, they are localized and isolated surface waves with large amplitude, which apparently appear from nowhere and disappear without a trace \cite{rw1,rw2,Today,Berlin,NK}. Subsequently, intensive researches about the characteristics of rogue waves have been implemented in other physical areas, such as  Bose-Einstein condensates\cite{E-1,E-2}, nonlinear optical system\cite{o-3,o-4,o-5}, superfluids\cite{su-1}, plasma\cite{pl-1,pl-2} even in nonlocal media \cite{arxi}.  In the recent past, some important studies indicated that the rogue waves possess some hallmark phenomenological features including following an unusual $L$-shaped statistics \cite{wuchi1,wuchi2,yin10,yin11}.
Mathematically, rogue wave solution was first time reported in the nonlinear Schr\"odinger (NLS) equation by Peregrine \cite{D.H.Peregrine}. The basic properties of this rogue wave can be summarized as: (i) (quasi-) rational solutions; (ii) localized in both time and space; and (iii) large amplitude (their peak has a height of at least three times of the
background) allocated with two  hollows. This is considered as the `definition' of the first-order $(1+1)$ dimensional
 rogue wave solution. Recently, higher-order rogue waves in the NLS equation were
reported in many other articles \cite{PD-3,NA-1,PD-2,AA-5,BG-6,DJ-6,ohta-nls,he}. To date, exact rogue wave solutions have been found in a variety of nonlinear integrable systems, such as three wave interaction equation \cite{three-wave-1,three-wave-2,three-wave-3}, the Sasa-Satsuma equation\cite{S-S-1,S-S-2,S-S-3,S-S-4}, the Hirota equation\cite{he1}, the derivative nonlinear Schr\"odinger equation \cite{he2}, even as some coupled nonlinear Schr\"odinger system\cite{three-wave-1,qinzhenyun1,Lin,chenshihua1,He,Chow2,2DFS}, etc \cite{yan-1,yan-2,yan-3,zhu1,zhu2,yong1}. In addition to the identification of rogue waves in integrable systems, rogue waves in dissipative dynamical systems has also been reported \cite{yin10,falk}.

Though  several interesting results are reported for the one dimensional nonlinear equations, to the best of our knowledge, only few results are available in  higher dimensional nonlinear systems.This is mainly because of the complications in solving higher dimensional systems and tedious numerical analysis.  For example,  several  higher dimensional nonlinear equations have been derived in ocean, Fluid dynamics, bulk materials and  ultrafast optics and also  the extension of nonlinear dynamics to a higher spatiotemporal dimensionality is inevitable.  Especially, ultrafast optics rogue waves are also higher dimensional, where spatial and temporal degrees of freedom cannot be treated separately. This fact has been established through several theoretical and experimental studies \cite{high-1,high-2,high-3,high-4,high-5,high-6,high-7}. Recently, Klein and Saut obtained a subclass of numerical solutions for the Kadomtsev-Petviashvili (KP) equation \cite{XX-1,XX-2}, these solutions describe that lumps form at a line soliton and also travel with higher speed than the soliton. In this work, motivated by the above facts, we consider one such a nonlinear higher dimensional problem to explore explicit solutions of $(2+1)$-dimensional soliton equations possessing those unique behaviors.

The third-type of Davey-Stewartson (DSIII) equation is given by
\begin{equation}\label{DSIII-a}
\begin{aligned}
iq_t-q_{xx}+q_{yy}-2{\lambda}q[(\int_{-\infty}^x|q|_y^2dx'+u_1(y,t))-(\int_{-\infty}^y|q|_x^2dy'+u_2(x,t))]=0, \lambda=\pm1, \\
\end{aligned}
\end{equation}
where the complex valued scalar function $q(x,y,t)$ represents a certain physical quantity and $u_1(y,t)$ and $u_2(x,t)$ are two arbitrary real-valued functions. Using the transformations
\begin{subequations}
\begin{equation}
\begin{aligned}
V_x=(|q|^2)_y,\hspace{20pt}V=\int_{-\infty}^x(|q|^2)_ydx'+u_1(y,t),
\end{aligned}
\end{equation}
\\and\\
\begin{equation}
\begin{aligned}
U_y=(|q|^2)_x,\hspace{20pt}U=\int_{-\infty}^y(|q|^2)_xdy'+u_2(x,t),
\end{aligned}
\end{equation}
\end{subequations}
equation (\ref{DSIII-a}) becomes the following system of coupled partial differential equations:
\begin{subequations}\label{DSIII-b}
\begin{equation}
\begin{aligned}
iq_t-q_{xx}+q_{yy}-2{\lambda}q(V-U)=0,
\end{aligned}
\end{equation}
\begin{equation}\label{V_x}
\begin{aligned}
V_x=(|q|^2)_y,
\end{aligned}
\end{equation}
\begin{equation}\label{U_y}
\begin{aligned}
U_y=(|q|^2)_x,
\end{aligned}
\end{equation}
\end{subequations}
where $V$ and $U$ are two real functions.
These  equations were obtained by Boiti $et$ $al$.\cite{Y-1} by employing ideas \cite{Y-2} of the  so-called direct linearisation  method of Fokas and Ablowitz \cite{Y-3}. However, as noted by Fokas\cite{Y-4}, it was first time derived earlier by Schulman\cite{Y-5} and by Fokas and
Santini\cite{Y-6,Y-7} using the symmetry approach.
%\indent

As mentioned above, the researches about rogue waves in $(2+1)$-dimensional soliton equations are only few \cite{qin-mk,DSI,DSII,YOC,Fokas}. These facts actually
motivated us to seek more solutions for the third-type Davey-Stewartson (DSIII) equation. Although Liu $et$ $al.$ have obtained parallel line rogue waves for the DSIII equation \cite{binbin}, but there still exist other kinds of rogue waves, i.e., multi-rogue waves, and higher order rogue waves.  Besides, a various forms of hybrid solutions
for the DSIII equation have not been studied before, such as solutions consisting of lumps and soltions, such as solutions describing rogue waves interacting with solitons and breathers, and so on.

This paper is organized as follows. In section \ref{2}, dark solitons and mixed solutions consisting of dark solitons and breathers of the DSIII equation are given in terms of Gram-type determinants of $N\times N$ matrices. In section \ref{3}, rational and semi-rational solutions which include
rogue waves, solitons and breathers are expressed in terms of determinant. The dynamical features of rogue waves and semi-rational solutions are demonstrated in section \ref{4}. Our results are summarized in section \ref{5}.

\section{Solitons, Breathers Via Determinants Of $N\times N$ Matrices}\label{2}
In this section, we derive the explicit form of solutions consisting of solitons and breathers for DSIII equation (\ref{DSIII-b}).
Using the following dependent variable bilinear transformations
\begin{equation}\label{dvtransformation}
\begin{aligned}
q=\frac{g}{f},\qquad  V=-\lambda(\log{f})_{yy},\qquad  U=-\lambda(\log{f})_{xx},
\end{aligned}
\end{equation}
the DSIII equation can be transformed into the following bilinear forms
\begin{equation} \label{bili:DSIII}
\begin{aligned}
(iD_{t}-{D_{x}}^2+{D_{y}}^2)g \cdot f &=0,\\
({\lambda}D_{x}D_{y}-2)f \cdot f &=-2{g \cdot g^{*}}\,.
\end{aligned}
\end{equation}
Here, $f$ is a real function, $g$ is a complex function, asterisk denotes complex conjugation, and the operator $D$ is the Hirota's bilinear differential operator\cite{Hirota} defined by
\begin{eqnarray*}
&P(D_{x},D_{y},D_{t}, )F(x,y,t\cdot\cdot\cdot)\cdot G(x,y,t,\cdot\cdot\cdot) \mbox{\hspace{6.3cm}}   \\
&=P(\partial_{x}-\partial_{x^{'}},\partial_{y}-\partial_{y^{'}},
\partial_{t}-\partial_{t^{'}},\cdot\cdot\cdot)F(x,y,t,\cdot\cdot\cdot)G(x^{'},y^{'},t^{'},
\cdot\cdot\cdot)|_{x^{'}=x,y^{'}=y,t^{'}=t},
\end{eqnarray*}
where $P$ is a polynomial of $D_{x}$,$D_{y}$,$D_{t},\cdot\cdot\cdot$.
\\
 \begin{theorem}\label{th1}
 {\sl The DSIII equation admits both soliton and breather solutions of the form
\begin{equation}\label{soliton}
\begin{aligned}
q=\frac{g}{f},\qquad  V=-\lambda(\log{f})_{yy},\qquad  U=-\lambda(\log{f})_{xx},
\end{aligned}
\end{equation}
where $f$ and $g$ are given by the following  Gram determinant
\begin{equation}\label{so-fg}
\begin{aligned}
f=\det\limits_{1\leq i,j\leq N}(m^{(0)}_{i,j})\,,g=\det\limits_{1\leq i,j\leq N}(m^{(1)}_{i,j}),
\end{aligned}
\end{equation}
and the matrix element $m_{i,j}^{(n)}$ is defined by
\begin{equation}\label{so-m}
\begin{aligned}
&m^{(n)}_{i,j}=\gamma_{ij}+\frac{1}{p_{i}+q_{j}}(-\frac{p_{i}}{p^{*}_{j}})^{n}\psi_{i}\phi_{j},
\end{aligned}
\end{equation}
with
\begin{equation}\label{xi1}
\begin{aligned}
&\psi_{i}=e^{\xi_{i}}\,,\phi_{j}=e^{\eta_{j}}\,,\\
&\xi_{i}=\lambda\,p_{i}\,x+\frac{1}{p_{i}}\,y-i\,(\frac{1}{p_{i}^{2}}+p_{i}^{2})\,t+\xi_{i0}\,,\\
&\eta_{j}=\lambda\,p_{j}^{*}\,x+\frac{1}{p_{j}^{*}}\,y+i\,(\frac{1}{p_{j}^{*2}}+p_{i}^{*2})\,t+\xi^{*}_{j0}\,.
\end{aligned}
\end{equation}
Here $i\,,j\,,N$ are arbitrary positive integers, $p_{i}\,$ and $\xi_{i0}$ are arbitrary complex constants, $\gamma_{ij}=1\,,0$\,.
Note that a grammian $G=\det\limits_{1\leq i,j\leq N}(g_{ij})$ is the determinant of a matrix with entries $g_{ij}=\int_{a_0}^{b_0}f_if_jdx$.
}
\\
\end{theorem}

\begin{figure}[!htbp]
\centering
\subfigure[]{\includegraphics[height=6cm,width=6cm]{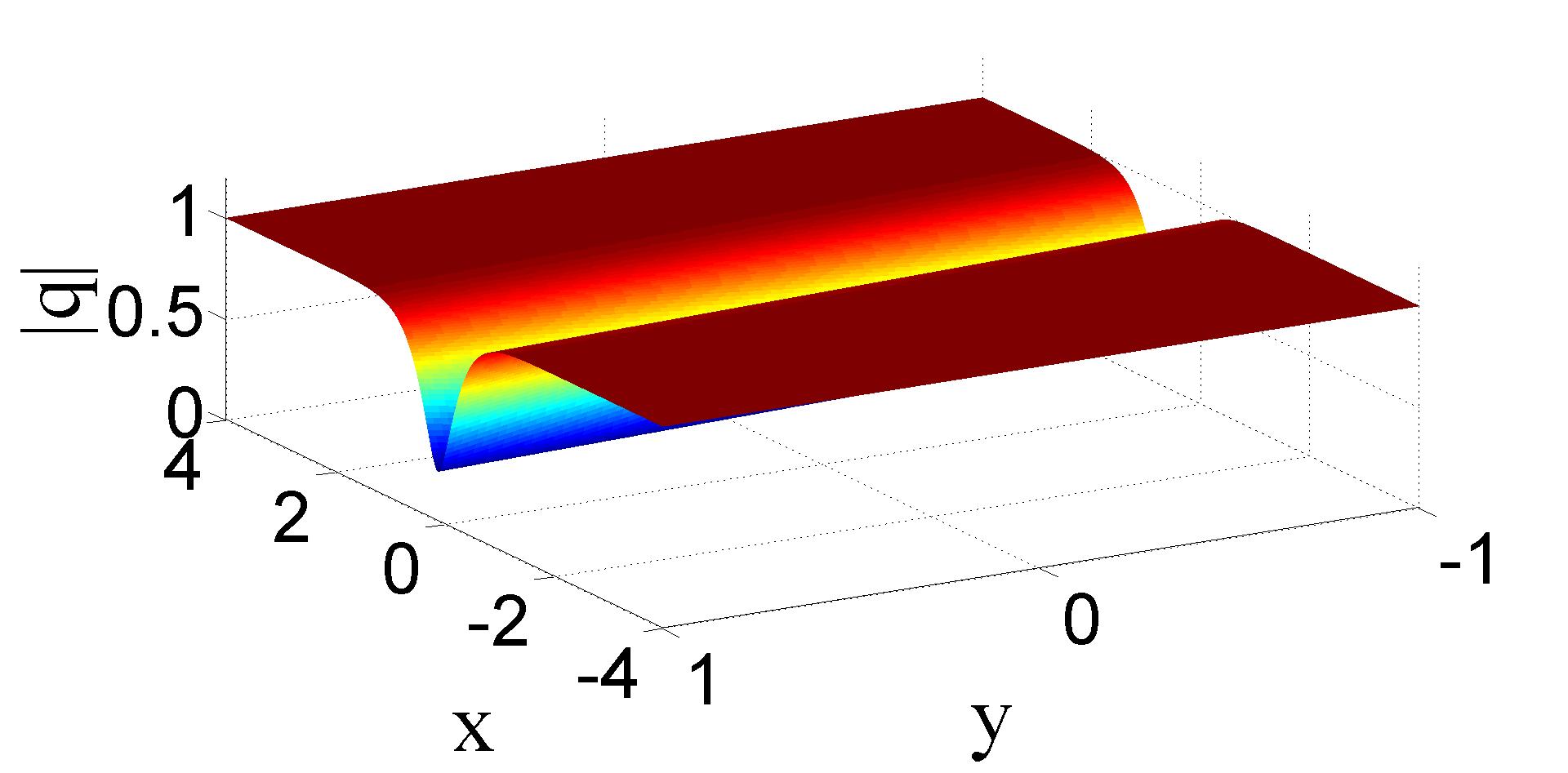}}\qquad\qquad
\subfigure[]{\includegraphics[height=6cm,width=6cm]{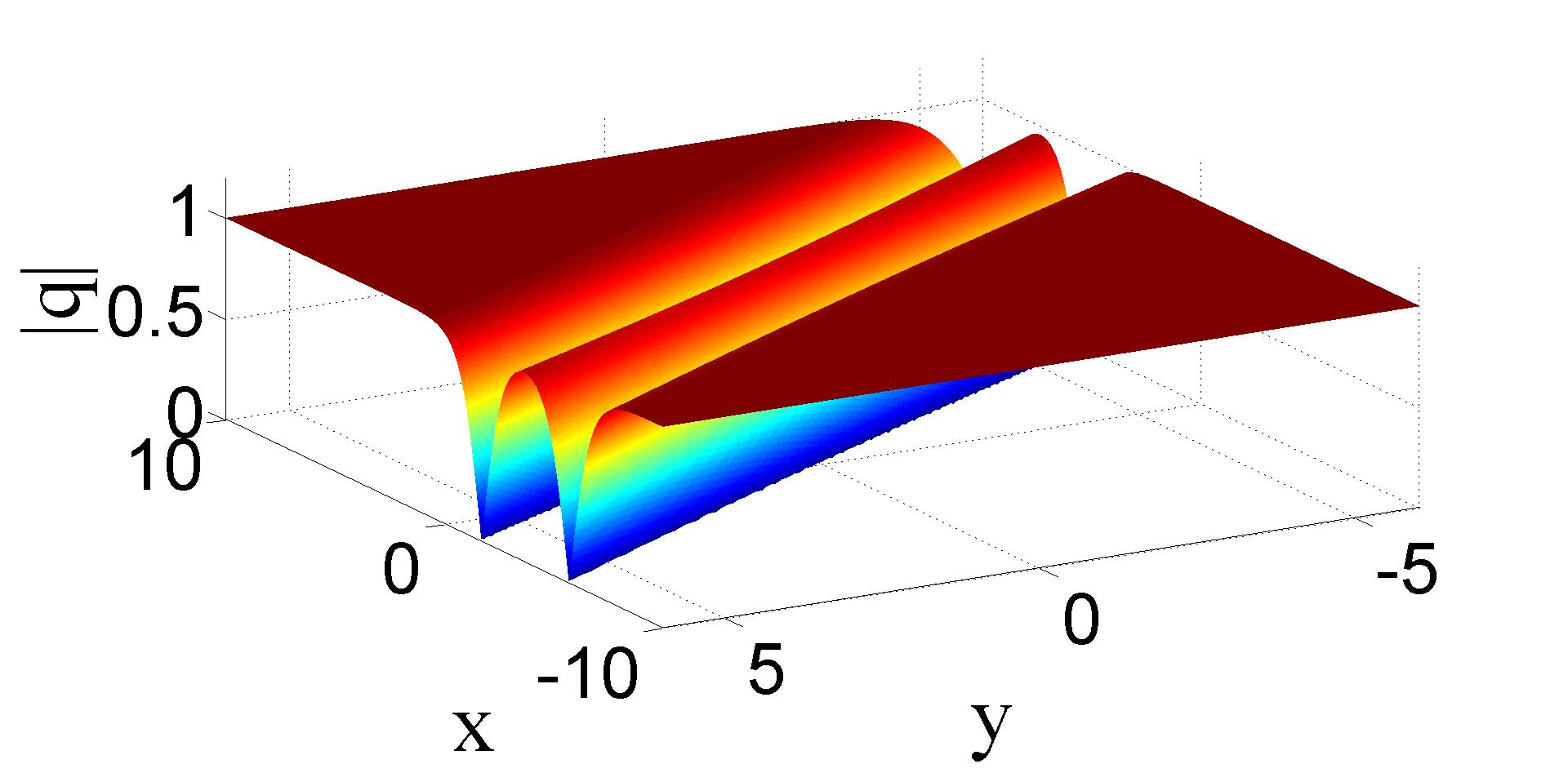}}
\subfigure[]{\includegraphics[height=6cm,width=6cm]{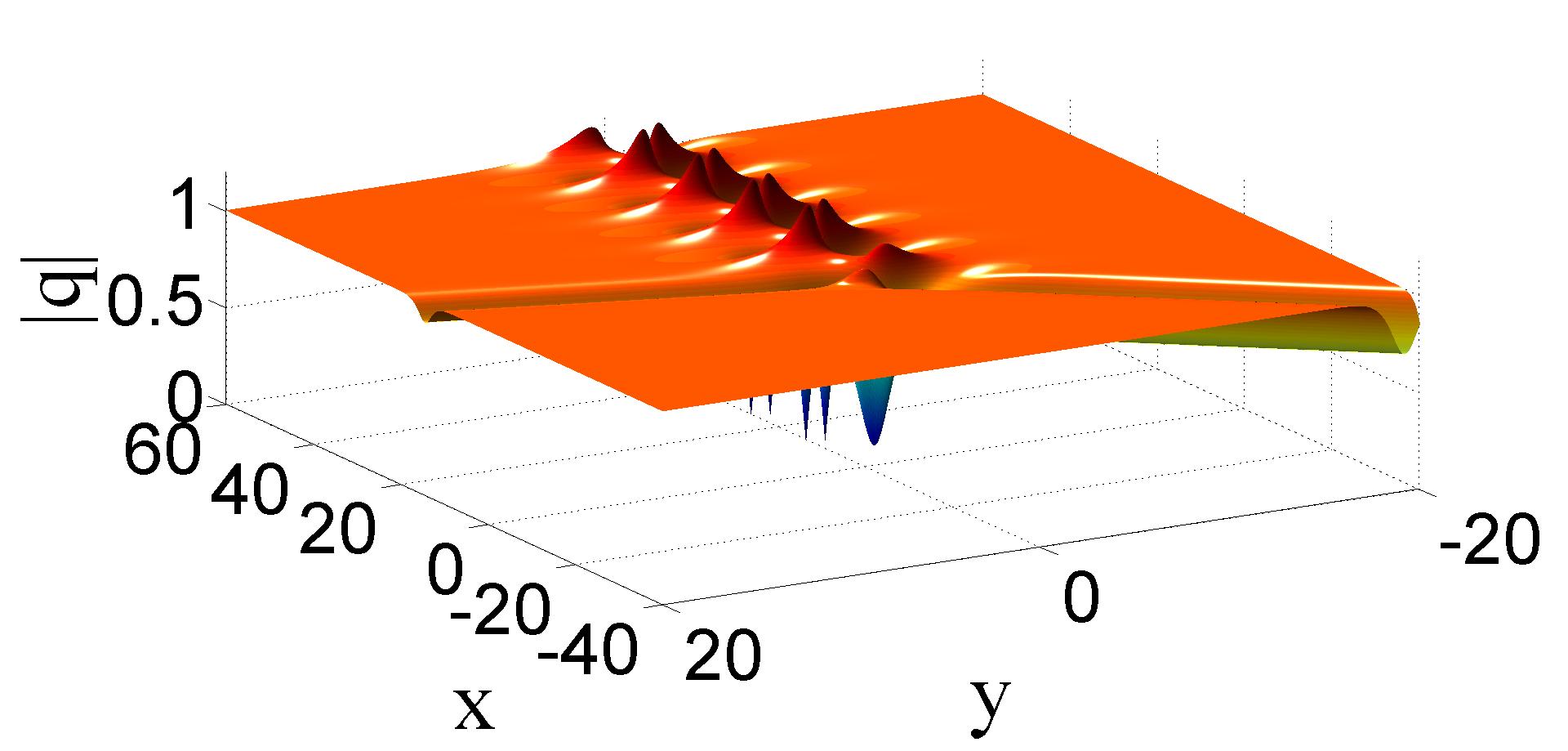}}\qquad\qquad
\subfigure[]{\includegraphics[height=6cm,width=6cm]{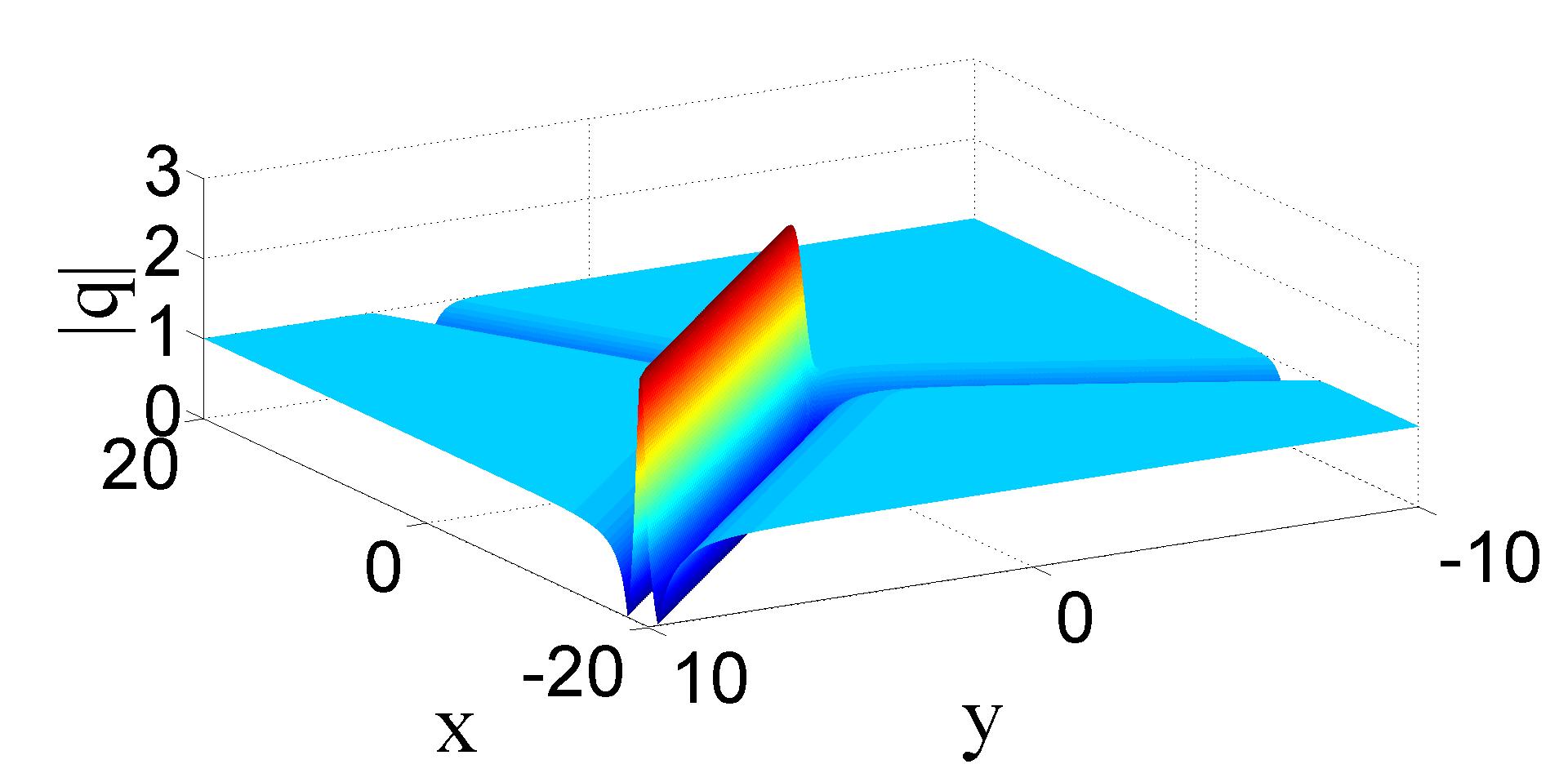}}
\caption{Dynamics of solutions of the DSIII equation defined in equation \eqref{soliton}. (a) One-dark soliton with parameters $N=1,\gamma_{11}=1,\lambda=1,p_{1}=2+i,\xi_{01}=0,t=0$.  (b) A second-order dark soliton with parameters $N=2,\gamma_{11}=1,\gamma_{12}=0,\gamma_{21}=0,\gamma_{22}=1,\lambda=1,p_{1}=1,p_{2}=\frac{9}{10}\,,\xi_{01}=0\,,\xi_{02}=0,t=0$. (c) A hybrid of
dark soliton and a breather with parameters $N=2,\gamma_{ij}=1\,(i,j=1,2),\lambda=1,p_{1}=\frac{1}{2}+i, p_{2}=\frac{1}{2}+\frac{1}{2}i\,,\xi_{01}=0\,,\xi_{02}=0,t=0$. (d) A hybrid of a
dark soliton and a soliton-type solution with parameters $N=2,\gamma_{ij}=1\,(i,j=1,2),\lambda=1,p_{1}=\frac{1}{2}, p_{2}=1\,,\xi_{01}=0\,,\xi_{02}=0,t=0$.~}\label{fig1}
\end{figure}

\noindent\textbf{Remark 1.} Setting $\gamma_{ij}=1$ when $i = j$,  and $\gamma_{ij}=0$ when $i \neq j$, solutions \eqref{soliton} are generting  $N$th-order dark solitons (see Fig.\ref{fig1}(a) and Fig.\ref{fig1}(b) for first two dark solitons.). An explicit form of the first order dark soliton is given in eq.(\ref{1-so}).

\noindent\textbf{Remark 2.} Setting $\gamma_{ij}=1\,,N\geq 2$, and  $p_{i}$ are real parameters, solutions \eqref{soliton} admit  the mixed solution of dark solitons and soliton-type solutions (see Fig.\ref{fig1}(d)).

\noindent\textbf{Remark 3.} Setting $\gamma_{ij}=1\,,N\geq 2$, and  some parameters $p_{i}$ are complex, but $p_{i}\neq\,p_{j}^{*} (i\neq j)$,  solutions \eqref{soliton}  are a hybrid of breathers and solitons (see Fig.\ref{fig1}(c)).

\noindent\textbf{Remark 4.} Setting $\gamma_{ij}=1\,,N\geq 2$, and  some parameters $p_{i}$ are complex and also satisfy $p_{i}=\,p_{j}^{*} (i\neq j)$, the resulting  solutions \eqref{soliton} are a hybrid of line breathers \cite{qianchao}, breathers and dark solitons (see Figs.\ref{fig1-case3}, \ref{fig2} ).
%%%%%%%%%%%%%%%%%%%%%%%%%%%%%%%%%%%%%%%%%%%%%%%%%%%%%%%%%%%%%%%%fig1
%%%%%%%%%%%%%%%%%%%%%%%%%%%%%%%%%%%%%%%%%%%%%%%%%%%%%%%%%%%%%%%%%%%%%%%%%%%%%%%%%%%%%%%%%%%%%%%%%%%%%%%%
%%%%%%%%%%%%%%%%%%%%%%%%%%%%%%%%%%%%%%%%%%%%%%%%%%%%%%%%%%%%%%%%%%%%%%%%%%%%%%%%%%%%%%%%%%%%%%%%%%%%%%%%%%
 %%%%%%%%%%%%%%%%%%%%%%%%%%%%%%%%%%%%%%%%%%%%%%%%%%%%%%%%%%%%%%%%fig2
\begin{figure}[!htbp]
\centering
\subfigure[$t=-10^{4}$]{\includegraphics[height=6cm,width=6cm]{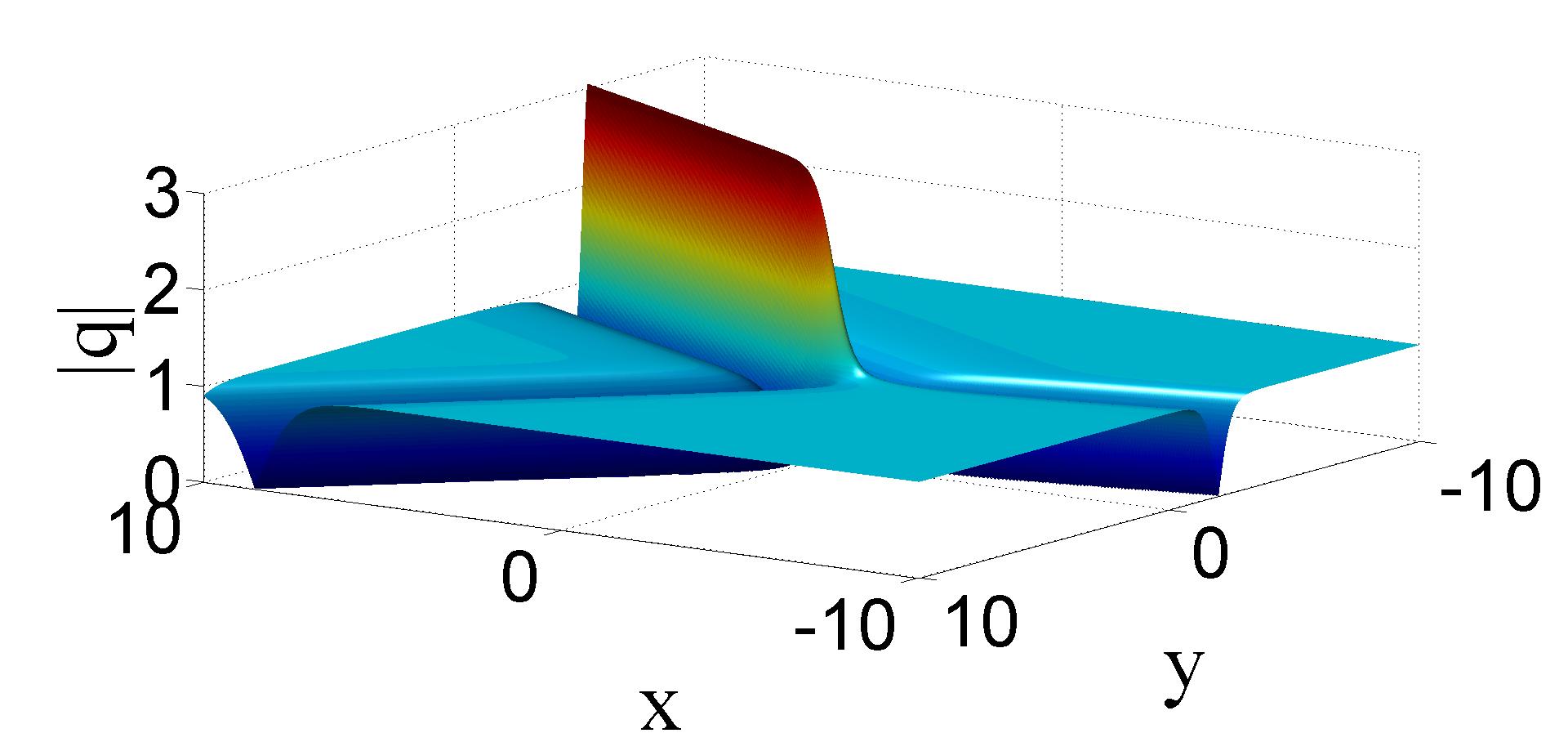}}\qquad\qquad
\subfigure[$t=-10$]{\includegraphics[height=6cm,width=6cm]{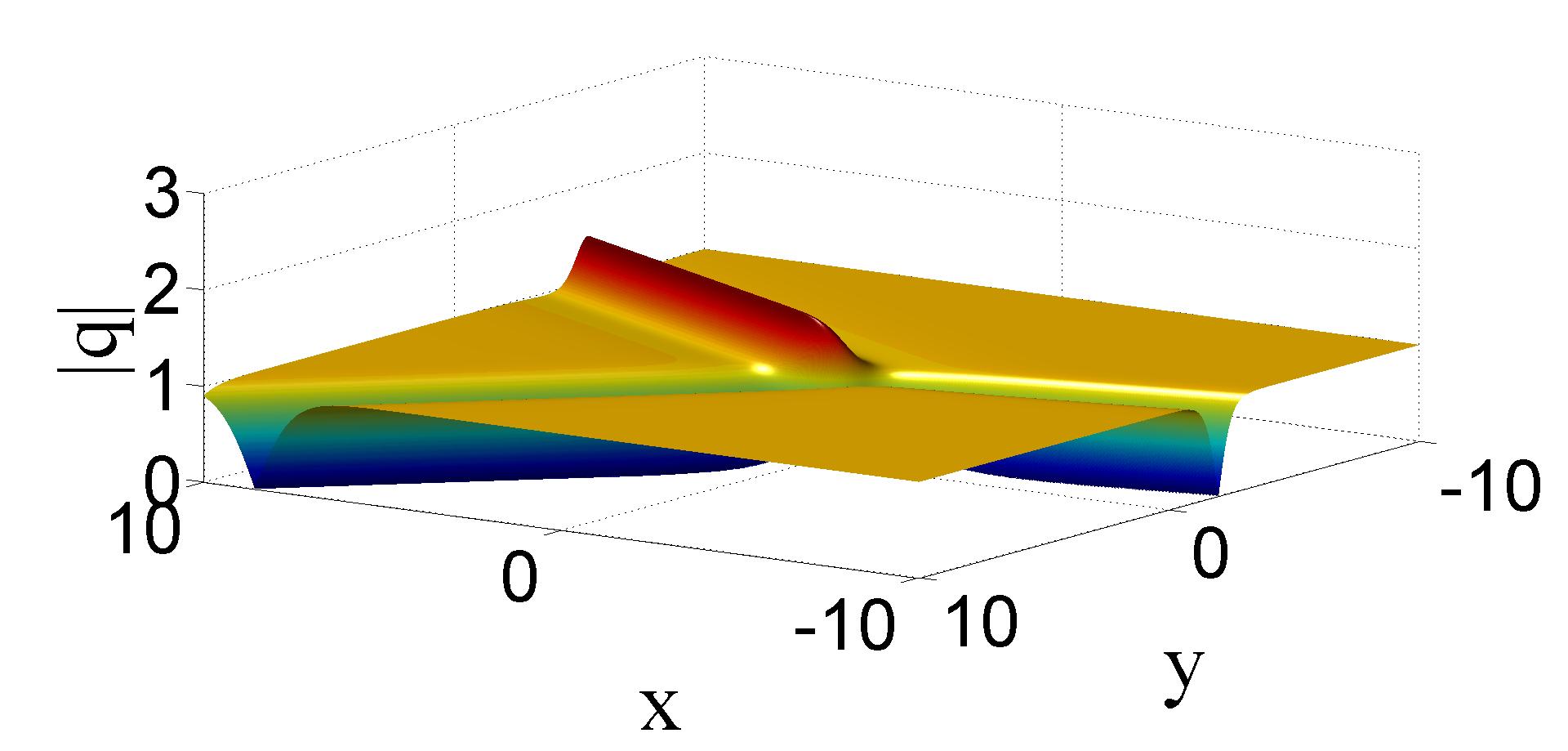}}
\subfigure[$t=10$]{\includegraphics[height=6cm,width=6cm]{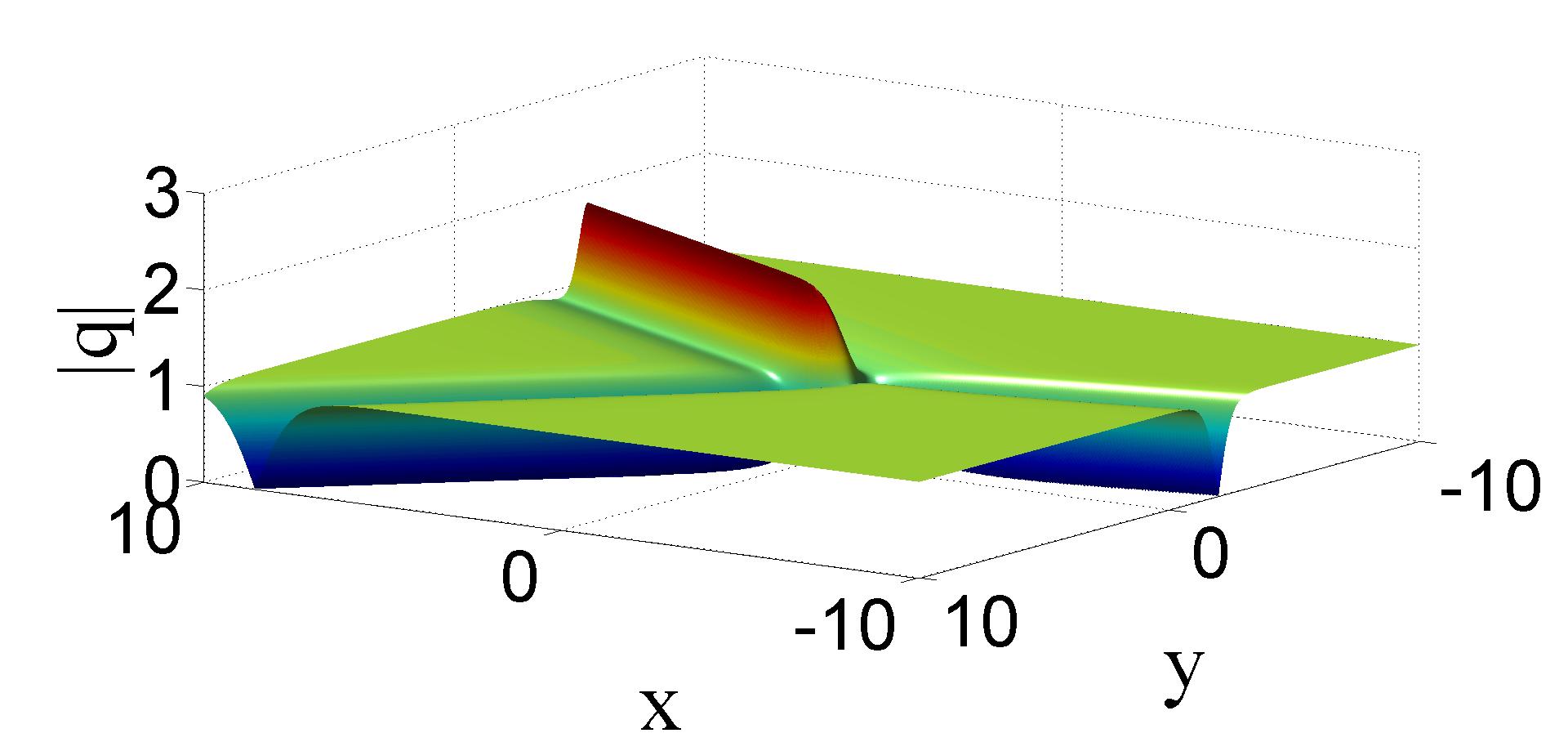}}\qquad\qquad
\subfigure[$t=10^{4}$]{\includegraphics[height=6cm,width=6cm]{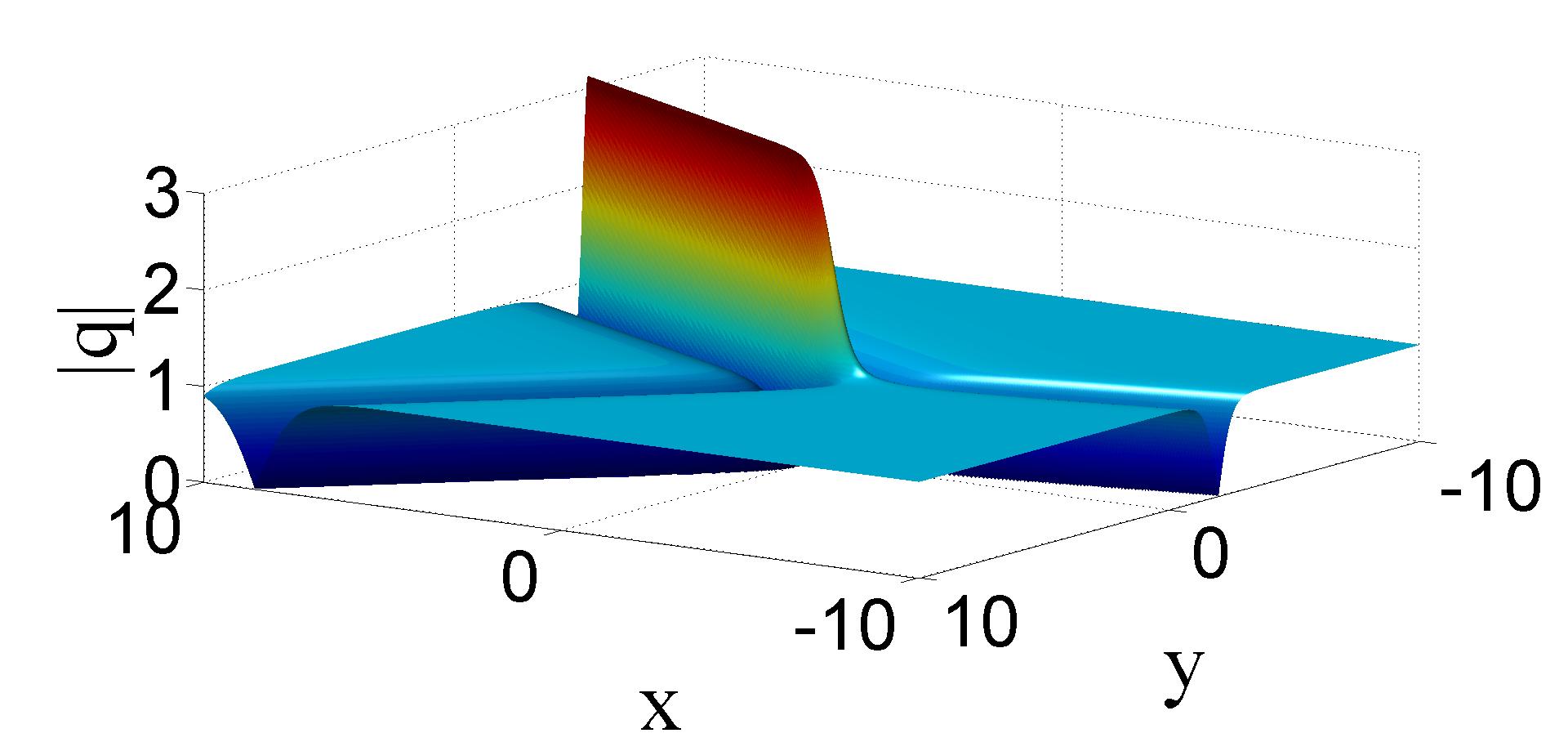}}
\caption{The time evolution of mixed solutions consisting of a dark soliton and soliton-type solution of the DSIII equation given by equation \eqref{soliton} with parameters $N=2,\gamma_{ij}=1\,(i,j=1,2),p_{1}=\frac{1}{2},p_{2}=1,\lambda=-1,\xi_{01}=0,\xi_{02}=0$.~}\label{fig1-2}
\end{figure}
%%%%%%%%%%%%%%%%%%%%%%%%%%%%%%%%%%%%%%%%%%%%%%%%%%%%%%%%%%%%%%%%%%%%%%%%%%%%%%%%%%%%%%%%%%%%%%%%%%%%%%%%%
 %%%%%%%%%%%%%%%%%%%%%%%%%%%%%%%%%%%%%%%%%%%%%%%%%%%%%%%%%%%%%%%%fig2
\begin{figure}[!htbp]
\centering
\subfigure[$t=-5$]{\includegraphics[height=6cm,width=6cm]{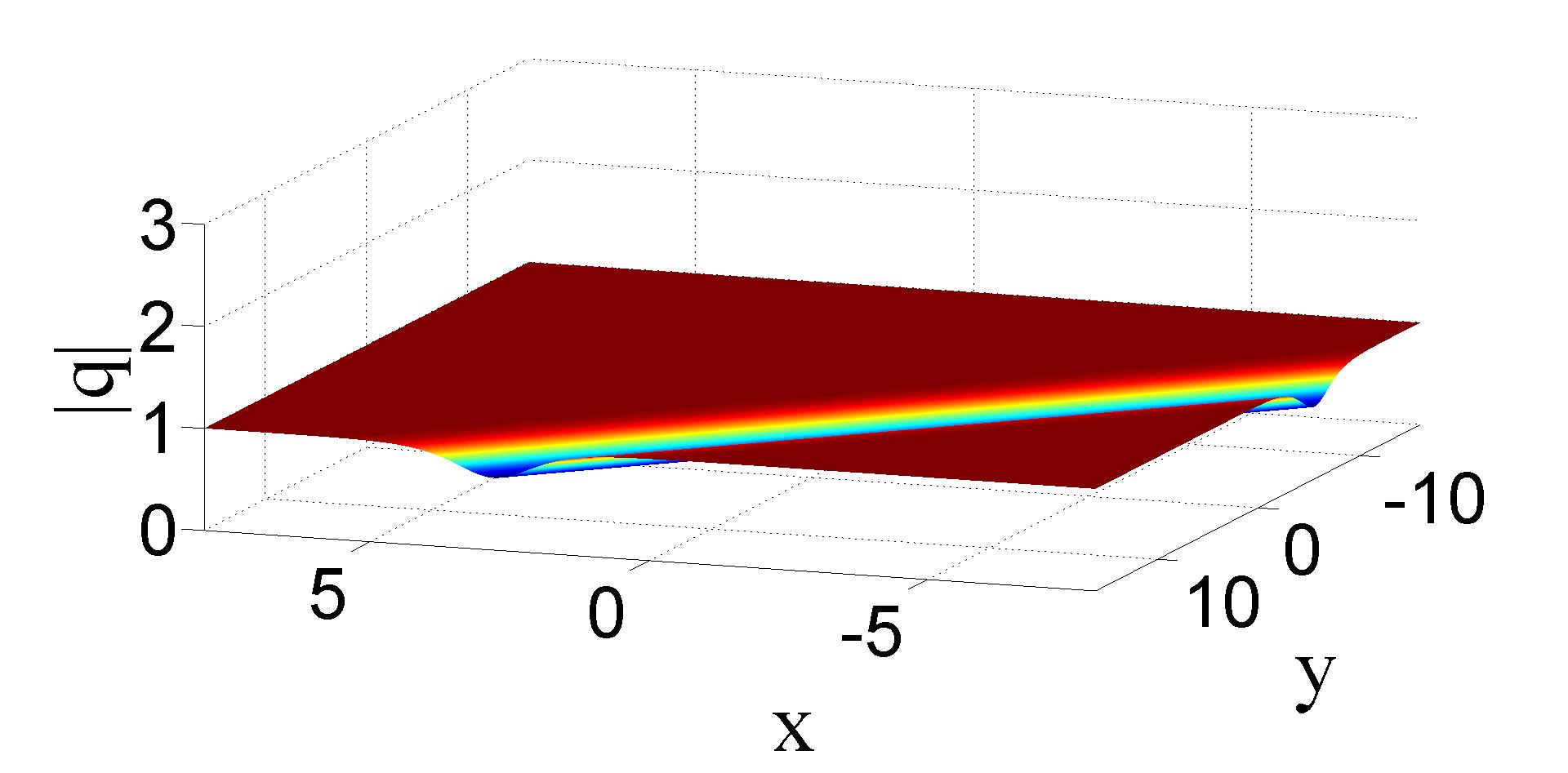}}\qquad\qquad
\subfigure[$t=-1$]{\includegraphics[height=6cm,width=6cm]{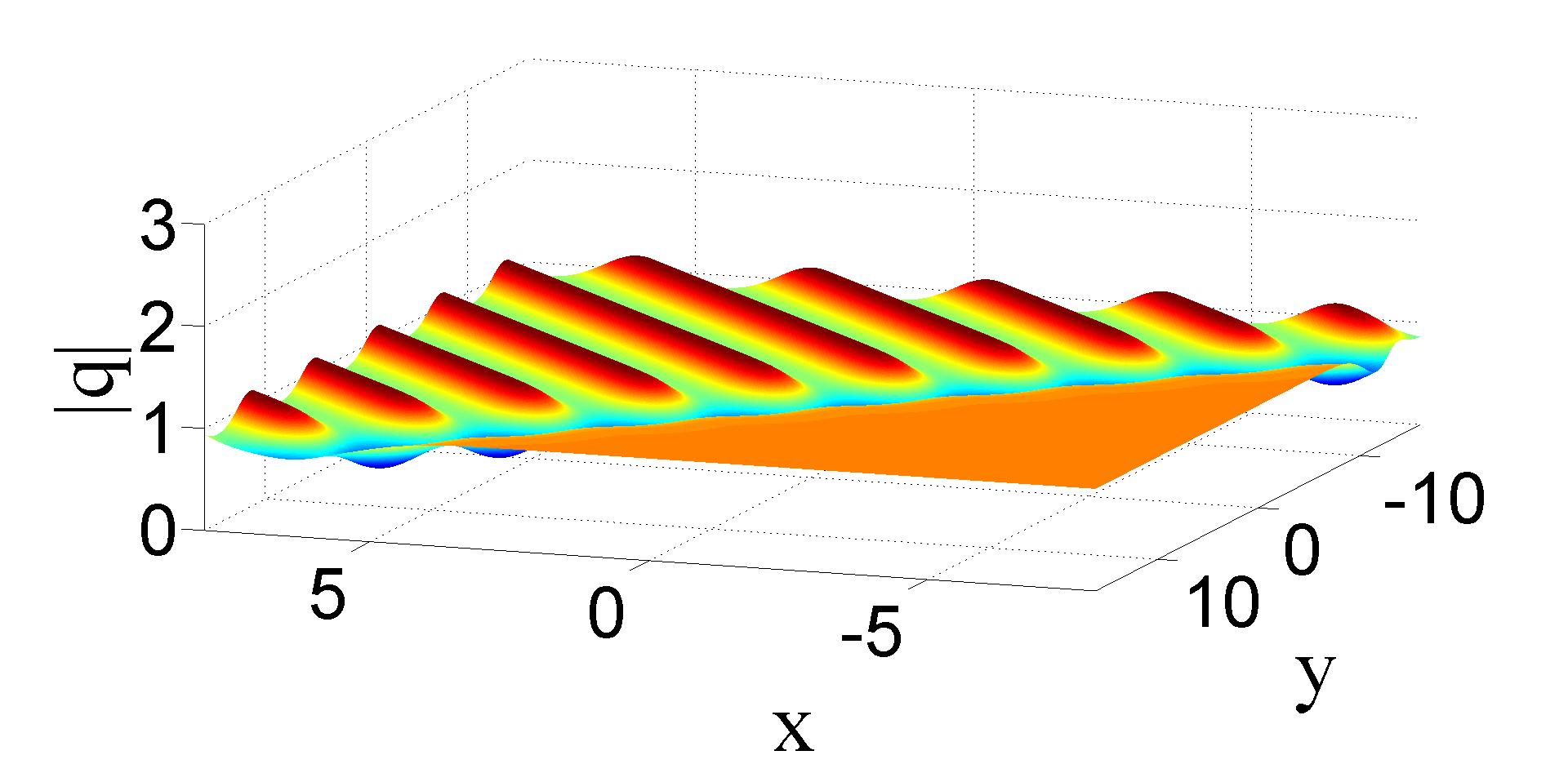}}
\subfigure[$t=0$]{\includegraphics[height=6cm,width=6cm]{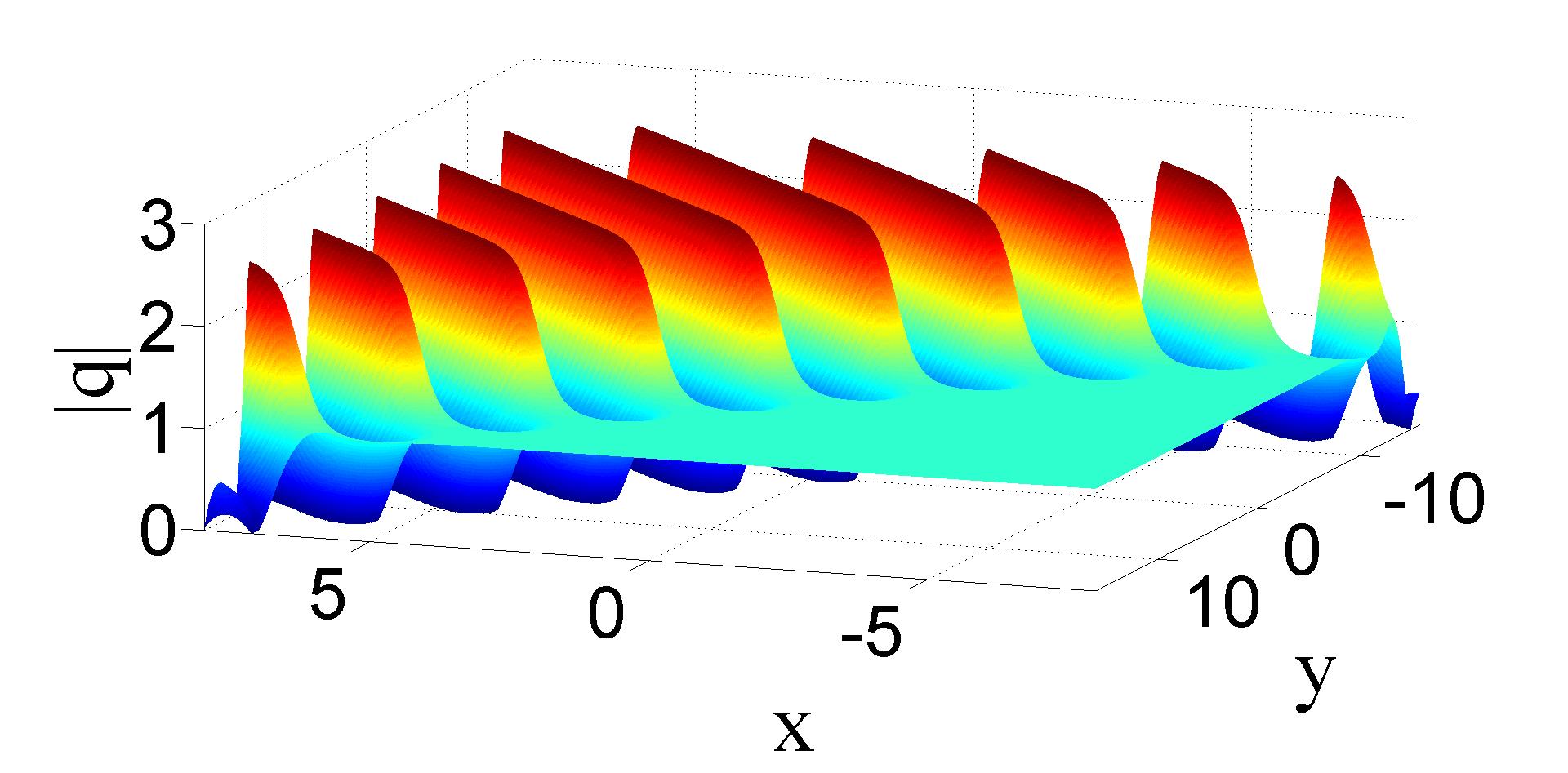}}\qquad\qquad
\subfigure[$t=5$]{\includegraphics[height=6cm,width=6cm]{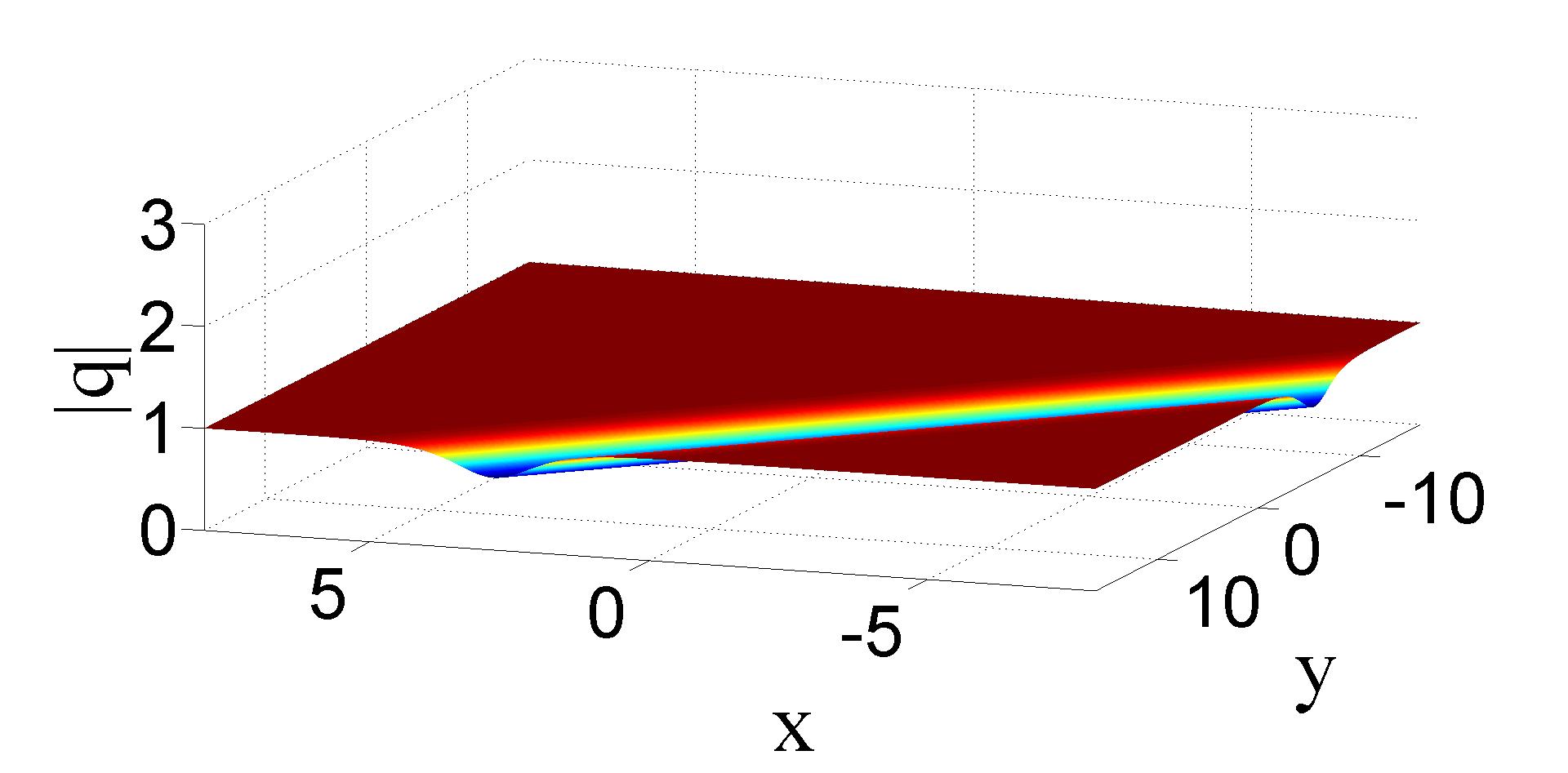}}
\caption{The time evolution of mixed solutions consisting of a dark soliton and periodic line wave of the DSIII equation given by equation \eqref{soliton} with parameters $N=2,\gamma_{ij}=1\,(i,j=1,2),p_{1}=1+i,p_{2}=1-i,\lambda=-1,\xi_{01}=0,\xi_{02}=0$.~}\label{fig1-case3}
\end{figure}

%%%%%%%%%%%%%%%%%%%%%%%%%%%%%%%%%%%%%%%%%%%%%%%%%%%%%%%%%%%%%%%%%%%%%%%%%%%%%%%%%%%%%%%%%%%%%%%%%%%%%%%%%%%%
%%%%%%%%%%%%%%%%%%%%%%%%%%%%%%%%%%%%%%%%%%%%%%%%%%%%%%%%%%%%%%%%%%%%%%%%%%%%%%%%%%%%%%%%%%%%%%%%%%%%%%%%%%%5
 %%%%%%%%%%%%%%%%%%%%%%%%%%%%%%%%%%%%%%%%%%%%%%%%%%%%%%%%%%%%%%%%fig2
\begin{figure}[!htbp]
\centering
\subfigure[$t=-10$]{\includegraphics[height=6cm,width=6cm]{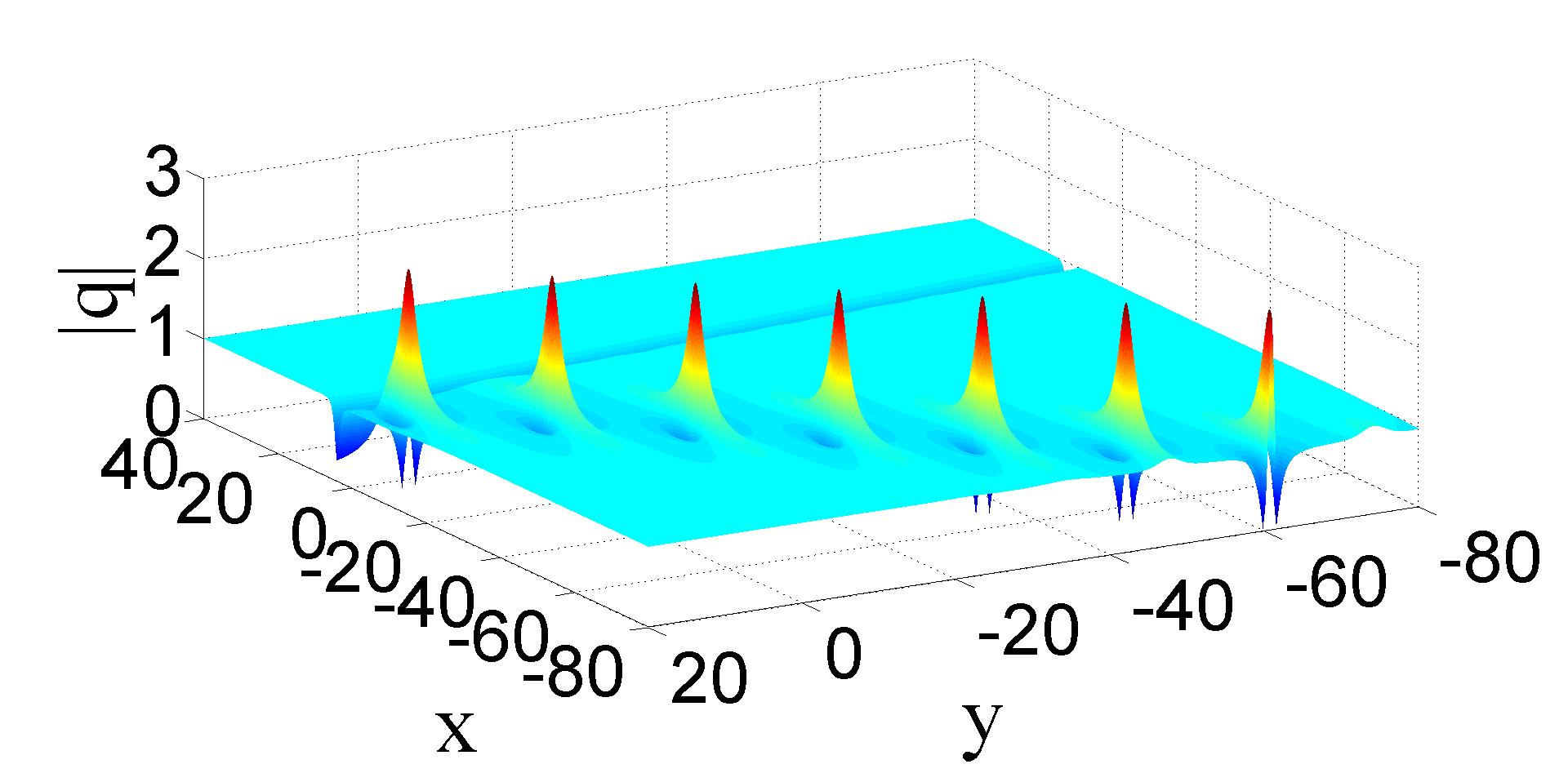}}\qquad\qquad
\subfigure[$t=-1$]{\includegraphics[height=6cm,width=6cm]{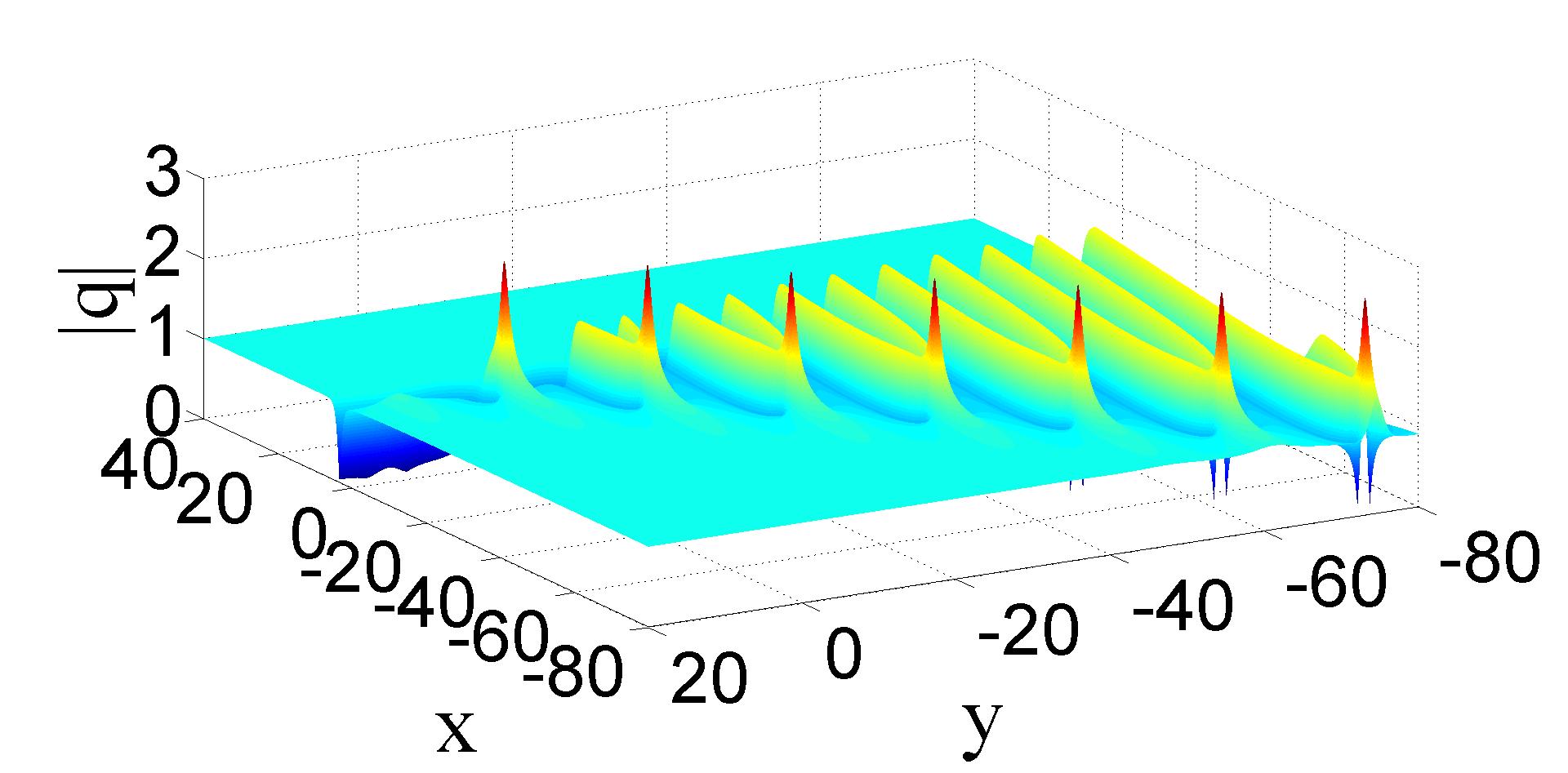}}
\subfigure[$t=0$]{\includegraphics[height=6cm,width=6cm]{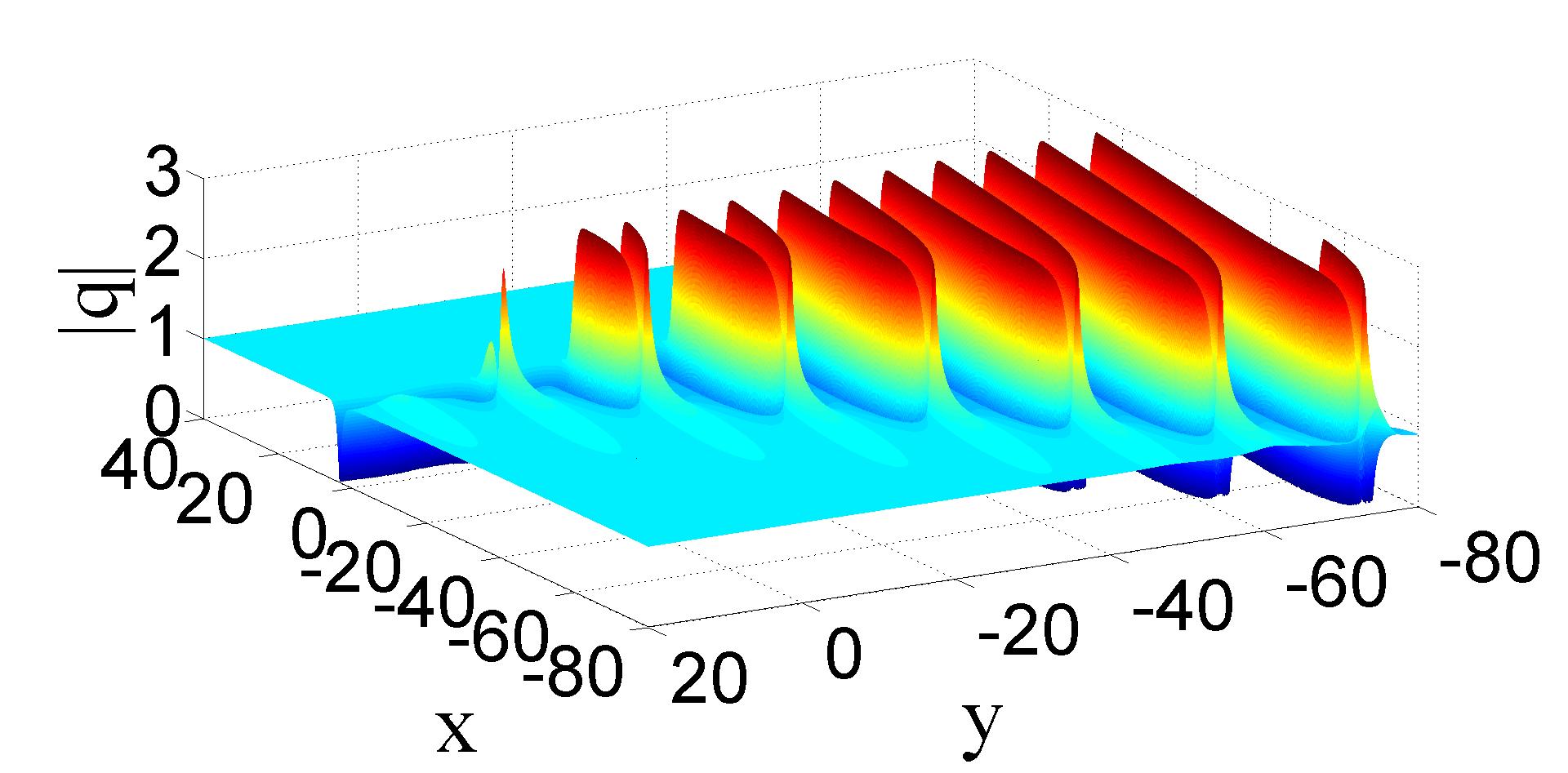}}\qquad\qquad
\subfigure[$t=10$]{\includegraphics[height=6cm,width=6cm]{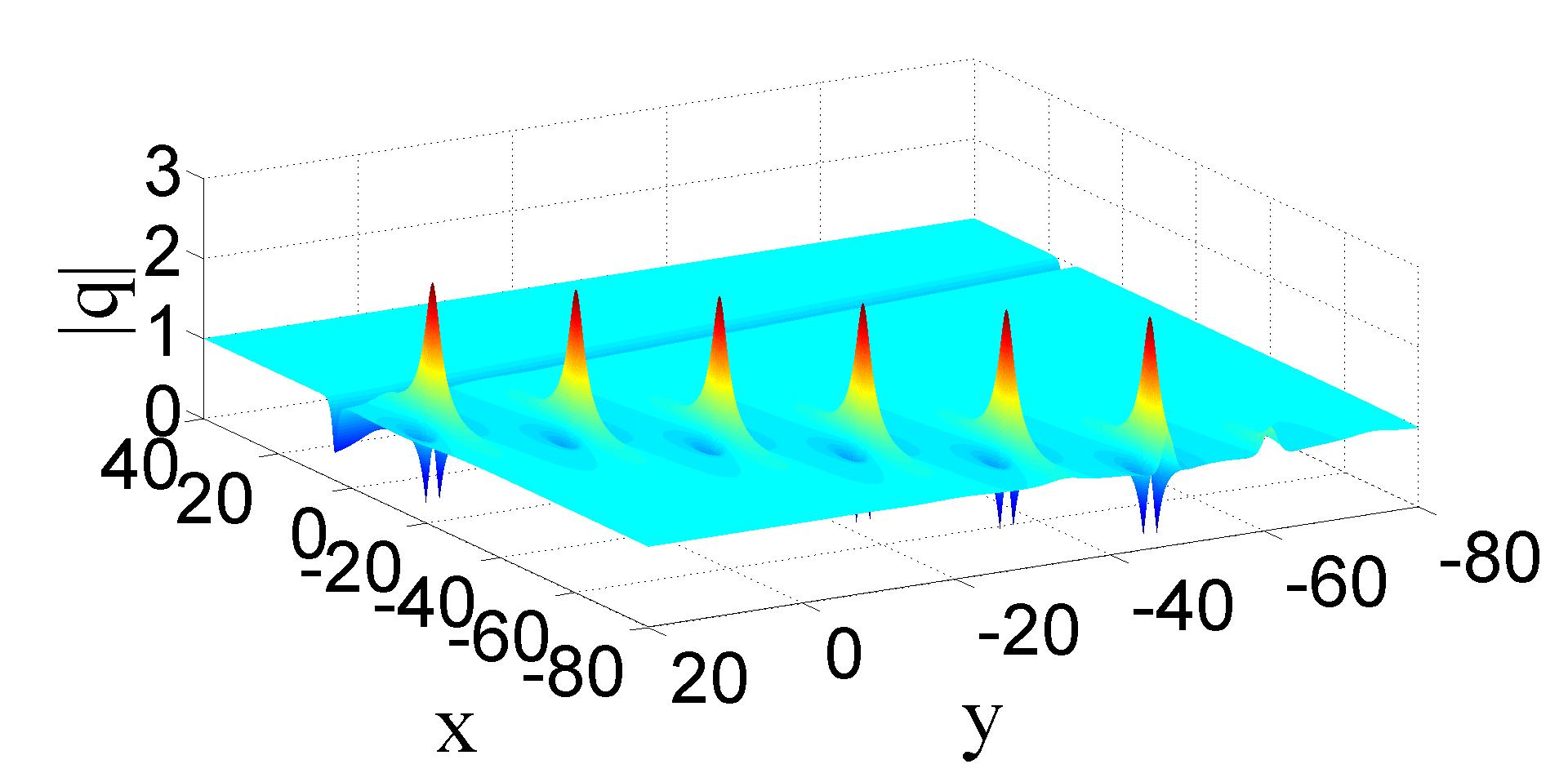}}
\caption{The time evolution of mixed solutions consisting of a breather, a dark soliton and periodic line waves of the DSIII equation given by equation \eqref{soliton} with parameters $N=3,\gamma_{ij}=1\,(i,j=1,2,3),p_{1}=1-\frac{i}{2}\,,p_{2}=1+\frac{i}{2}\,,p_{3}=1\,,\lambda=1,\xi_{01}=0\,,\xi_{02}=0\,,\xi_{03}=0$.~}\label{fig2}
\end{figure}
%%%%%%%%%%%%%%%%%%%%%%%%%%%%%%%%%%%%%%%%%%%%%%%%%%%%%%%%

The fact that these solutions are non-singular in nature which can be proved by the same method as for the non-singularity of solutions in Ref \cite{dark-dark}, thus its detailed  proof is omitted here. Next we provide a short proof for
this theorem based on the lemma below.

\begin{lemma}
{\sl Let $m_{ij}^{(n)}, \psi_{i}^{(n)}$ and $\phi_{j}^{(n)}$ be functions of variables $x_{-1}\,,x_{-2}\,,x_{1}\,,x_{2}$ satisfying the following differential and difference relations,
\begin{equation}\label{eq:m}
\begin{aligned}
\partial_{x_{1}}m_{ij}^{(n)}&=\psi_{i}^{(n)}\phi_{j}^{(n)},\\
\partial_{x_{-1}}m_{ij}^{(n)}&=-\psi_{i}^{(n-1)}\phi_{j}^{(n+1)},\\
\partial_{x_{2}}m_{ij}^{(n)}&=\psi_{i}^{(n+1)}\phi_{j}^{(n)}+\psi_{i}^{(n)}\phi_{j}^{(n-1)},\\
\partial_{x_{-2}}m_{ij}^{(n)}&=-\psi_{i}^{(n-2)}\phi_{j}^{(n+1)}-\psi_{i}^{(n-1)}\phi_{j}^{(n+2)},\\
\partial_{x_{v}}\psi_{i}&=\psi_{i}^{(n+v)},\\
\partial_{x_{v}}\phi_{j}&=-\phi_{j}^{(n-v)}\quad (v=-2,-1,1,2).\\
\end{aligned}
\end{equation}
Then the determinant
\begin{equation}\label{eq:m1}
\begin{aligned}
\tau_{n}=\det\limits_{1\leq i,j\leq N}(m_{ij}^{(n)})
\end{aligned}
\end{equation}
satisfies the following bilinear equations in $KP$ hierarchy
\begin{equation}\label{eq:KP}
\begin{aligned}
(D^{2}_{x_{1}} - D_{x_{2}})\tau_{n+1} \cdot \tau_{n} &=0,\\
(D^{2}_{x_{-1}} + D_{x_{-2}})\tau_{n+1} \cdot \tau_{n} &=0,\\
(D_{x_{1}}D_{x_{-1}}-2)\tau_{n}\cdot\tau_{n}&=-2\tau_{n+1}\tau_{n-1}.
\end{aligned}
\end{equation}
}
\end{lemma}
This Lemma can be proved by a similar way as the proof of the Lemma 1 in Ref. \cite{dark-dark} or the proof of the Lemma 3.2 in Ref \cite{ohta-nls}, thus we omit its proof  in this paper. Next we provide a short proof of  Theorem \ref{th1} based on this Lemma.

\begin{Proof}
{\sl In order to prove Theorem \ref{th1}, functions $m_{ij}^{(n)},\,\psi_{i}^{(n)}$ and $\phi_{j}^{(n)}$ are selected by the following forms
 \begin{equation}\label{psi}
\begin{aligned}
m_{ij}^{(n)}&=\gamma_{ij}+\frac{1}{p_{i}+q_{j}}\psi^{(n)}_{i}\phi^{(n)}_{j},\\
\psi^{(n)}_{i}&=p_{i}^{n}e^{\xi_{i}}\,,\\
\phi^{(n)}_{j}&=(-q_{j})^{-n}e^{\eta_{j}}\,,
\end{aligned}
\end{equation}
where
%\begin{subequations}\label{eq:xe}
%\begin{align}
$\xi_{i}=\frac{1}{p_{i}^{2}}x_{-2}+\frac{1}{p_{i}}x_{-1}+p_{i}x_{1}+p_{i}^{2}x_{2}+\xi_{i0},
\eta_{j}=-\frac{1}{q_{j}^{2}}x_{-2}+\frac{1}{q_{j}}x_{-1}+q_{j}x_{1}-q_{j}^{2}x_{2}+\eta_{j0}$,
%\end{align}
%\end{subequations}
and  $p_{i}\,,q_{j}\,,\xi_{i0}\,,\eta_{j0}$ are arbitrary complex constants.

Furthermore, taking the  parametric  condition
\begin{equation}\label{constraint}
\begin{aligned}
q_{j}=p_{j}^{*},\qquad \eta_{j0}=\xi_{i0}^{*},
\end{aligned}
\end{equation}
and assuming $x_{-1}\,,x_{1}$ are real, $x_{-2}\,,x_{2}$ are pure imaginary, we have
\begin{equation}
\begin{aligned}
\eta_{j}^{'}=\xi_{j}^{'*},\quad m_{ij}^{*}(n)=m_{ji}(-n),\quad \tau_{n}^{*}=\tau_{-n}.
\end{aligned}
\end{equation}
Applying the change of independent variables
$x_{-2}=-i\,t\,,x_{-1}=y\,, x_{1}=\lambda\,x\,,x_{2}=-i\,t\,$,
and setting $\tau_{0}=f\,,\tau_{1}=g\,,\tau_{-1}=g^{*},$ then the bilinear equation \eqref{eq:KP} can be transformed
into the bilinear equation \eqref{bili:DSIII}. Thus solutions of DSIII equation \eqref{DSIII-b} expressed in Theorem 1 can be generated from solutions of bilinear equation \eqref{eq:KP}. Hence the Theorem 1 has been proved. $\Box$
}
\end{Proof}

To get the single soliton solution $q$ of the DSIII equation, we take $N=1$ in formula \eqref{so-fg}. After simple algebra, single soliton solution $q$ can be expressed as
\begin{equation}\label{1-so}
\begin{aligned}
q=\frac{1}{2}[1+y_1+(y_1-1)\tanh(\frac{\xi_{1}+\xi^{*}_{1}+\rho_1}{2})],
\end{aligned}
\end{equation}
where $$\xi_1=\lambda p_1x+\frac{y}{p_1}-i(\frac{1}{p_1^2}+p_1^2)t+\xi_{i0},\,y_{1}=-\frac{p_1}{p^*_{1}},\,e^{\rho_1}=\frac{1}{p_1+p^{*}_{1}}.$$
It is easy to see that $|q|$  approaches to the constant background amplitudes $1$ as $x,y\rightarrow \pm\infty$, see Fig. \ref{fig1}(a). Besides, at any fixed time,  as $x$ varies from $-\infty$ to $+\infty$, the phases of $|q|$ acquires shifts the amount of $2\phi$, where
$$y_1=e^{2i\phi}.$$
At the center of the soliton where $\xi_{1}+\xi_{1}^*+\rho_1=0$, intensity of the single soliton is
$$
|q|_{center}=\cos\phi<1,
$$
here the constant amplitude is $1$. Thus the center intensity is lower than the constant background intensity, which means the single soliton is a dark soliton.

In addition to dark-dark soliton solutions, these solutions also possess other dynamical behaviours.
To analysis the typical behaviours of these mixed solutions, we consider the case of $N=2,\gamma_{ij}=1\,(i,j=1,2)$. In this case, the corresponding solution can be classify into three case:

case 1: $p_{1},p_{2}$ are real.$\\$
In this case, the corresponding solution is a mixture of dark soliton and soliton-type solution. This solution with parameters
\begin{equation}\label{2-w-d}
\begin{aligned}
p_1=\frac{1}{2},p_2=1,\lambda=-1,\xi_{10}=\xi_{20}=0,
\end{aligned}
\end{equation}
is expressed as
\begin{equation}\label{2w-case1}
\begin{aligned}
q={\frac {-{{\rm e}^{4\,y}}-24\,{{\rm e}^{\frac{9}{4}\,it+y+\frac{3}{2}\,x}}-6\,{{\rm e}
^{-\frac{9}{4}\,it+y+\frac{3}{2}\,x}}+18\,{{\rm e}^{2\,x+2\,y}}+9\,{{\rm e}^{x}}}{24\,
{{\rm e}^{y+\frac{3}{2}\,x}}\cos \left( \frac{9}{4}\,t \right) -18\,{{\rm e}^{2\,x+2\,
y}}-{{\rm e}^{4\,y}}-9\,{{\rm e}^{x}}}}.
\end{aligned}
\end{equation}
The time evolution of this solution is shown in Fig. \ref{fig1-2}. It is seen that this solution is composed of a dark soliton and a soliton-type solution. The soliton-type solution behaves as a $W$-shaped soliton and ends  on the dark soliton. However, different from the soliton solution, this soliton-type solution possesses altering amplitudes.

case 2: $p_{1},p_{2}$ are complex, and $p_{1}^{*}\neq p_{2}$.$\\$
The corresponding solutions consist of a dark soliton and a breather, and the breather ends on the dark soliton. This solution with parameters
\begin{equation}\label{2-w-d}
\begin{aligned}
p_1=\frac{1}{2}+i,p_2=\frac{1}{2}+\frac{1}{2}i,\lambda=-1,\xi_{10}=\xi_{20}=0,
\end{aligned}
\end{equation}
is expressed as
\begin{equation}\label{2-w-case2}
\begin{aligned}
q=\frac{g}{f},
\end{aligned}
\end{equation}
where
%%%%%%%%%%%%%%%%%%%%%%%%%%%%%%%%%%%%%%%%%%%
\begin{equation}
\begin{aligned}
f=&5\,{{\rm e}^{-3\,t-x+2\,y}}+{{\rm e}^{-{\frac {57}{25}t}+{\frac {14
\,y}{5}}-2\,x}}+5\,{{\rm e}^{{\frac {18}{25}t}+\frac{4}{5}\,y-x}}- \left( 4-
2\,i \right) {{\rm e}^{-{\frac {57}{50}t}+\frac{7}{5}\,y+\frac{i}{5}y-x-\frac{i}{2}x+{\frac
{123\,i}{100}}t}}\\
&- \left( 4+2\,i \right) {{\rm e}^{-{\frac {57}{50}t
}-{\frac {123\,i}{100}}t+\frac{7}{5}\,y-\frac{i}{5}y-x+\frac{i}{2}x}},\\
g=&-5\,i{{\rm e}^{-3\,t-x+2\,y}}- \left( \frac{4}{5}+\frac{3}{5}\,i \right) {{\rm e}^{-{
\frac {57}{25}t}+{\frac {14\,y}{5}}-2\,x}}+ \left( 3-4\,i \right) {
{\rm e}^{{\frac {18\,}{25}t}+\frac{4}{5}\,y-x}}+\\
& \left( 1+7\,i \right) {
{\rm e}^{-{\frac {57\,t}{50}}+\frac{7}{5}\,y+\frac{i}{5}y-x-\frac{i}{2}x+{\frac {123\,i}{100}}
t}}- \left( 2-2\,i \right) {{\rm e}^{-{\frac {57}{50}t}-{\frac {123
\,i}{100}}t+\frac{7}{5}\,y-\frac{i}{5}y-x+\frac{i}{2}x}}.
\end{aligned}
\end{equation}
This solution is shown in Fig. \ref{fig1} (c).

case 3: $p_{1},p_{2}$ are complex, and $p_{1}^{*}= p_{2}$.$\\$
The corresponding solutions consist of a dark soliton and a line breather, and the line breather also ends on the dark soliton. This solution with parameters
\begin{equation}\label{2-w-d}
\begin{aligned}
p_1=1+i,p_2=1-i,\lambda=-1,\xi_{10}=\xi_{20}=0,
\end{aligned}
\end{equation}
is expressed as
\begin{equation}\label{2-w-case3}
\begin{aligned}
q=\frac{-\left( -2+2\,i \right) {{\rm e}^{-i \left( 2\,x+y \right) }}+ \left( 2
+2\,i \right) {{\rm e}^{i \left( 2\,x+y \right) }}+{{\rm e}^{-2\,x+y}}+4\,i({{\rm e}^{-3\,t
}}-{{\rm e}^{3\,t}})}{\left( -2+2\,i \right) {{\rm e}^{-i \left( 2\,x+y \right) }}- \left(
2+2\,i \right) {{\rm e}^{i \left( 2\,x+y \right) }}+{{\rm e}^{-2\,x+y}
}+4\,({{\rm e}^{-3\,t}}+\,{{\rm e}^{3\,t}})}.
\end{aligned}
\end{equation}
This solution is shown in Fig. \ref{fig1-case3}. As seen that, this solution is a dark soliton when $t\rightarrow\pm \infty$. In the intermediate time, periodic line waves arise from the constant background and then decay back to the same background (see the panels at $t=-1,0$). As line rogue wave can be treated as a limited case of these periodic line wave, hereafter we refer to these periodic line waves as line breather \cite{qianchao}.

  For larger $N$ and different choices of $p_{j}$, these solutions given by equation \eqref{soliton} possess various dynamical behaviours, such as a hybrid of multi-breathers and multi-dark solitons, a hybrid of usual breathers, line breathers and multi-dark solitons. Fig. \ref{fig2} demonstrates one type of these mixed solutions. As can be seen, this solution
 is only the combination of a dark soliton and a breather when $t\rightarrow-\infty$ ( see the $ t=-10$ panel  ). In the intermediate time, a line breather arises from the constant background, thus the solution is made up of one breather, one line breather and a dark soliton (see the $t=0$ panel). At larger time, the line breather disappears again to this background, and the solution is again mixing of a usual breather and a dark soliton (see the $t=10$ panel).
Specifically, the line breather just exists on the region where the dark soliton and the breather cross over each other within a short period of time.

The above discussions just focus on solutions generated by exponential functions, next,  we consider the construction of  rational and semi-rational solutions of the DSIII equation.

\section{Rational and semi-rational solutions in the determinant form}\label{3}

Inspired by the works of Y. Ohta and J. Yang \cite {DSI, DSII}, we construct rational and semi-rational solutions for the DSIII equation by
introducing the following differential operators
\begin{equation}\label{AB}
\begin{aligned}
A_{i}=\sum\limits_{k=0}^{n_{i}} c_{ik}(p_{i}\partial_{p_{i}})^{n_{i}-k}\,,\qquad
B_{j}=\sum\limits_{l=0}^{n_{j}} c_{jl}^{*}(p_{j}^{*}\partial_{p_{j}^{*}})^{n_{j}-l},\nonumber
\end{aligned}
\end{equation}
where $c_{ik}\,,c_{jl}$ are arbitrary complex constants. Besides,
we define functions $M^{(n)}_{ij},\Psi_{i}^{(n)},\Phi^{(n)}_{j}$ as
\begin{equation}\label{ABM}
\begin{aligned}
M^{(n)}_{ij}=&A_{i}B_{j}m_{i,j}^{(n)},\\
\Psi_{i}^{(n)}=&A_{i}\psi^{(n)}_{i},\\
\Phi^{(n)}_{j}=&B_{j}\phi^{(n)}_{j},
\end{aligned}
\end{equation}
then it is easy to find that $M^{(n)}_{ij},\Psi_{i}^{(n)},\Phi^{(n)}_{j}$ also satisfy the differential and difference relations defined in \eqref{eq:m},
\begin{equation}\label{eq:ABM}
\begin{aligned}
\partial_{x_{1}}M_{ij}^{(n)}&=\Psi_{i}^{(n)}\Phi_{j}^{(n)},\\
\partial_{x_{-1}}M_{ij}^{(n)}&=-\Psi_{i}^{(n-1)}\Phi_{j}^{(n+1)},\\
\partial_{x_{2}}M_{ij}^{(n)}&=\Psi_{i}^{(n+1)}\Phi_{j}^{(n)}+\Psi_{i}^{(n)}\Phi_{j}^{(n-1)},\\
\partial_{x_{-2}}M_{ij}^{(n)}&=-\Psi_{i}^{(n-2)}\Phi_{j}^{(n+1)}-\Psi_{i}^{(n-1)}\Phi_{j}^{(n+2)},\\
\partial_{x_{v}}\Psi_{i}&=\Psi_{i}^{(n+v)},\\
\partial_{x_{v}}\Phi_{j}&=-\Phi_{j}^{(n-v)}\quad (v=-2,-1,1,2).\\
\end{aligned}
\end{equation}
Thus the solutions
\begin{equation}\label{ra-f}
\begin{aligned}
f= \det\limits_{1\leq i,j\leq N}(M^{(0)}_{i,j})\,,g= \det\limits_{1\leq i,j\leq N}(M^{(1)}_{i,j}),
\end{aligned}
\end{equation}
still satisfy bilinear equation \eqref{bili:DSIII}.

Through the operator relations
\begin{equation}\label{af}
\begin{aligned}
A_{i}p_{i}^{n}e^{\xi_{i}}&=\sum\limits_{k=0}^{n_{i}} c_{ik}(p_{i}\partial_{p_{i}})^{n_{i}-k}p_{i}^{n}e^{\xi_{i}}=p_{i}^{n}e^{\xi_{i}}\sum\limits_{k=0}^{n_{i}} c_{ik}(p_{i}\partial_{p_{i}}+\xi_{i}^{'}+n)^{n_{i}-k}\,,\\
B_{j}(-p_{j})^{-n}e^{\xi_{j}}&=\sum\limits_{l=0}^{n_{j}} c^{*}_{jl}(p^{*}_{j}\partial_{p^{*}_{j}})^{n_{j}-l}(-p^{*}_{j})^{-n}e^{\xi^{*}_{j}}=(-p^{*}_{j})^{-n}e^{\xi^{*}_{j}}\sum\limits_{l=0}^{n_{j}} c^{*}_{jl}(p^{*}_{j}\partial_{p^{*}_{j}}+\xi_{j}^{'*}-n)^{n_{j}-l}\,,
\end{aligned}
\end{equation}
where
\begin{equation}\label{xi-11}
\begin{aligned}
\xi_{i}^{'}&=\lambda\,p_{i}\,x-\frac{1}{p_{i}}\,y-2i\,(\frac{1}{p{i}^{2}}-p_{i}^{2})\,t,
\end{aligned}
\end{equation}
and $\xi_{i}$ are given by equation \eqref{xi1}.

The matrix element $M_{ij}$ defined in equation \eqref{ra-f} becomes
\begin{equation}\label{rM}
\begin{aligned}
M_{ij}^{(n)}=&A_{i}B_{j}m_{ij}^{(n)}\\
=&(-\frac{p_{i}}{p^{*}_{j}})^{n}e^{\xi_{i}+\xi^{*}_{j}}\sum_{k=0}^{n_{i}}c_{ik}(p_{i}\partial_{p_{i}}+\xi_{i}^{'}+n)^{n_{i}-k}\\
&\times\sum_{l=0}^{n_{j}}c^{*}_{jl}(p_{j}^{*}
\partial_{p_{j}^{*}}+\xi_{j}^{'*}-n)^{n_{j}-l}\frac{1}{p_{i}+p_{j}^{*}}+\gamma_{ij}c_{i\,n_{i}}d_{j\,n_{j}},
\end{aligned}
\end{equation}
hence the semi-rational solutions of bilinear equation \eqref{bili:DSIII} are obtained. Taking $\delta_{ij}=0$ and by the gauge freedom of $f$ and $g$, the rational solutions are generated.

Based on the above results, semi-rational solutions of the DSIII equation can be proposed in the following Theorem.

\begin{theorem}\label{th2}
{\sl  The DSIII equation \eqref{DSIII-b} has rational and semi-rational solutions of the form
\begin{equation}\label{rational}
\begin{aligned}
q=\frac{g}{f},\qquad  V=-\lambda(\log{f})_{yy},\qquad  U=-\lambda(\log{f})_{xx},
\end{aligned}
\end{equation}
with
\begin{equation}\label{BT-1}
\begin{aligned}
f= \det\limits_{1\leq i,j\leq N}(M^{(0)}_{i,j})\,,g= \det\limits_{1\leq i,j\leq N}(M^{(1)}_{i,j}),
\end{aligned}
\end{equation}
and the matrix elements in $f$ and $g$ are defined by
\begin{equation}\label{rMM}
\begin{aligned}
M_{ij}^{(n)}=&(-\frac{p_{i}}{p^{*}_{j}})^{n}e^{\xi_{i}+\xi^{*}_{j}}\sum_{k=0}^{n_{i}}c_{ik}(p_{i}\partial_{p_{i}}+\xi_{i}^{'}+n)^{n_{i}-k}\\
&\times\sum_{l=0}^{n_{j}}c^{*}_{jl}(p_{j}^{*}
\partial_{p_{j}^{*}}+\xi_{j}^{'*}-n)^{n_{j}-l}\frac{1}{p_{i}+p_{j}^{*}}+\gamma_{ij}c_{i\,n_{i}}c^*_{j\,n_{j}},
\end{aligned}
\end{equation}
and $\xi_i\,,\xi_i^{'}$ are given by equations \eqref{xi1} and \eqref{xi-11}\,,
and $i\,,j\,,n_{i}\,,n_{j}$ are  arbitrary positive integers, $c_{ik}\,,c_{jl}\,,p_{i}\,,p_{j}$ are arbitrary complex constants, and  $\gamma_{ij}$ are arbitrary real constants.}
\\
\end{theorem}

\noindent\textbf{Remark 5.} Setting $\gamma_{ij}=0$ in equation \eqref{rMM} and by the gauge freedom of $f$ and $g$ in equation \eqref{rational}, then the semi-rational solutions \eqref{rational} become rational solutions for the DSIII equation. These rational solutions can be classified into two types: rogue waves and lumps.

\noindent\textbf{Remark 6.} The parameters $\gamma_{ij}$ play a key role in the combination of semi-rational solutions \eqref{rational}. When $\gamma_{ii}=1\,,\gamma_{ij}=0\, (i\neq j)$, the obtained  semi-rational solutions consist of rogue waves and dark solitons or lumps and dark solitons. When $\gamma_{ij}=1$, the semi-rational solutions are a combination of rogue waves, breathers and dark solitons.

\noindent\textbf{Remark 7.} By a scaling of $M_{ij}$ given by equation (\ref{rMM}), we can normalize $c_{i0}=1$ without loss of generality, and thus hereafter we set $c_{i0}=1$.

\noindent\textbf{Remark 8.} For any non-zero column vector $\mu=(\mu_{1}\,,\mu_{2}\,...\mu_{N})^{T}$ and $\overline{\mu}$ being its complex transpose, we have
\begin{equation}\label{non}
\begin{aligned}
\overline{\mu} f\mu&=\sum\limits_{i,j=1}^{N}\overline{\mu_{i}} M_{i,j}^{(0)}\mu_{j}\geq\sum\limits_{i,j=1}^{N}\overline{\mu_{i}}\mu_{j}A_{i}B_{j}\frac{1}{p_{i}+q_{j}}e^{\xi_{i}+\eta_{j}}\mid_{q_{j}=p_{j}^*}\\
&=\sum\limits_{i,j=1}^{N}\overline{\mu_{i}}\mu_{j}A_{i}B_{j}\int\limits_{-\infty}^{x} e^{\xi_{i}+\eta_{j}}dx\mid_{q_{j}=p_{j}^*}\\
&=\int\limits_{-\infty}^{x}(\sum\limits_{i,j=1}^{N}\overline{\mu_{i}}\mu_{j}A_{i}B_{j}e^{\xi_{i}+\eta_{j}}\mid_{q_{j}=p_{j}^*})dx\\
&=\int\limits_{-\infty}^{x}|\sum\limits_{i=1}^{N}\overline{\mu_{i}}A_{i}e^{\xi_{i}}|^{2}dx > 0\,,\nonumber
\end{aligned}
\end{equation}
thus we have proved that $f$ is a  positive definitive. Therefore, the solutions $q, U$, and  $V$ given in this Theorem are non-singular in nature.

The above obtained semi-rational solutions possess various other  dynamical behaviours which have never been reported before, such as interactions between rogue waves and dark solitons, interactions between lumps, breathers and dark solitons, and so on. In the section 4, we concentrate on demonstrating these unique dynamics and properties.

\section{Dynamics of rogue waves}\label{4}
In this section, we analyse the dynamics of rational and semi-rational solutions of the DSIII equation, and also illustrate the  dynamics of rogue waves.  We first consider the fundamental (simplest) semi-rational solutions.

\subsection{Dynamics of the fundamental semirational solutions}$\\$

The fundamental semi-rational solutions of the DSIII equation are generated from equation \eqref{rational} by taking $N=1\,,n_{i}=1$. In this case, they are expressed explicitly by the following formulae :
\begin{equation}\label{rational-1}
\begin{aligned}
q=\frac{g}{f},\qquad  V=-\lambda(\log{f})_{yy},\qquad  U=-\lambda(\log{f})_{xx}\,,
\end{aligned}
\end{equation}
where
\begin{equation}\label{rf}
\begin{aligned}
&f=e^{\xi_{1}+\xi_{1}^{*}}\left(\sum_{k=0}^{1}c_{1k}(p_{1}\partial_{p_{1}}+\xi_{1}^{'})^{1-k}\sum_{l=0}^{1}c_{1l}^{*}(p_{1}^{*}\partial_{p_{1}^{*}}+\xi_{1}^{'*})^{1-l} \right)\frac{1}{p_{1}+p_{1}^{*}}+\gamma_{11}c_{11}c_{11}^{*}\\
&=e^{\xi_{1}+\xi_{1}^{*}}(p_{1}\partial_{p_{1}}+\xi^{'}_{1}+c_{11})(p_{1}^{*}\partial_{p^{*}_{1}}+\xi_{1}^{'*}+c^{*}_{11})\frac{1}{p_{1}+p_{1}^{*}}+\gamma_{11}c_{11}c_{11}^{*}\\
&=\frac{e^{\xi_{1}+\xi_{1}^{*}}}{p_{1}+p_{1}^{*}}[(\xi_{1}^{'}-\frac{p_{1}}{p_{1}+p_{1}^{*}}+c_{11})(\xi_{1}^{'*}-\frac{p_{1}^{*}}{p_{1}+p_{1}^{*}}+c_{11}^{*})+\frac{p_{1}p_{1}^{*}}{(p_{1}+p_{1}^{*})^{2}}]+\gamma_{11}c_{11}c_{11}^{*},\\
\end{aligned}
\end{equation}
\\
\begin{equation}\label{rg}
\begin{aligned}
&g=e^{\xi_{1}+\xi_{1}^{*}}\left(\sum_{k=0}^{1}c_{1k}(p_{1}\partial_{p_{1}}+\xi_{1}^{'}+1)\sum_{l=0}^{1}c_{1l}^{*}(p_{1}^{*}\partial_{p_{1}^{*}}+\xi_{1}^{'*}-1\right ) \frac{1} {p_{1}+p_{1}^{*}}+\gamma_{11}c_{11}c_{11}^{*}\\
&=e^{\xi_{1}+\xi_{1}^{*}}(p_{1}\partial_{p_{1}}+\xi^{'}_{1}+c_{11}+1)(p_{1}^{*}\partial_{p^{*}_{1}}+\xi_{1}^{'*}+c^{*}_{11}-1)\frac{1}{p_{1}+p_{1}^{*}}+\gamma_{11}c_{11}c_{11}^{*}\\
&=\frac{e^{\xi_{1}+\xi_{1}^{*}}}{p_{1}+p_{1}^{*}}[(\xi_{1}^{'}-\frac{p_{1}}{p_{1}+p_{1}^{*}}+c_{11}+1)(\xi_{1}^{'*}-\frac{p_{1}^{*}}{p_{1}+p_{1}^{*}}+c_{11}^{*}-1)+\frac{p_{1}p_{1}^{*}}{(p_{1}+p_{1}^{*})^{2}}]+\gamma_{11}c_{11}c_{11}^{*},\\
\end{aligned}
\end{equation}
and
\begin{equation}\label{xi11}
\begin{aligned}
\xi_{1}&=\lambda\,p_{1}\,x+\frac{1}{p_{1}}\,y-i\,(\frac{1}{p_{1}^{2}}+p_{1}^{2})\,t+\xi_{10}\,,\\
\xi_{i}^{'}&=\lambda\,p_{1}\,x-\frac{1}{p_{1}}\,y-2i\,(\frac{1}{p{1}^{2}}-p_{1}^{2})\,t\,,\nonumber
\end{aligned}
\end{equation}
$p_{1}\,,\xi_{10}$ and $c_{11}$ are arbitrary complex parameters. Without loss of generality,
we set $p_{1}=p_{1R}+i\,p_{1I}\,,c_{11}=c_{1R}+i\,c_{1I}$ and then rewrite the above functions $f$ and $g$ as
\begin{equation}\label{rf-f}
\begin{aligned}
f=&\frac{e^{\xi_{1}+\xi^{*}_{1}}}{2p_{1R}}(\theta\,\theta^{*}+\theta_{0})+\gamma_{11}(c_{1R}^{2}+c_{1I}^{2}), \\
g=&\frac{e^{\xi_{1}+\xi^{*}_{1}}}{2p_{1R}}[(\theta+1)\,(\theta^{*}-1)+\theta_{0}]+\gamma_{11}(c_{1R}^{2}+c_{1I}^{2}),
\end{aligned}
\end{equation}
where $$\theta=l_{1}+i\,l_{2}\,,\theta_{0}=\frac{p_{1R}^{2}+p_{1I}^{2}}{4p_{1R}^{2}},$$
\begin{equation}
\begin{aligned}
l_{1}=&\lambda\,p_{1R}\,x+\frac{p_{1R}}{p_{1R}^{2}+p_{1I}^{2}}y-(\frac{4p_{1R}p_{1I}}{(p_{1R}^{2}+p_{1I}^{2})^{2}}+2p_{1R}p_{1I})t+c_{1R}+\frac{1}{2},\\
l_{2}=&\lambda\,p_{1I}\,x-\frac{p_{1I}}{p_{1R}^{2}+p_{1I}^{2}}y-[\frac{2(p_{1R}^{2}-p_{1I}^{2})}{(p_{1R}^{2}+p_{1I}^{2})^{2}}-2(p_{1R}^{2}-p_{1I}^{2})]t+c_{1I}-\frac{p_{1I}}{2p_{1R}}.\nonumber
\end{aligned}
\end{equation}
Thus the final expression of the semi-rational solutions is obtained as
\begin{equation}\label{final-1}
\begin{aligned}
q=&1-\frac{2i\,l_{2}+1}{l_{1}^{2}+l_{2}^{2}+\theta_{0}+2p_{1R}\gamma_{11}(c_{1R}^{2}+c_{1I}^{2})e^{-\xi_{1}-\xi_{1}^{*}}}\,,\\
V=&-\lambda \frac{4p_{1R}\delta_{11}(c_{1R}^{2}+c_{1I}^{2})e^{-\xi_{1}-\xi_{1}^{*}}A+B}{[l_{1}^2+l_{2}^{2}+\theta_{0}+2p_{1R}\gamma_{11}(c_{1R}^{2}+c_{1I}^{2})e^{-\xi_{1}-\xi_{1}^{*}}]^{2}}\,,\\
U=&-\lambda\frac{4p_{1R}\delta_{11}(c_{1R}^{2}+c_{1I}^{2})e^{-\xi_{1}-\xi_{1}^{*}}C+D}{[l_{1}^2+l_{2}^{2}+\theta_{0}+2p_{1R}\gamma_{11}(c_{1R}^{2}+c_{1I}^{2})e^{-\xi_{1}-\xi_{1}^{*}}]^{2}}\,,\\
\end{aligned}
\end{equation}
where
 \begin{equation}
\begin{aligned}
A=&\frac{4p_{1R}^{2}}{(p_{1R}^{2}+p_{1I}^{2})^{2}}(2l_{1}^{2}+2l_{2}^{2}+2\theta_{0}-4l_{1}+1)+\frac{4p_{1R}p_{1I}}{(p_{1R}^{2}+p_{1I}^{2})^{2}}+\frac{p_{1I}^{2}}{(p_{1R}^{2}+p_{1I}^{2})^{2}}\,,\nonumber\\
B=&\frac{2p_{1R}^{2}}{(p_{1R}^{2}+p_{1I}^{2})^{2}}(-l_{1}^{2}+l_{2}^{2}+\theta_{0})-\frac{8p_{1R}p_{1I}}{(p_{1R}^{2}+p_{1I}^{2})^{2}}l_{1}l_{2}+\frac{2p_{1I}^{2}}{(p_{1R}^{2}+p_{1I}^{2})^{2}}(l_{1}^{2}-l_{2}^{2}+\theta_{0})\,,\\
C=&(2l_{1}^{2}+2l_{2}^{2}+2\theta_{0}+4\lambda_{1}l_{1}+1)p_{1R}^{2}+4\lambda\,p_{1R}\,p_{1I}l_{2}+2\lambda^{2}(p_{1R}^{2}+p_{1I}^{2})\,,\\
D=&2p_{1R}^{2}(-1_{1}^{2}+l_{2}^{2}+\theta_{0})-8l_{1}l_{2}p_{1R}p_{1I}+2p_{1I}^{2}(l_{1}^{2}-l_{2}^{2}+\theta_{0}).
\end{aligned}
\end{equation}
In particular, when $\gamma_{11}=0$, these semi-rational solutions become rational solutions for the DSIII equation, which possess two different dynamical behaviors:

(i) Lump solution. When $p_{1I}\neq0$, i.e., $p_{1}$ is complex, one can see that $q, U$, and  $V$ are constants along the trajectory where
 \begin{equation}\label{tra}
\begin{aligned}
&\lambda\,p_{1R}\,x+\frac{p_{1R}}{p_{1R}^{2}+p_{1I}^{2}}y-(\frac{4p_{1R}p_{1I}}{(p_{1R}^{2}+p_{1I}^{2})^{2}}+2p_{1R}p_{1I})t=0\,,\\
&\lambda\,p_{1I}\,x-\frac{p_{1I}}{p_{1R}^{2}+p_{1I}^{2}}y-[\frac{2(p_{1R}^{2}-p_{1I}^{2})}{(p_{1R}^{2}+p_{1I}^{2})^{2}}-2(p_{1R}^{2}-p_{1I}^{2})]t=0\,.
\end{aligned}
\end{equation}
Indeed, at any given time, $(q,U,V)\rightarrow (1\,,0\,,0)$ when $(x,y)\rightarrow \infty$, thus these solutions are permanent lumps moving on the constant background.

\begin{figure}[!htbp]
\centering
\subfigure[]{\includegraphics[height=5.0cm,width=5.0cm]{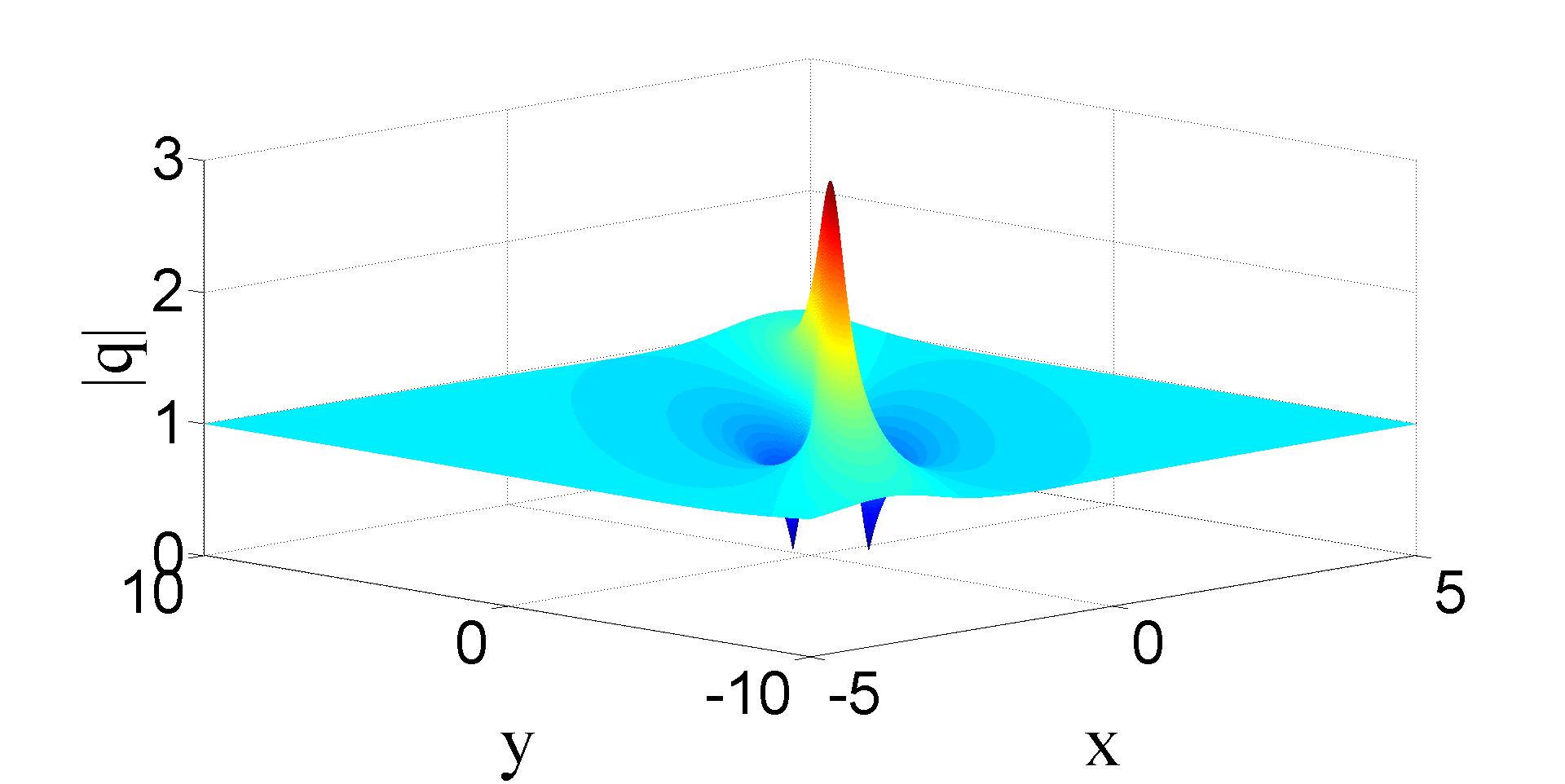}}\qquad
\subfigure[]{\includegraphics[height=5.0cm,width=5.0cm]{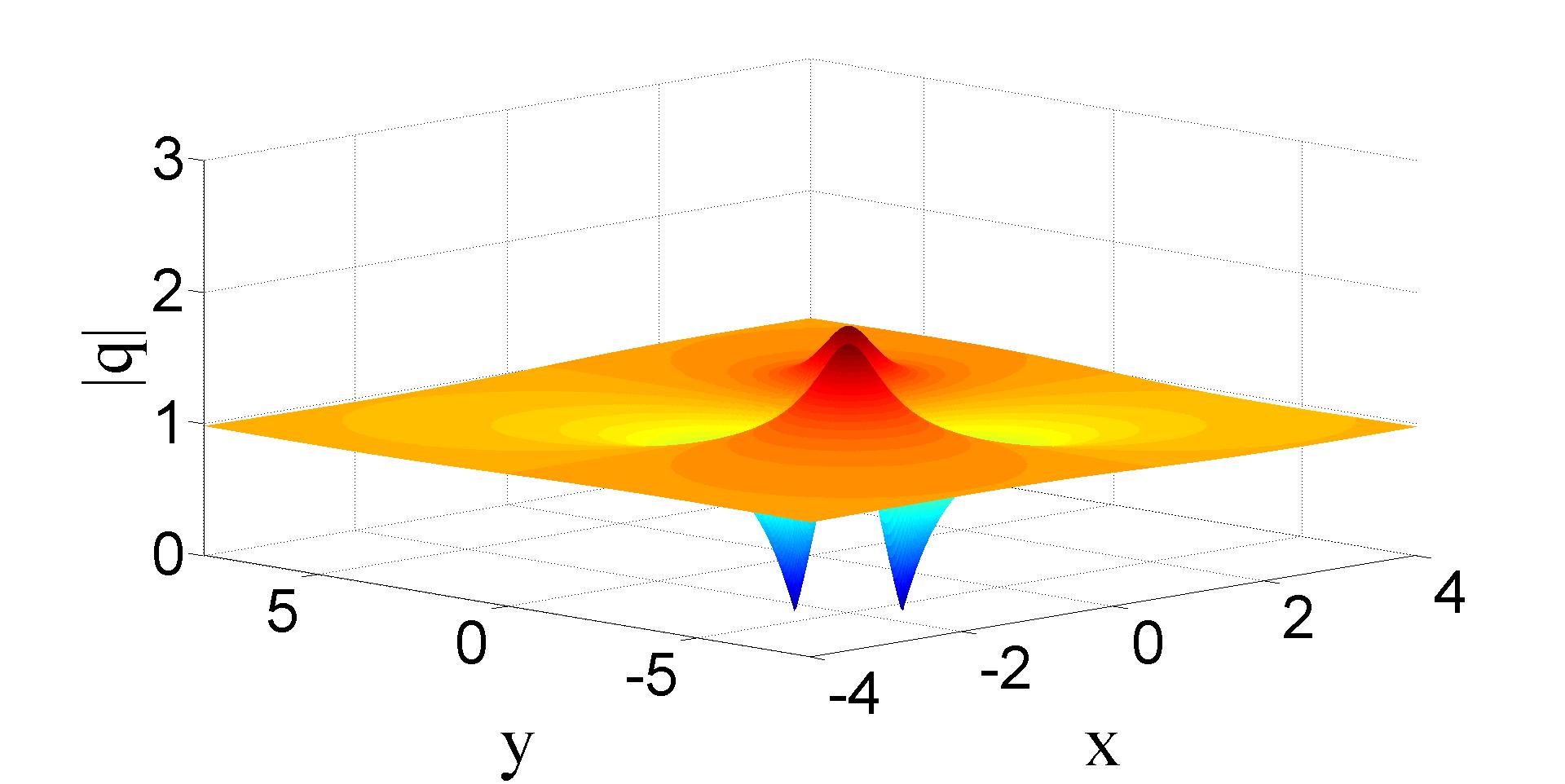}}\qquad
\subfigure[]{\includegraphics[height=5.0cm,width=5.0cm]{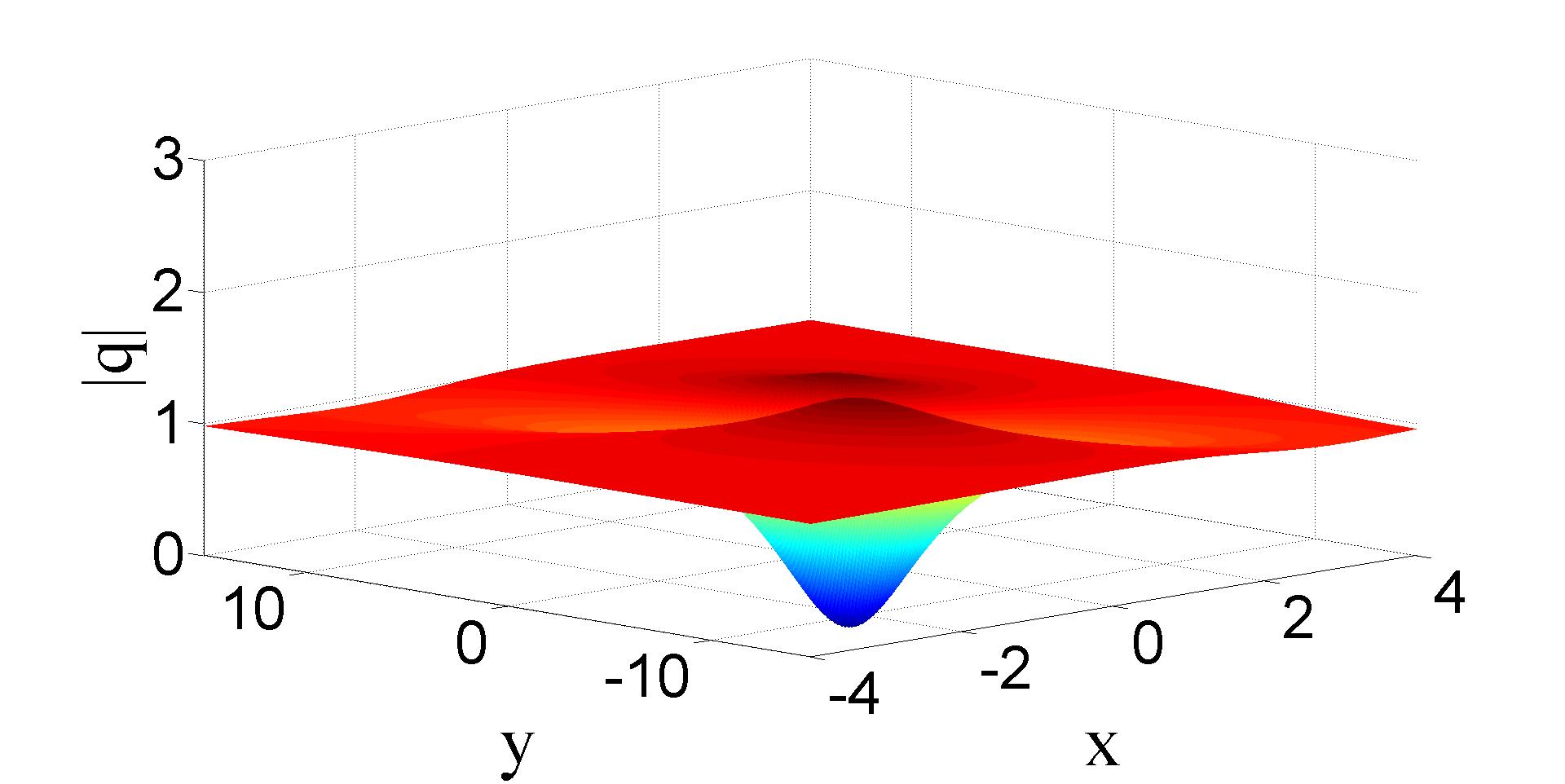}}
\caption{Three different patterns of fundamental lumps of the DSIII equation given by equation \eqref{final-1}  with parameters $\gamma_{11}=0\,,\lambda=1\,,c_{11}=0$: (a) A bright lump with parameters $p_{1}=2+i$; (b) A bi-model lump with parameters $p_{1}=1+i$\,;(c) A dark lump with parameters $p_{1}=1+2i$.~}\label{fig3}
\end{figure}

Based on the analysis of critical points for rational solutions \eqref{final-1}, these lumps can be classified into three patterns:

(a) Bright lump: when $0 <p_{1I}^{2}\leq\frac{1}{3}p_{1R}^{2}$, one upward hump  and two downward humps  (i.e, $q$ has one global maximum point and two global minimum points), see Fig. \ref{fig3}(a) ;

(b) Bi-model lump: when $ \frac{1}{3}p_{1R}^{2}< p_{1I}^{2}< 3p_{1R}^{2}$, two upward humps  and two downward humps  (i.e, $q$ has two global maximum points and  two global minimum points), see Fig. \ref{fig3}(b) ;

(c) Dark lump: when $  p_{1I}^{2}\geq 3p_{1R}^{2}$, two upward humps  and one downward hump (i.e, $q$ has one global maximum point and two global minimum points), see Fig. \ref{fig3}(c). \\

Three patterns of the first-order lump are plotted in Fig. \ref{fig3}. This classification is also suitable for higher order lump solutions. \\
 %%%%%%%%%%%%%%%%%%%%%%%%%%%%%%%%%%%%%%%%%%%%%%%%%%%%%%%%%%%%%%%%fig3
%%%%%%%%%%%%%%%%%%%%%%%%%%%%%%%%%%%%%%%%%%%%%%%%%%%%%%%%%%%%%%%%%%%%%%%%%%%%%%%%%%%%%%%%%%%%%%%%%%%%%%%%

\begin{figure}[!htbp]
\centering
\subfigure[$t=-10$]{\includegraphics[height=6.0cm,width=7.5cm]{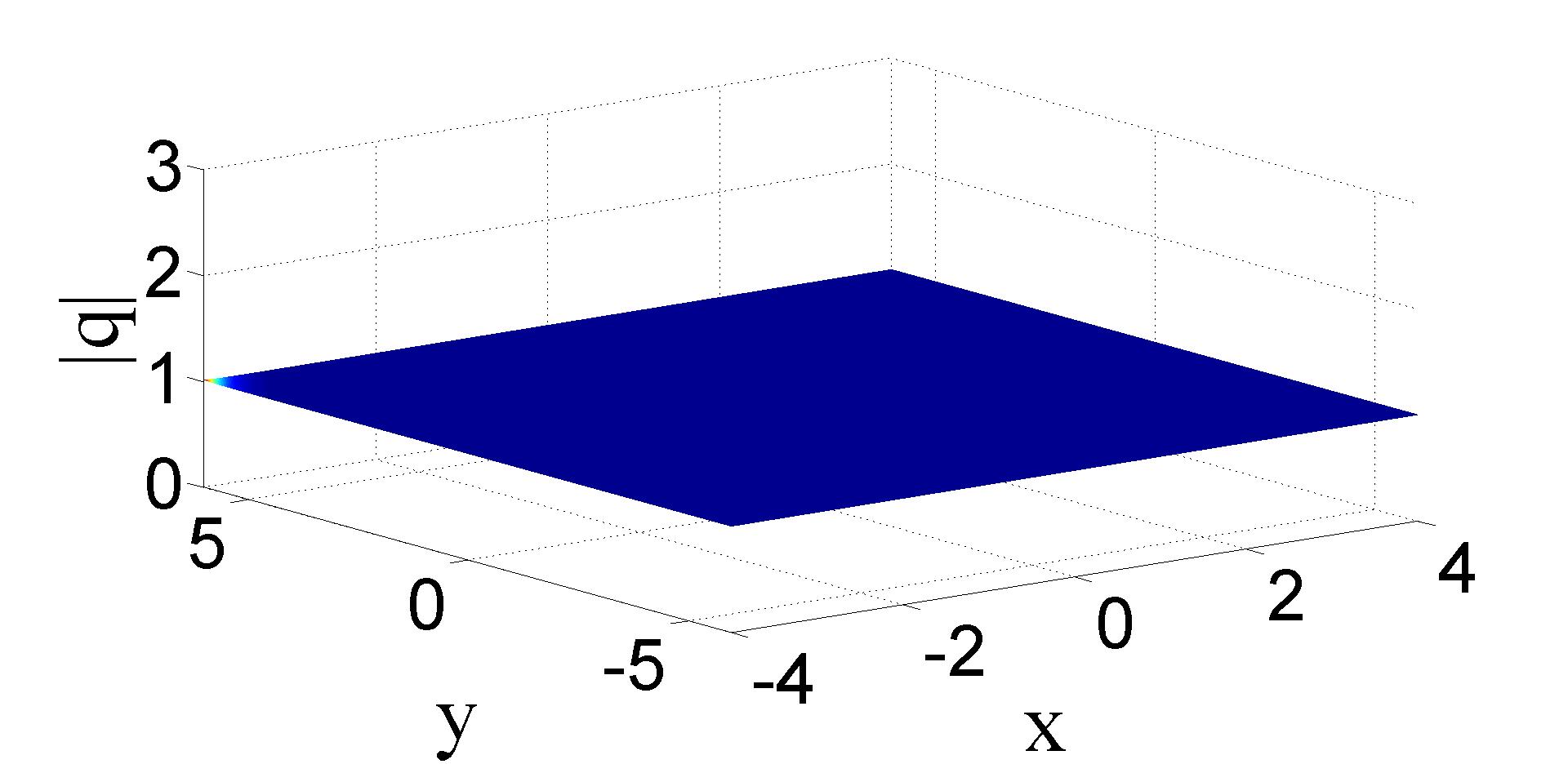}}\qquad
\subfigure[$t=-1$]{\includegraphics[height=6.0cm,width=7.5cm]{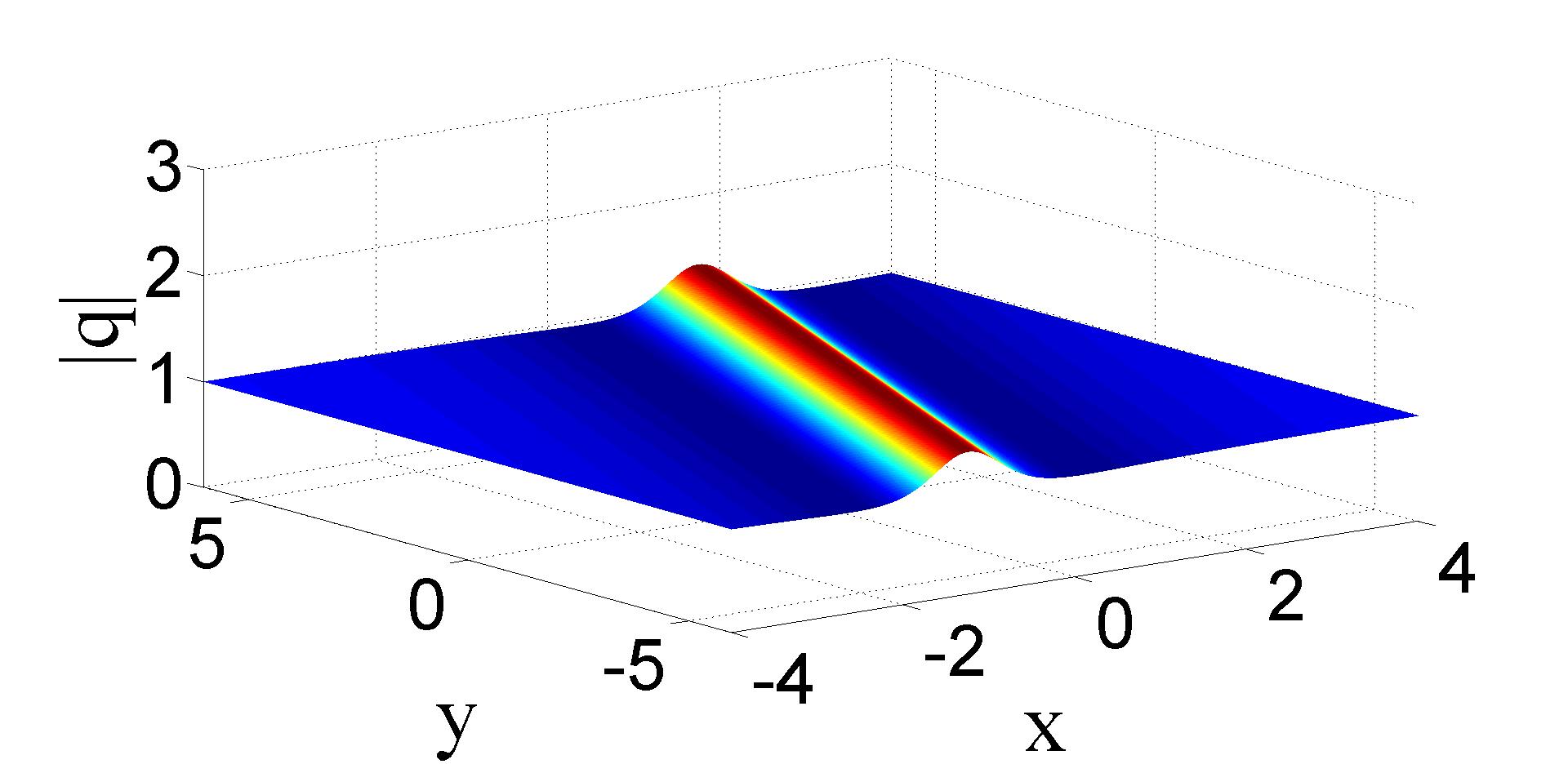}}
\subfigure[$t=0$]{\includegraphics[height=6.0cm,width=7.5cm]{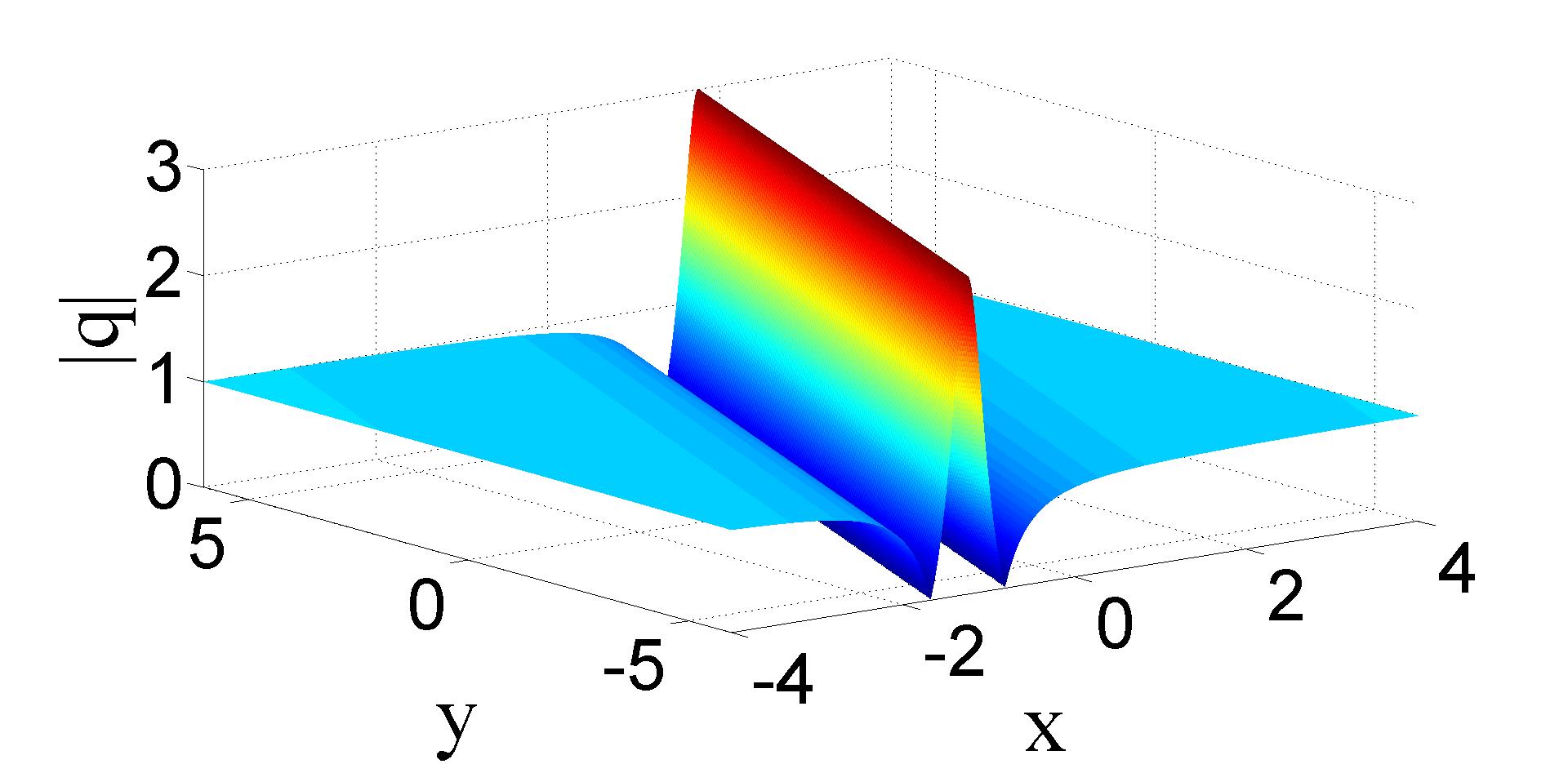}}\qquad
\subfigure[$t=10$]{\includegraphics[height=6.0cm,width=7.5cm]{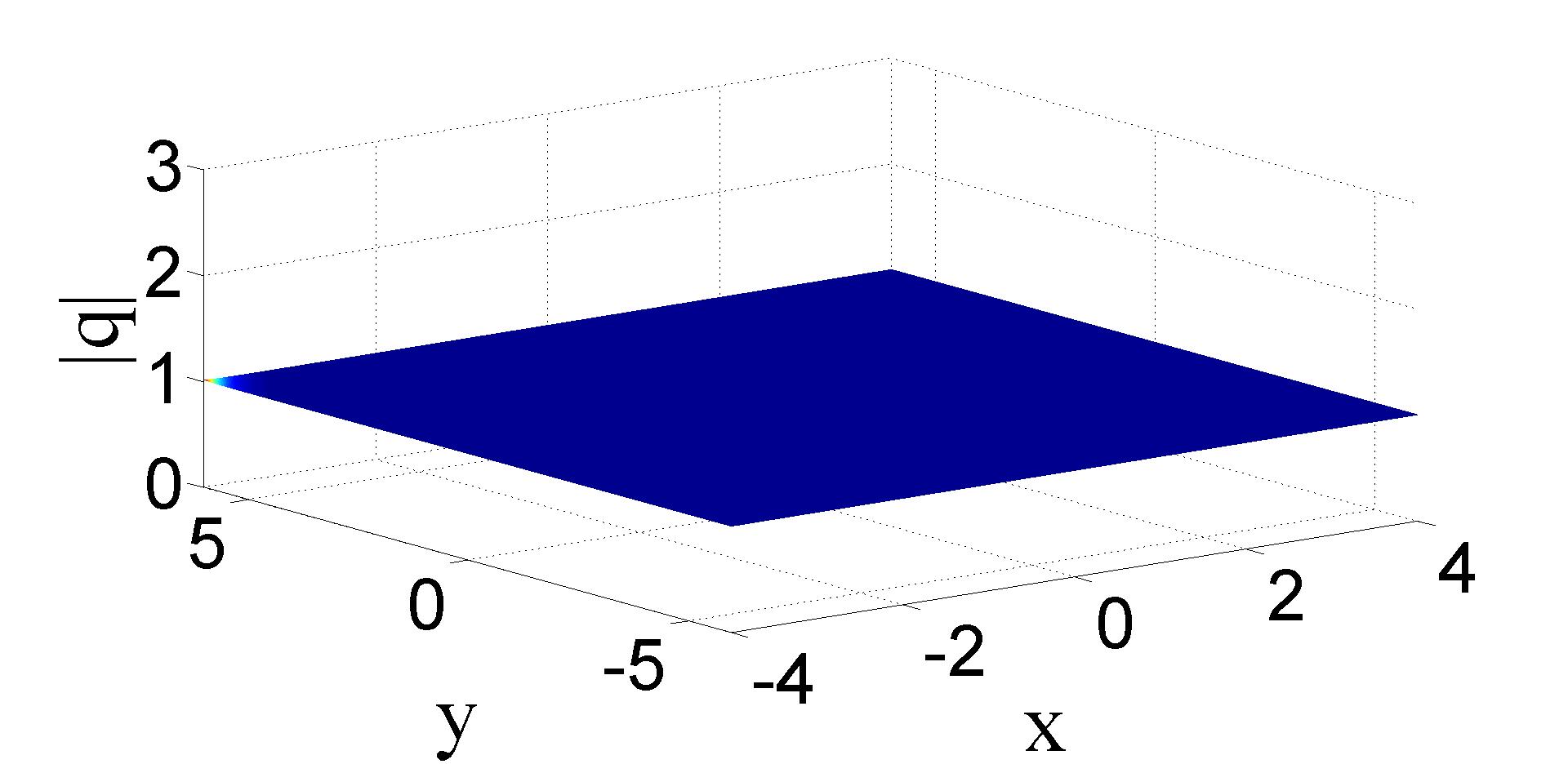}}
\caption{The time evolution of fundamental line  rogue waves of the DSIII equation given by equation \eqref{final-1} with parameters $\gamma_{11}=0\,,\lambda=1\,,p_{1}=2\,,c_{11}=0\,$.~}\label{fig4}
\end{figure}

(ii) Rogue wave solution. When $p_{1I}=0\,,p_{1R}\neq \pm 1$, i.e., $p_{1}$ is an arbitrary real parameter except $\pm1$, this solution is a line rogue wave. As can be seen in Fig. \ref{fig4}, this solution describes a line wave, which possesses a varying height. When $t\rightarrow \pm  \infty$, this line wave approaches towards the constant background (see $t=\pm 10$ panel ). In the intermediate time, it attains a  higher amplitude (see $t=0$ panel ). Specially, the maximum amplitude is three times the constant background amplitude, which is  same with the Peregrine rogue wave solutions of $(1+1)$-dimensional systems.

In contrast to the orientation direction of fundamental line rogue waves of the DSI equation, the orientation direction of this line rogue wave is almost arbitrary, while the former one has a limited range of orientations. It is
noted that solutions defined in equation \eqref{final-1} are independent of $t$ when $p_{1}=\pm1$, and they maintain a perfect line profile during their propagation in the $(x,y)$-plane without any decay. In this case, this solution is a rational soliton (i.e., W-shaped soliton) of the DSIII equation.

Next, we consider the case of $\gamma_{11}=1$ in equation (\ref{final-1}). In this case, this solution also has two typical behaviours.

\begin{figure}[!htbp]
\centering
\subfigure[$t=-10$]{\includegraphics[height=6cm,width=7.5cm]{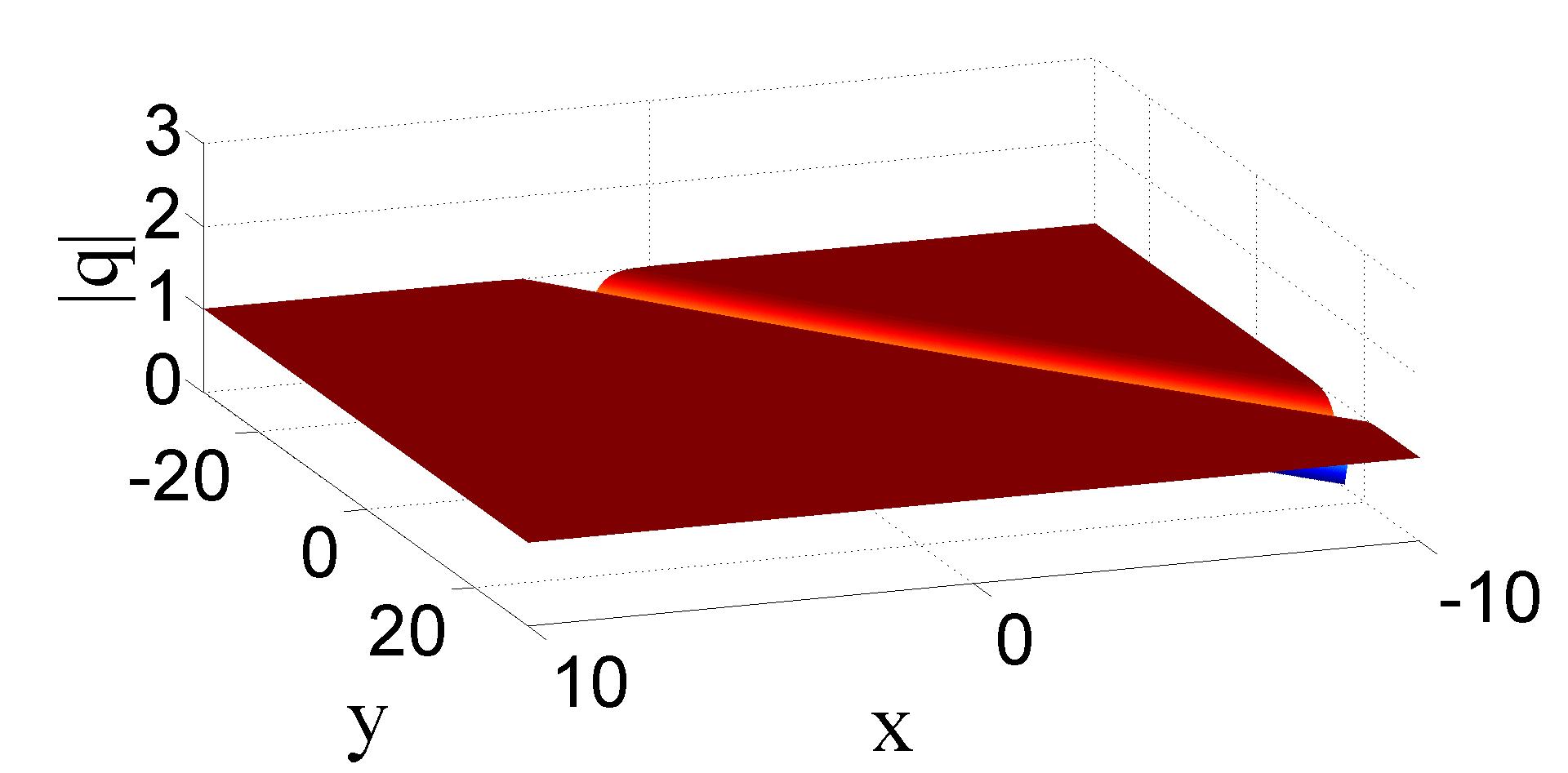}}\qquad\qquad
\subfigure[$t=-1$]{\includegraphics[height=6cm,width=7.5cm]{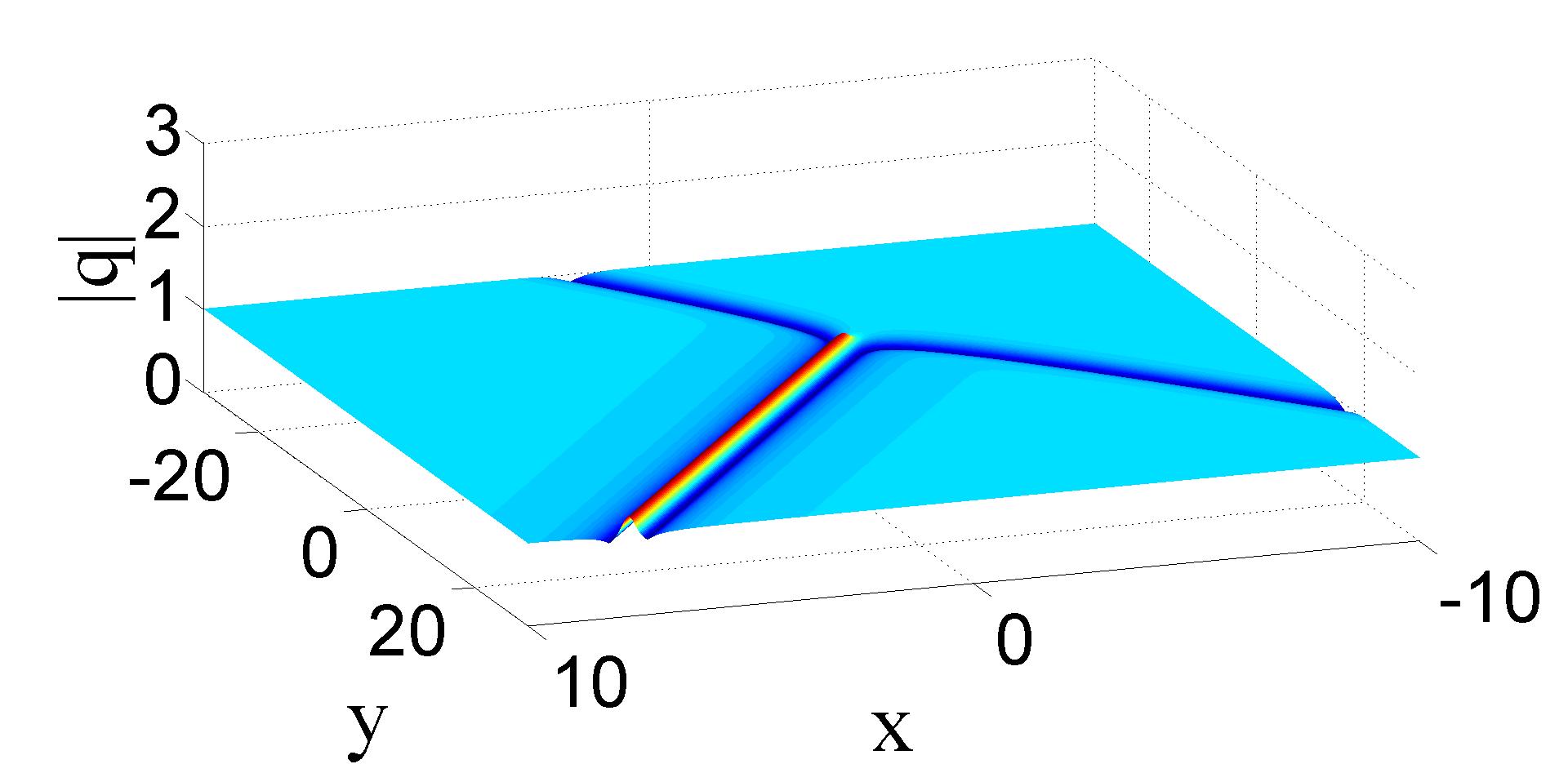}}
\subfigure[$t=0$]{\includegraphics[height=6cm,width=7.5cm]{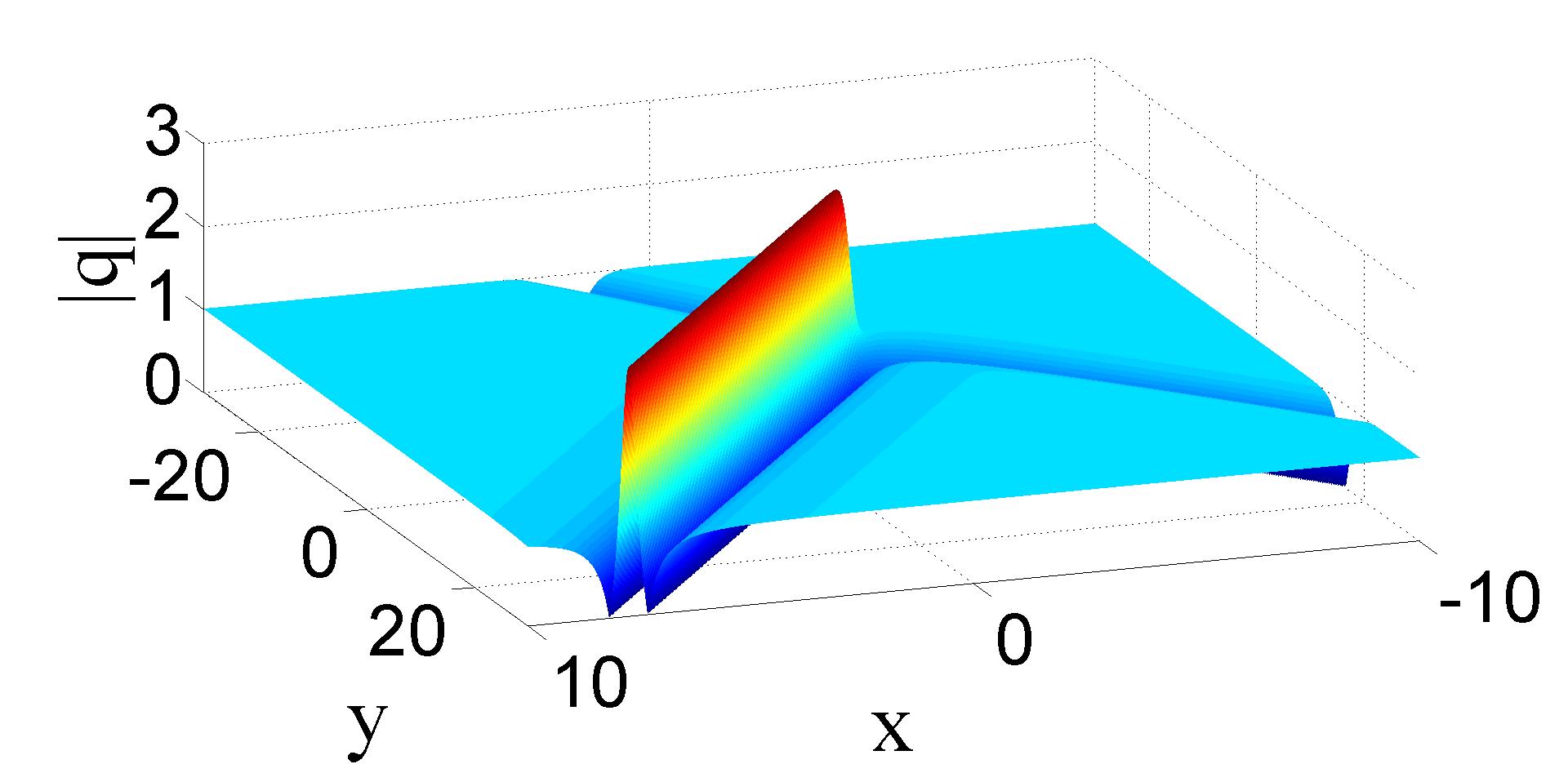}}\qquad\qquad
\subfigure[$t=10$]{\includegraphics[height=6cm,width=7.5cm]{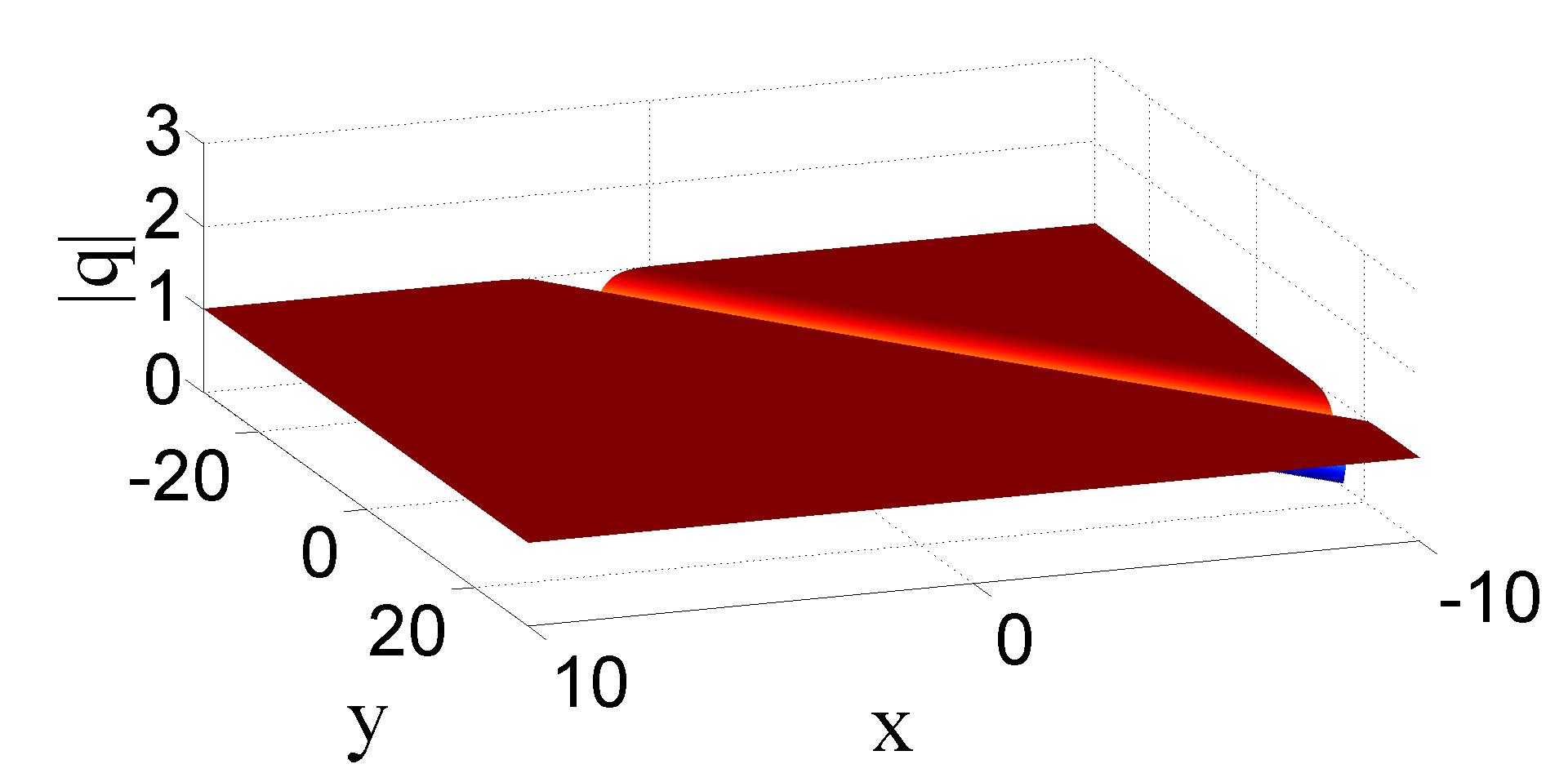}}
\caption{The time evolution of semi-rational solutions \eqref{final-1} consisting of a fundamental line rogue wave and a dark soliton  in the DSIII equation with parameters $\gamma_{11}=1\,,\lambda=1\,,p_{1}=2\,,c_{11}=10^{-4}\,,\xi_{10}=0$.~}\label{fig5}
\end{figure}

(a) A hybrid of rogue waves and dark solitons. When $p_{1I}=0\,,p_{1R}\neq \pm 1$, i.e., $p_{1}$ is an arbitrary real parameter except $\pm1$, this solution describes the interaction of a line rogue wave and a dark soliton. As
shown in Fig. \ref{fig5}, when $t\rightarrow \pm\infty$, this solution describes a dark soliton, and the rogue wave approaches to the constant background (see $t=\pm 10$ panel ). In the intermediate time, a line rogue wave arises from
the constant background ( see the panel at $t=0$ ), and then disappears into the constant background at larger time.  It is worthful to note that the fundamental rogue wave does not exist on the whole background, but only one side of the dark solion (see the panel at $t=0$).  May be the interaction between the dark soliton and the rogue wave destroy the wave structure of the rogue wave, and energy transfer between the dark soliton and the line rogue is also observed.
Specifically, when $p_1=\pm1$, the corresponding solution defined in \eqref{final-1} describes the interaction between a W-shaped soliton and a dark soliton, which features similar to the wave patterns in Fig.\ref{fig1} (c).

%%%%%%%%%%%%%%%%%%%%%%%%%%%%%%%%%%%%%%%%%%%%%%%%fig5

\begin{figure}[!htbp]
\centering
\subfigure[$t=-10$]{\includegraphics[height=6.0cm,width=7.5cm]{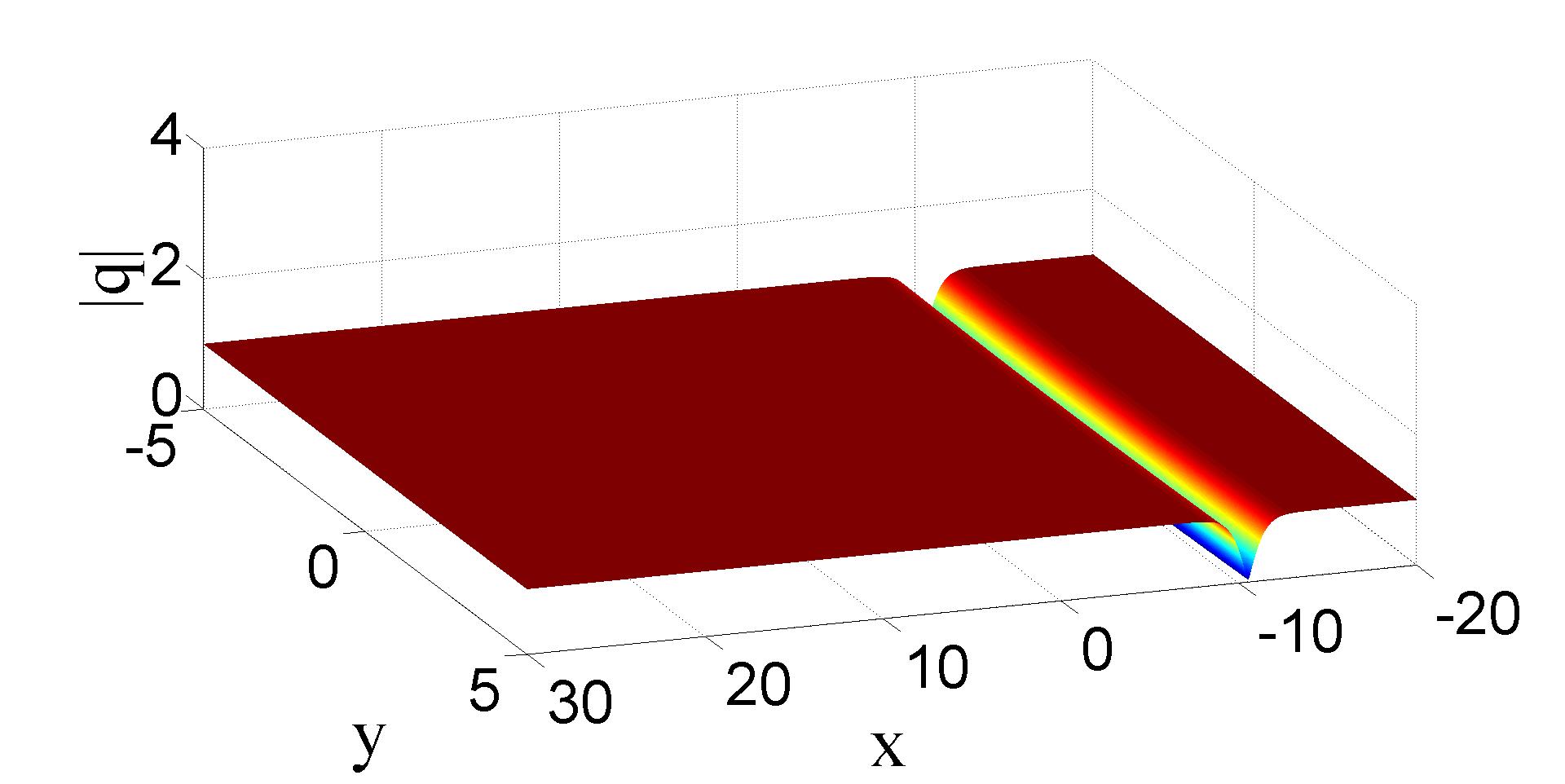}}\quad\quad
\subfigure[$t=0$]{\includegraphics[height=6.0cm,width=7.5cm]{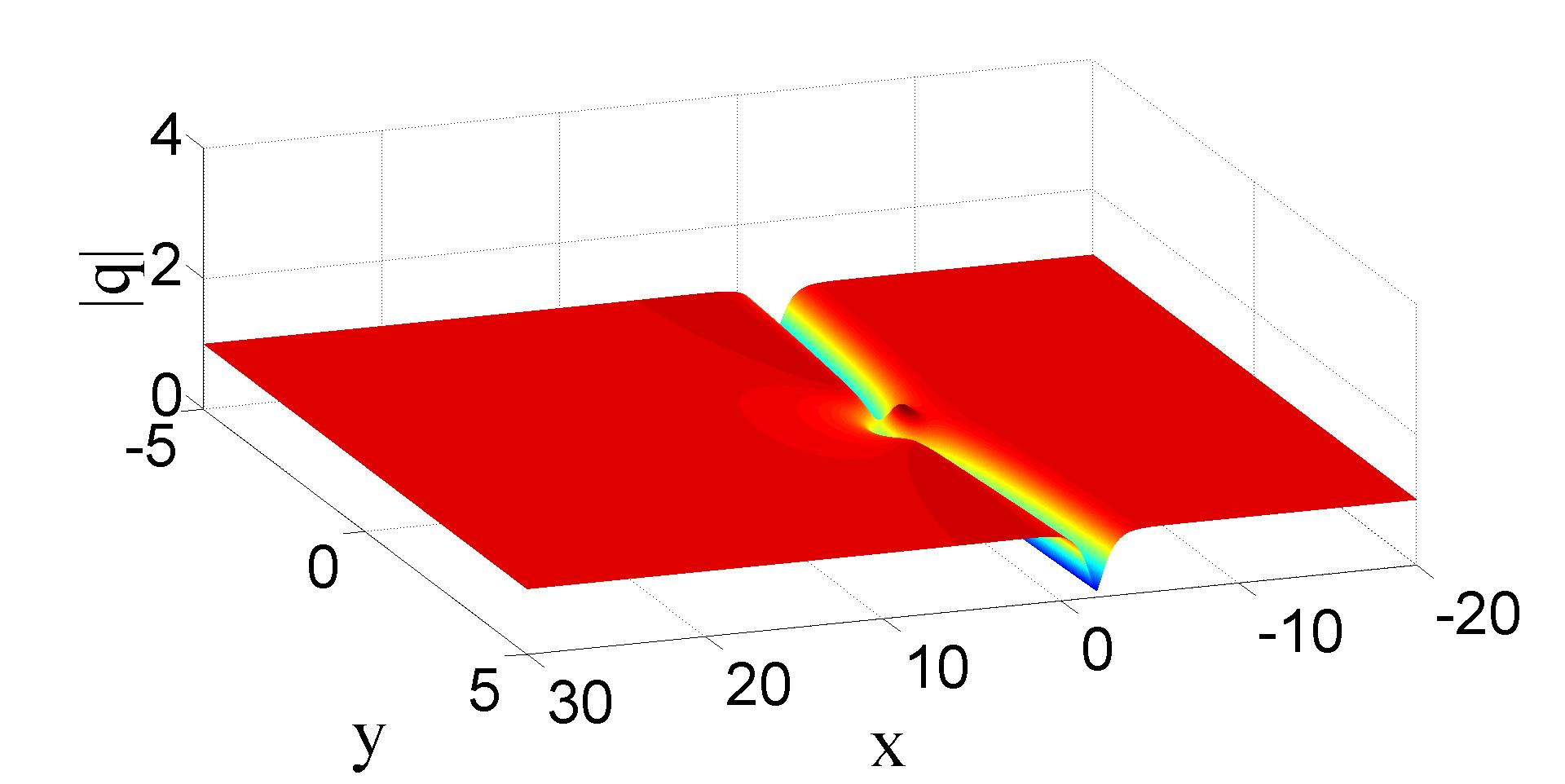}}
\subfigure[$t=1$]{\includegraphics[height=6.0cm,width=7.5cm]{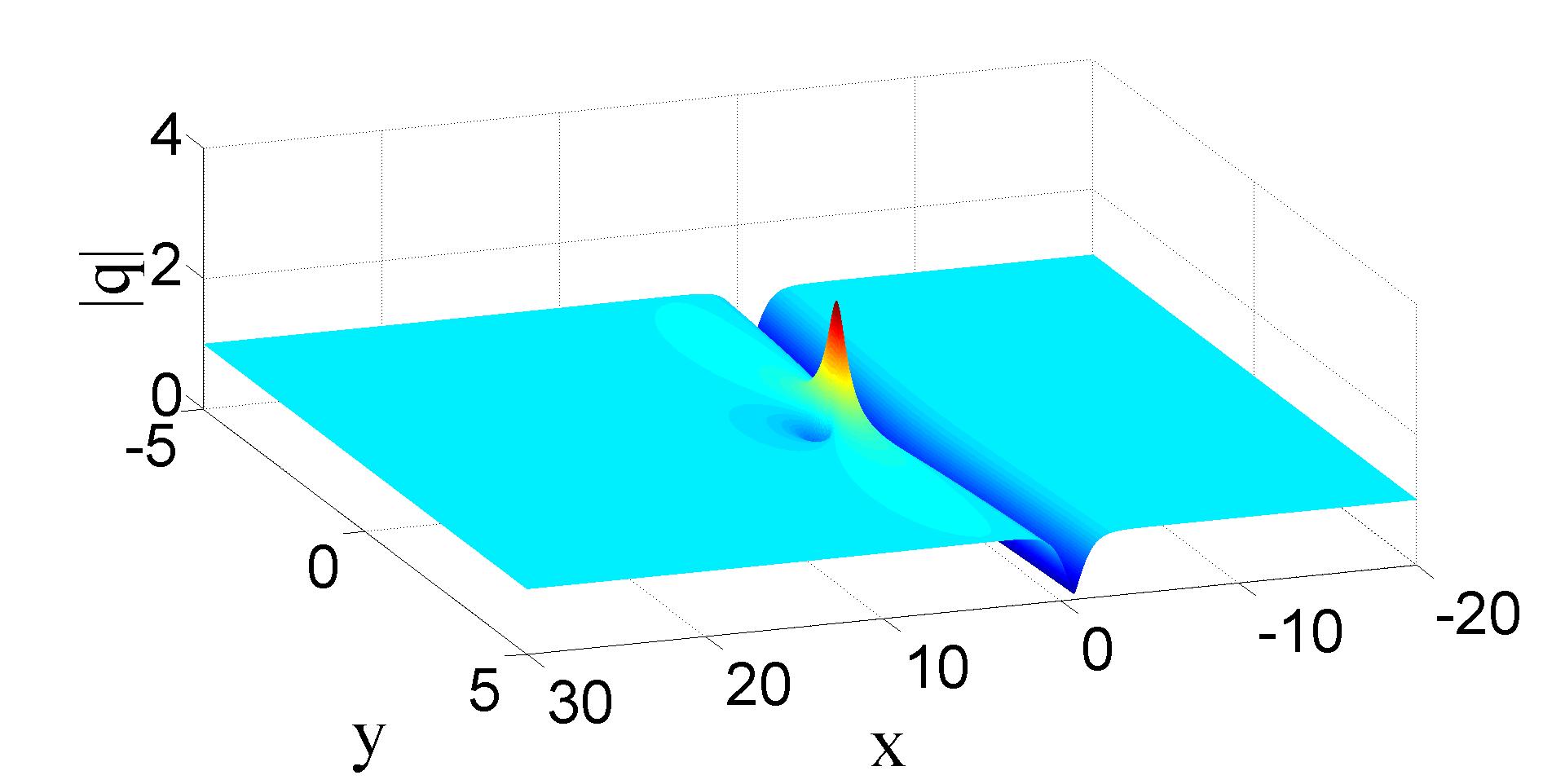}}\quad\quad
\subfigure[$t=10$]{\includegraphics[height=6.0cm,width=7.5cm]{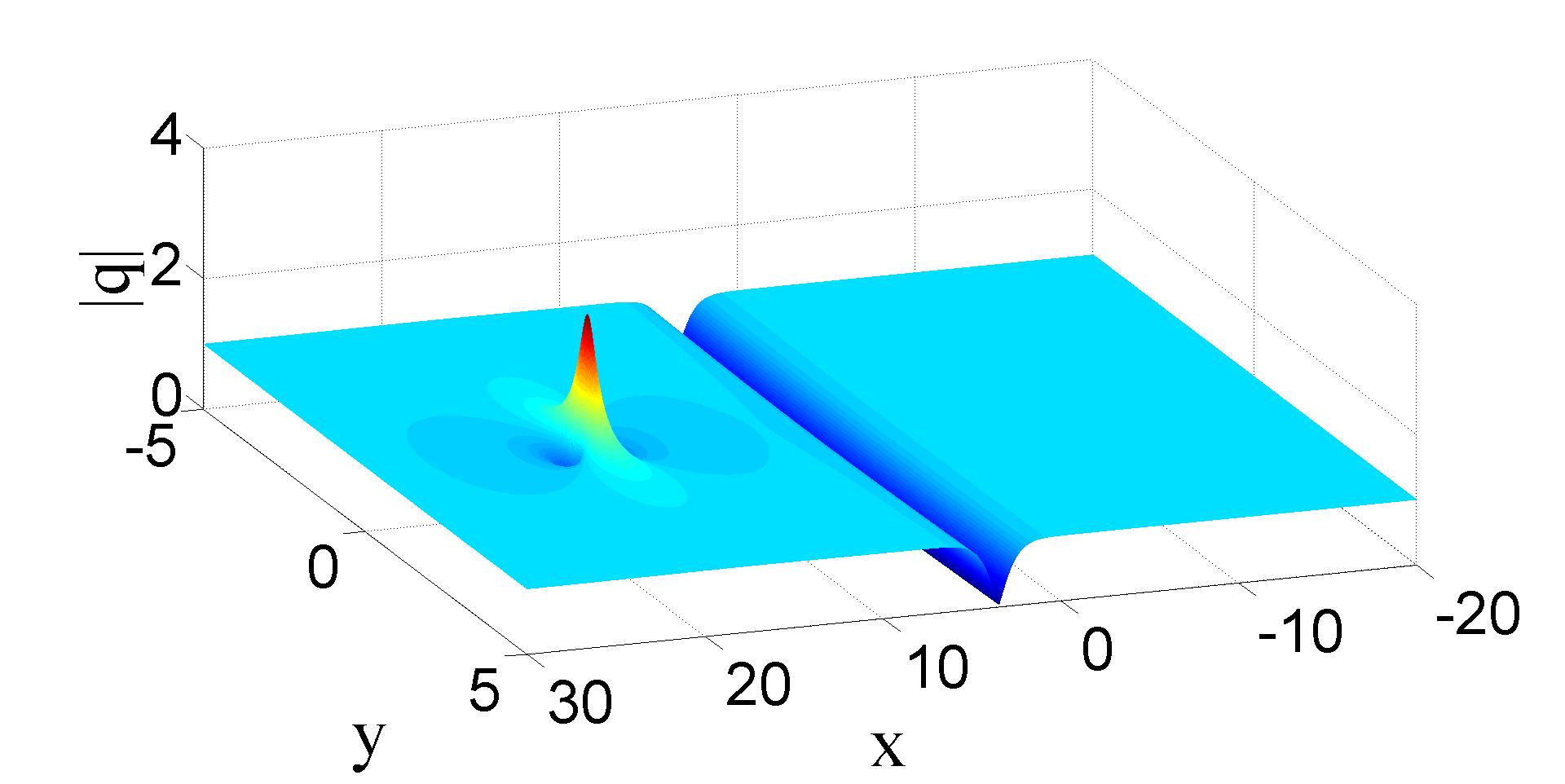}}
\caption{The time evolution of semi-rational solutions \eqref{final-1}  consisting of a lump and a dark soliton of the DSIII equation with parameters $\gamma_{11}=1\,,\lambda=1\,,p_{1}=\frac{2}{3}+\frac{1}{3}i\,,c_{11}=10^{-4}\,,\xi_{10}=0$.~}\label{fig6}
\end{figure}

(b) A hybrid of lumps and dark solitons.  When $p_{1I}\neq0$, i.e., $p_{1}$ is complex, the corresponding solution is a mixture of lumps and dark solitons. As can be seen in Fig. \ref{fig6}, this solution is a dark soliton when $t\rightarrow-\infty$ (see the $t=-10$ panel). In the intermediate time, a lump forms on the dark soliton, and travels with higher speed than the line soliton. Hence, at larger time, the original soliton disintegrates into a lump and a soliton (see the panel at $t=10$ ).  Obviously, these explicit solutions possess the unique dynamics simulated by numerical studies in Ref. \cite{XX-1,XX-2}.
Besides, at this juncture,  this kind of lump-line soliton solutions is quite different from the lump-line soliton solutions of the DSII equation derived by Fokas $et$ $al$. \cite{XX}, as the latter describes lump and line soliton for all times.
%%%%%%%%%%%%%%%%%%%%%%%%%%%%%%%%%%%%%%%%%%%%%%%%%%%%%%%%%%%%%%fig6

Two subclasses of non-fundamental semi-rational solutions can be derived from equation \eqref{BT-1} by taking $N>1\,,n_{i}=1$ or $N=1\,,n_{i}>1$ respectively. Below, we focus on these two subclasses of non-fundamental semi-rational solutions.

 \subsection{Dynamics of Multi-semi-rational solutions}$\\$

The first subclass of non-fundamental semi-rational solutions is obtained by taking $N>1\,,n_{i}=1$ in equation \eqref{BT-1}, which are multi-semi-rational solutions. For different parameter choices of $\gamma_{ij}\,,p_{i}$, these multi-semi-rational solutions can be
classified into three different patterns:

(1) Multi-rogue waves.  When $p_{i}$ are real except $\pm1$ and $\gamma_{ij}=0$, this semi-rational solution becomes non-fundamental rational solution for the DSIII equation,
which is multi-rogue waves of the DSIII equation.

 (2) A hybrid of multi-rogue-waves and dark solitons.  When $\gamma_{ii}=1\,,\gamma_{ij}=0\,(i\neq\,j )$, the corresponding semi-rational solution is a combination  of rogue waves and dark solitons.

 (3) A hybrid of multi-rogue-waves, breathers and  dark solitons.  When $\gamma_{ij}=1$, the corresponding solution is a mixture of multi-rogue waves, breathers and solitons.

To demonstrate this subclass of semi-rational solutions in detail, we first consider the case of $N=2$. In this case, solutions can be derived from Theorem 2 as
\begin{equation}\label{multi-2}
\begin{aligned}
q=\frac{\begin{vmatrix} M_{11}^{(1)} & M_{12}^{(1)} \,\\ M_{21}^{(1)}& M_{22}^{(1)} \end{vmatrix}}{\begin{vmatrix} M_{11}^{(0)} & M_{12}^{(0)} \,\\ M_{21}^{(0)}& M_{22}^{(0)} \end{vmatrix}}\,,
V={\rm -\lambda\,log} (\begin{vmatrix} M_{11}^{(0)} & M_{12}^{(0)} \,\\ M_{21}^{(0)}& M_{22}^{(0)} \end{vmatrix})_{yy}\,,
U={\rm -\lambda\,log} (\begin{vmatrix} M_{11}^{(0)} & M_{12}^{(0)} \,\\ M_{21}^{(0)}& M_{22}^{(0)} \end{vmatrix})_{xx},
\end{aligned}
\end{equation}
where
\begin{equation}\label{mij-2}
\begin{aligned}
M_{ij}^{(n)}=\frac{e^{\xi_{i}+\xi_{j}^{*}}}{p_{i}+p_{j}^{*}}[(\xi_{i}^{'}-\frac{p_{i}}{p_{i}+p_{j}^{*}}+c_{i1}+n)(\xi_{j}^{'*}-\frac{p_{j}^{*}}{p_{i}+p_{j}^{*}}+c_{i1}^{*}-n)+\frac{p_{j}^{*}}{(p_{i}+p_{j}^{*})^{2}}]+\gamma_{ij}c_{i1}c_{j1}^{*},
\end{aligned}
\end{equation}
and $\xi_{i}\,,\xi_{i}^{'}$ are given in Theorem 2. These solutions possess different dynamics determined by parameters $\gamma_{ij}$ and $p_{i}$.

\begin{figure}[!htbp]
\centering
\subfigure[$t=-10$]{\includegraphics[height=6.0cm,width=7.5cm]{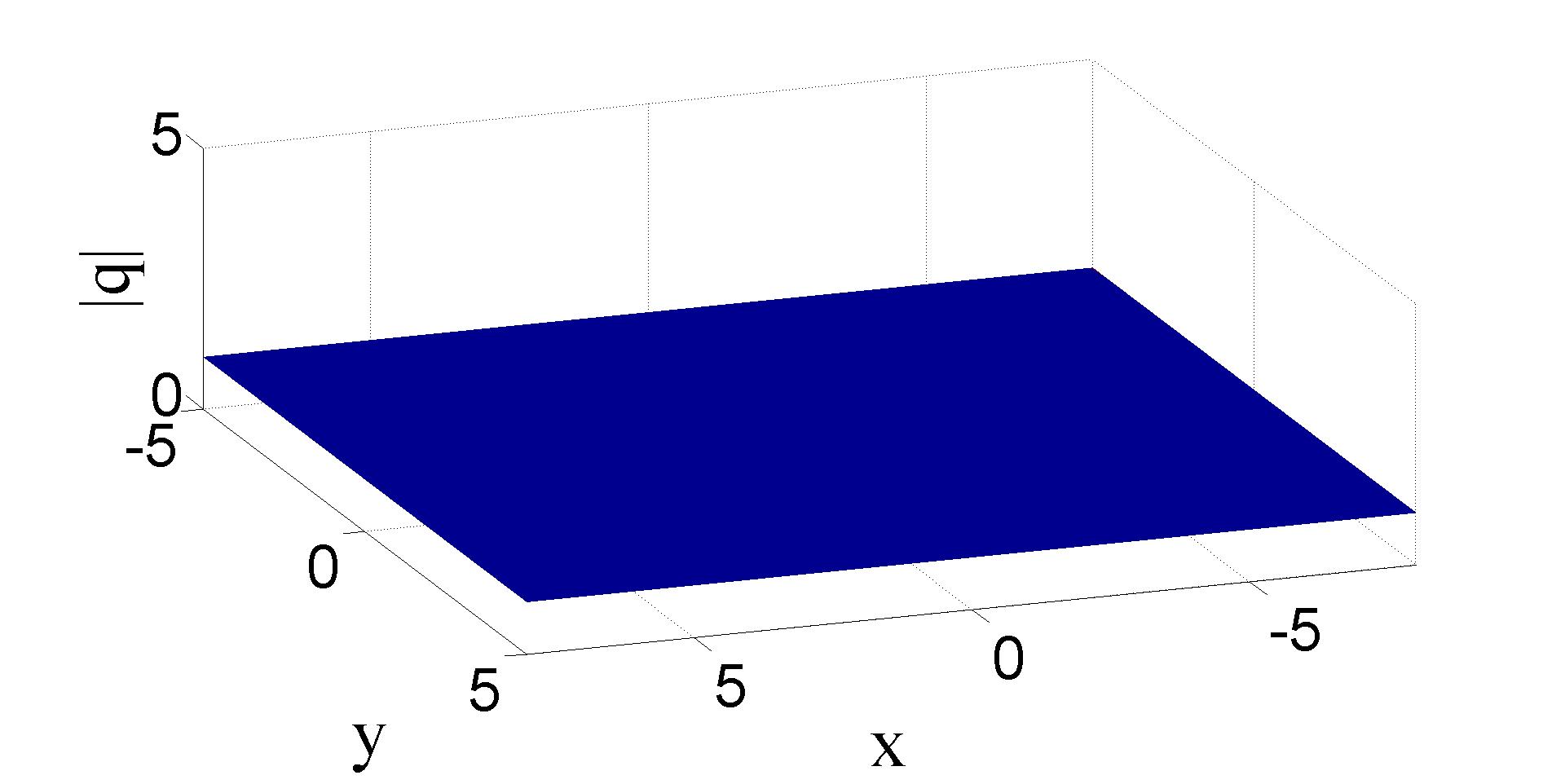}}
\subfigure[$t=-1$]{\includegraphics[height=6.0cm,width=7.5cm]{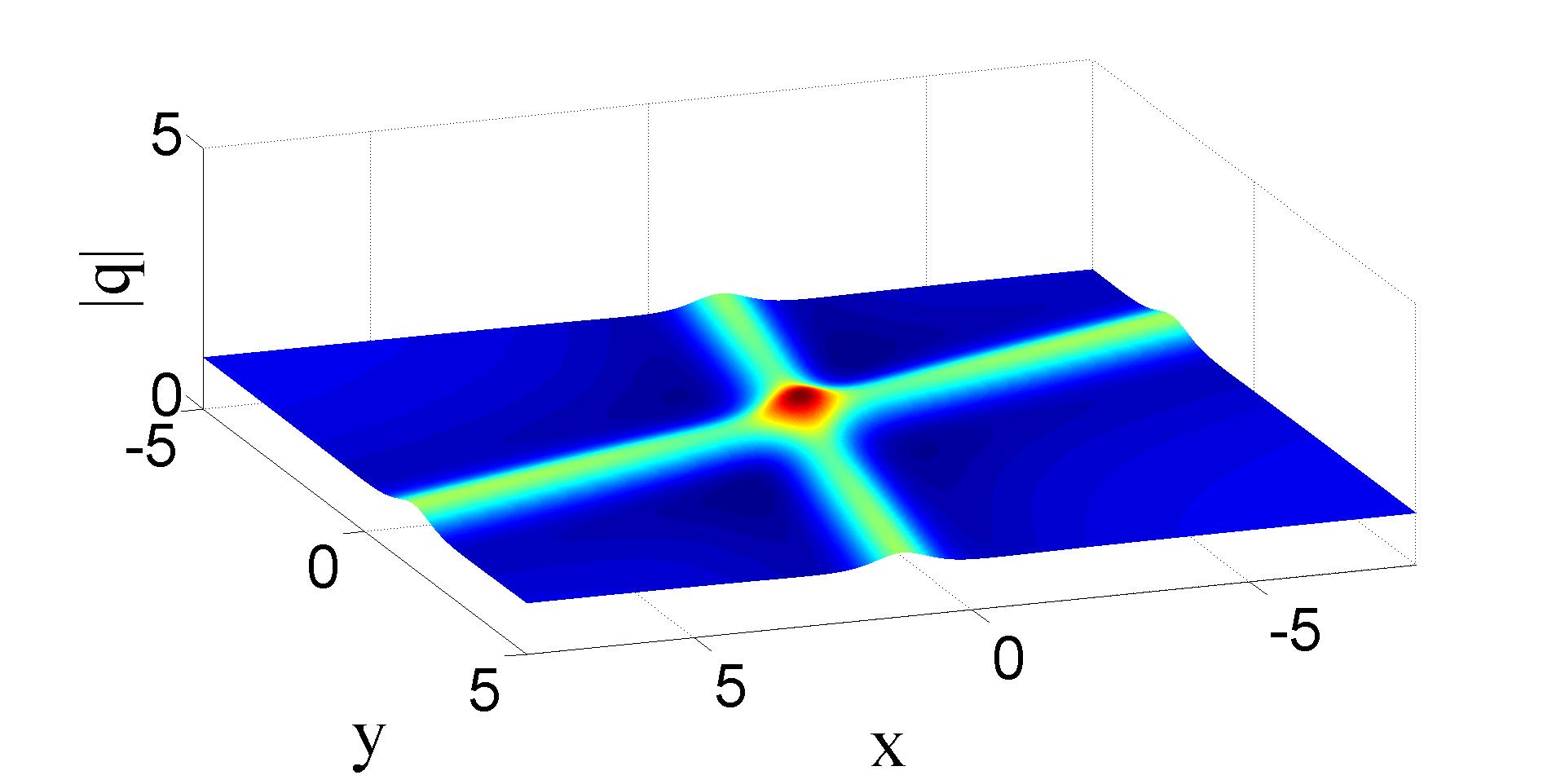}}\qquad
\subfigure[$t=0$]{\includegraphics[height=6.0cm,width=7.5cm]{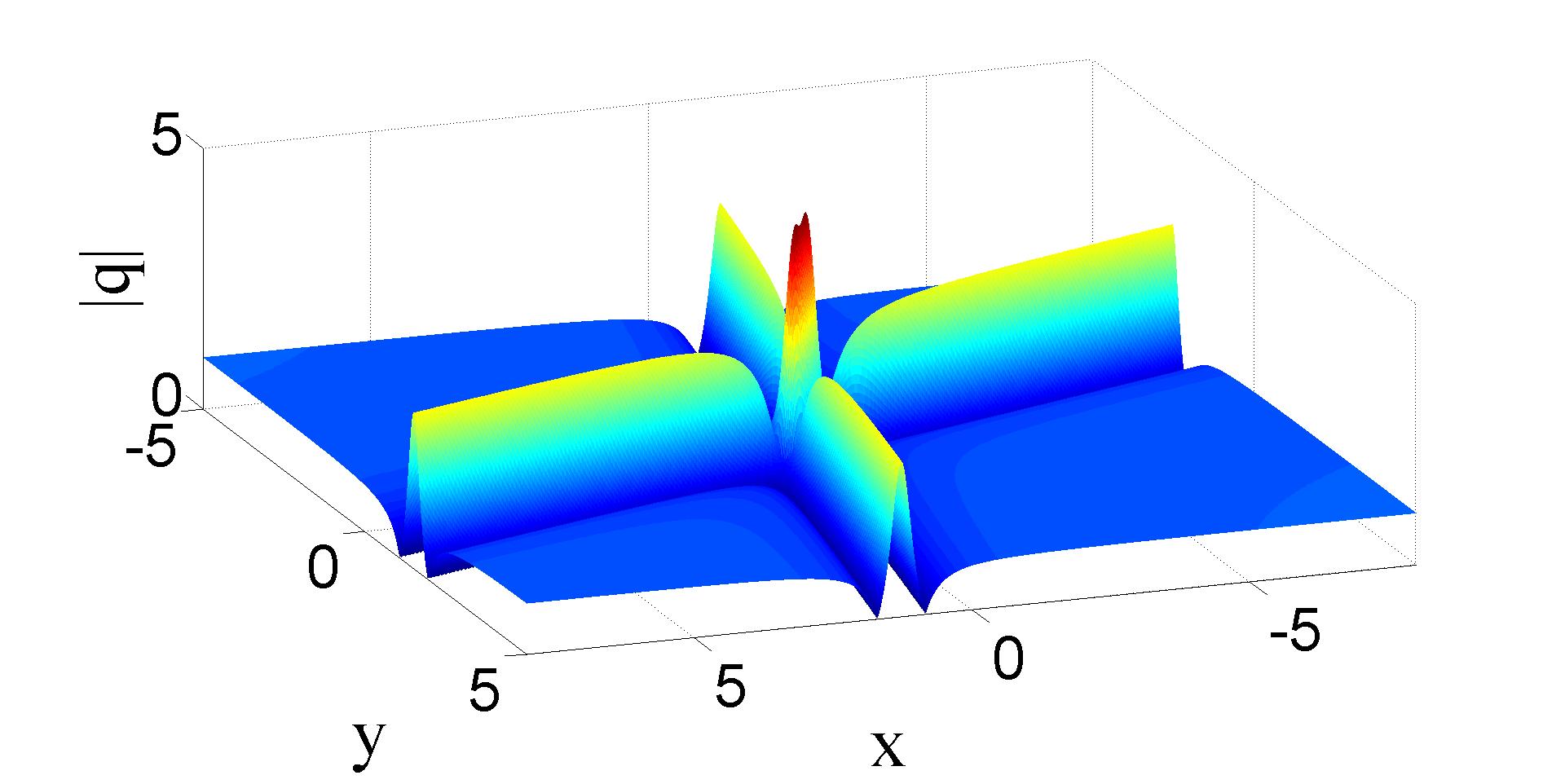}}
\subfigure[$t=10$]{\includegraphics[height=6.0cm,width=7.5cm]{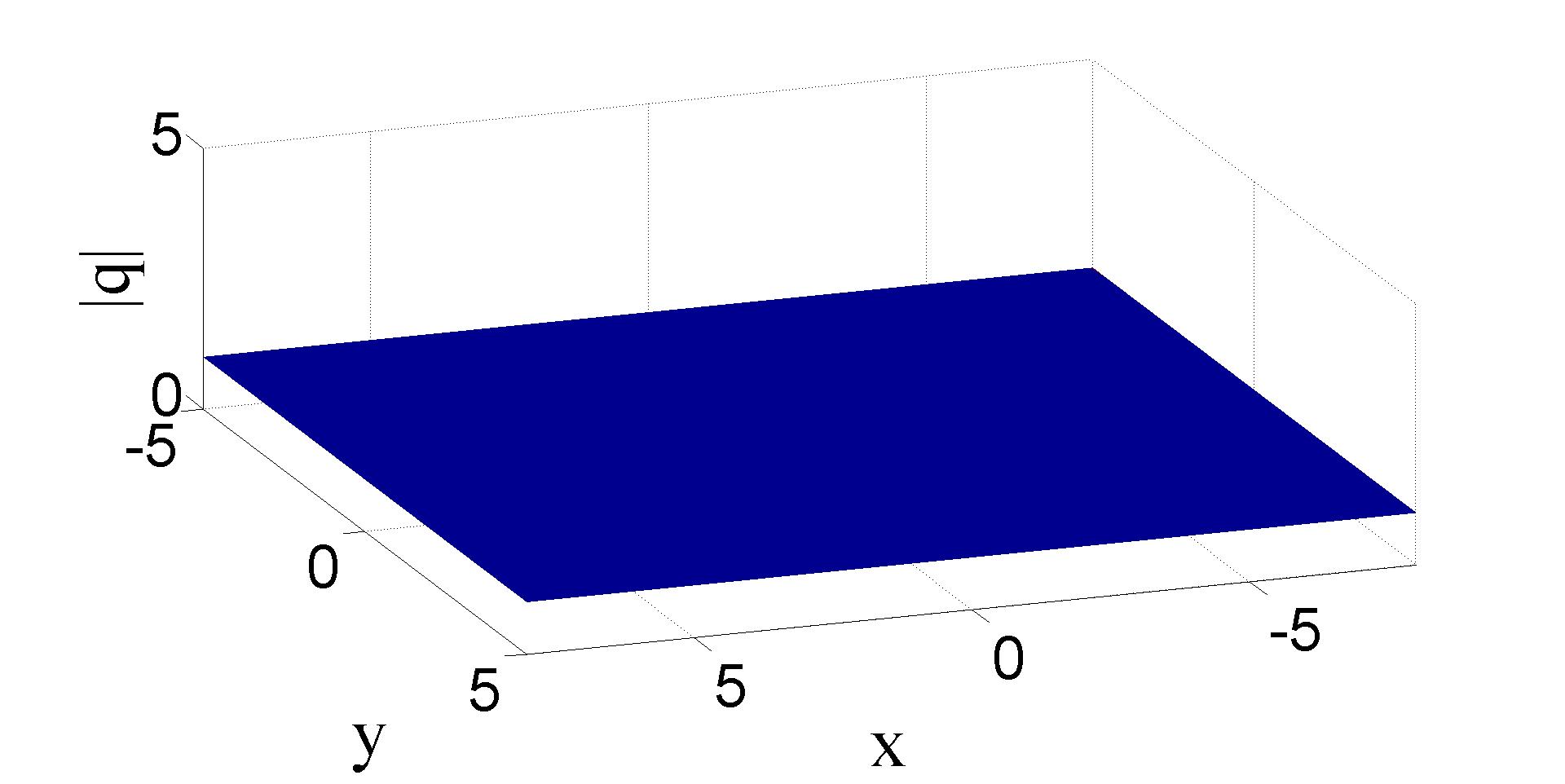}}\qquad
\caption{The time evolution of multi-rogue waves of the DSIII equation defined in equation \eqref{multi-2} with parameters \eqref{pa-4} and $\gamma_{11}=0\,,\gamma_{12}=0\,,\gamma_{21}=0\,,\gamma_{22}=0\,,c_{11}=0\,,c_{21}=0$.~}\label{fig7}
\end{figure}

(i) When one takes $\gamma_{ij}=0$ and real parameters $p_{i}$ ( $p_{i}\neq1$ ), this solution is actually the two-rogue wave solution of the DSIII equation. This solution with parameters
\begin{equation}\label{pa-4}
\begin{aligned}
p_{1}=2\,,p_{2}=\frac{3}{2}\,,c_{11}=0\,,c_{21}=0
\end{aligned}
\end{equation}
is shown in Fig.\ref{fig7}.  As can be seen from figure, two crossed line rogue waves arise from the constant background, and the interaction region attains much higher amplitudes, the wave patterns are no longer straight lines until these line waves disappear into the constant background. It is noted that these two line rogue wave would not be separated for all times. As non-fundamental rogue waves of the DSIII equation in Ref.\cite{binbin} are parallel line rogue waves, thus this two-rogue wave of the DSIII equation in our paper is quite different.
 %%%%%%%%%%%%%%%%%%%%%%%%%%%%%%%%%%%%%%%%%%%%%%%%%%%%%%%%%%%%%%%%%fig7

\begin{figure}[!htbp]
\centering
\subfigure[$t=-10$]{\includegraphics[height=6.0cm,width=7.5cm]{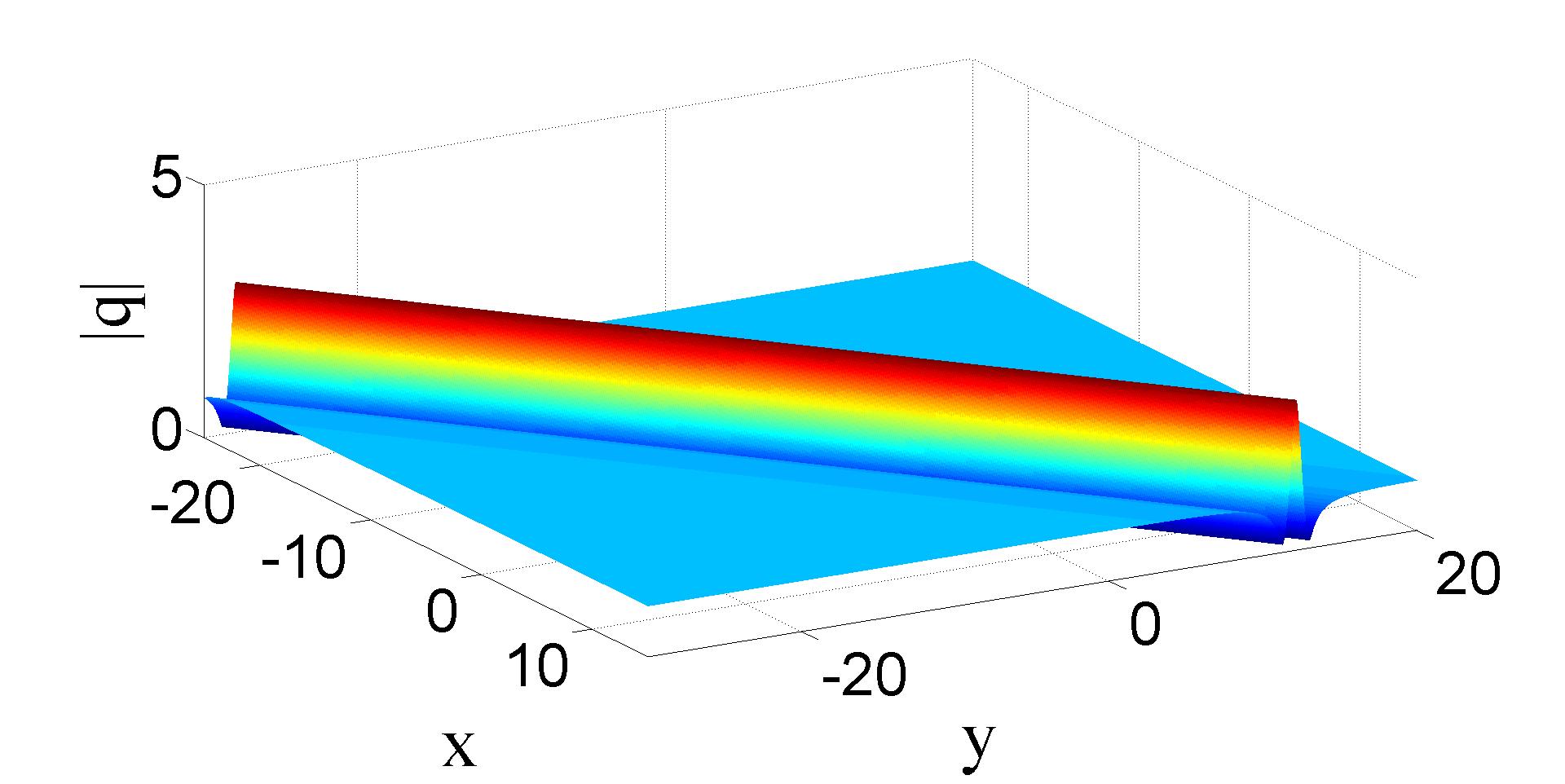}}\qquad
\subfigure[$t=-1$]{\includegraphics[height=6.0cm,width=7.5cm]{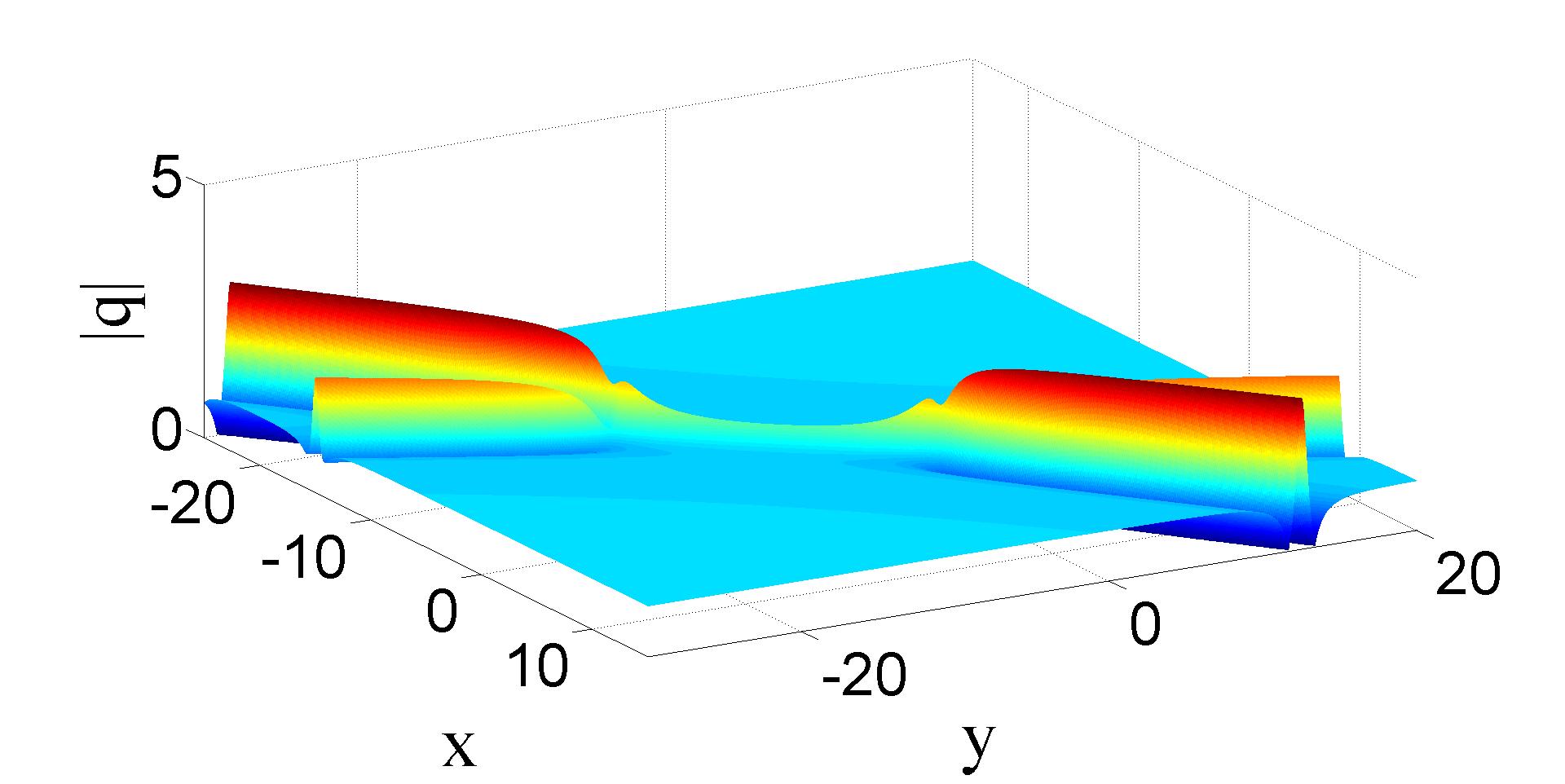}}
\subfigure[$t=0$]{\includegraphics[height=6.0cm,width=7.5cm]{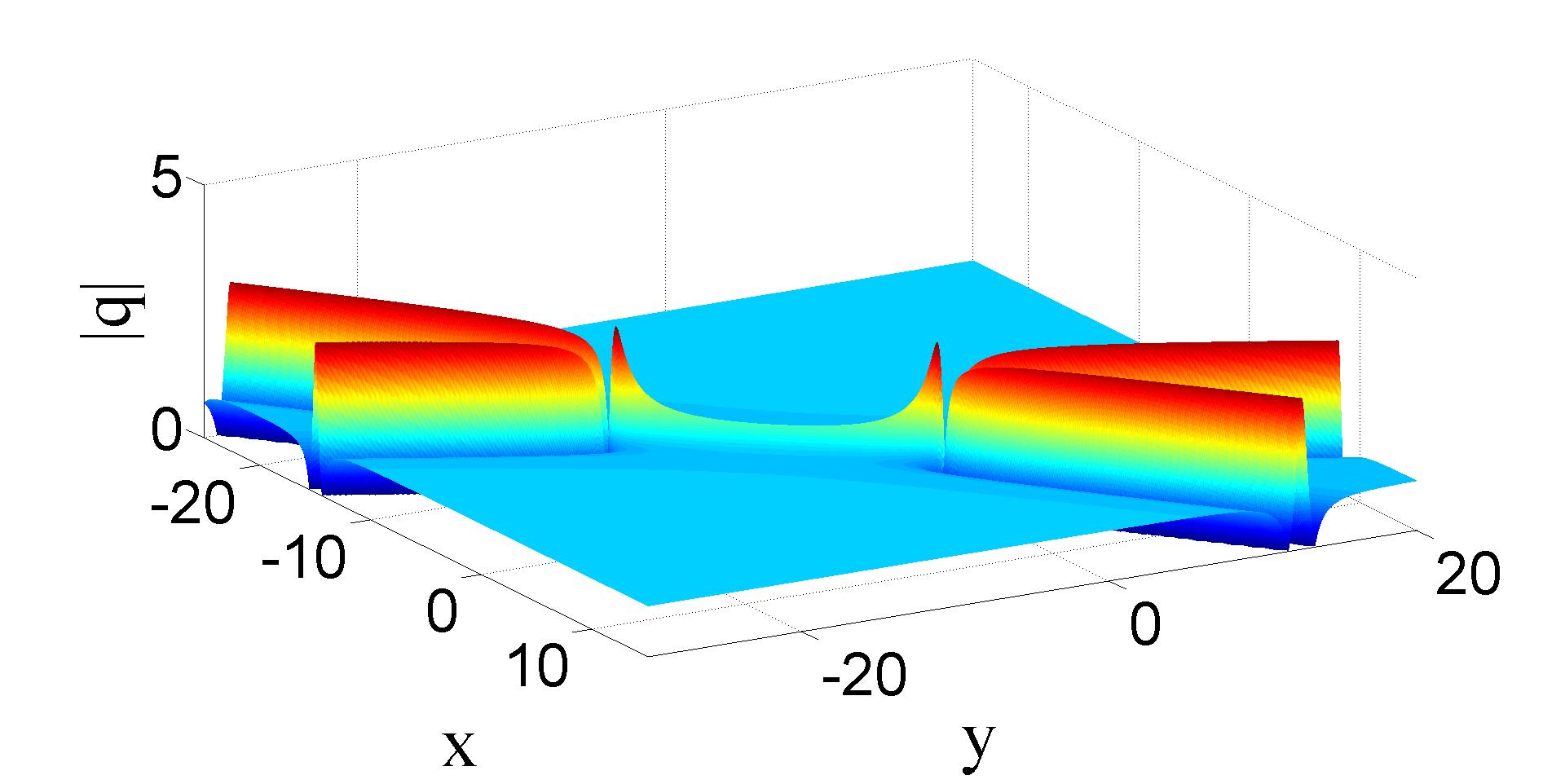}}\qquad
\subfigure[$t=10$]{\includegraphics[height=6.0cm,width=7.5cm]{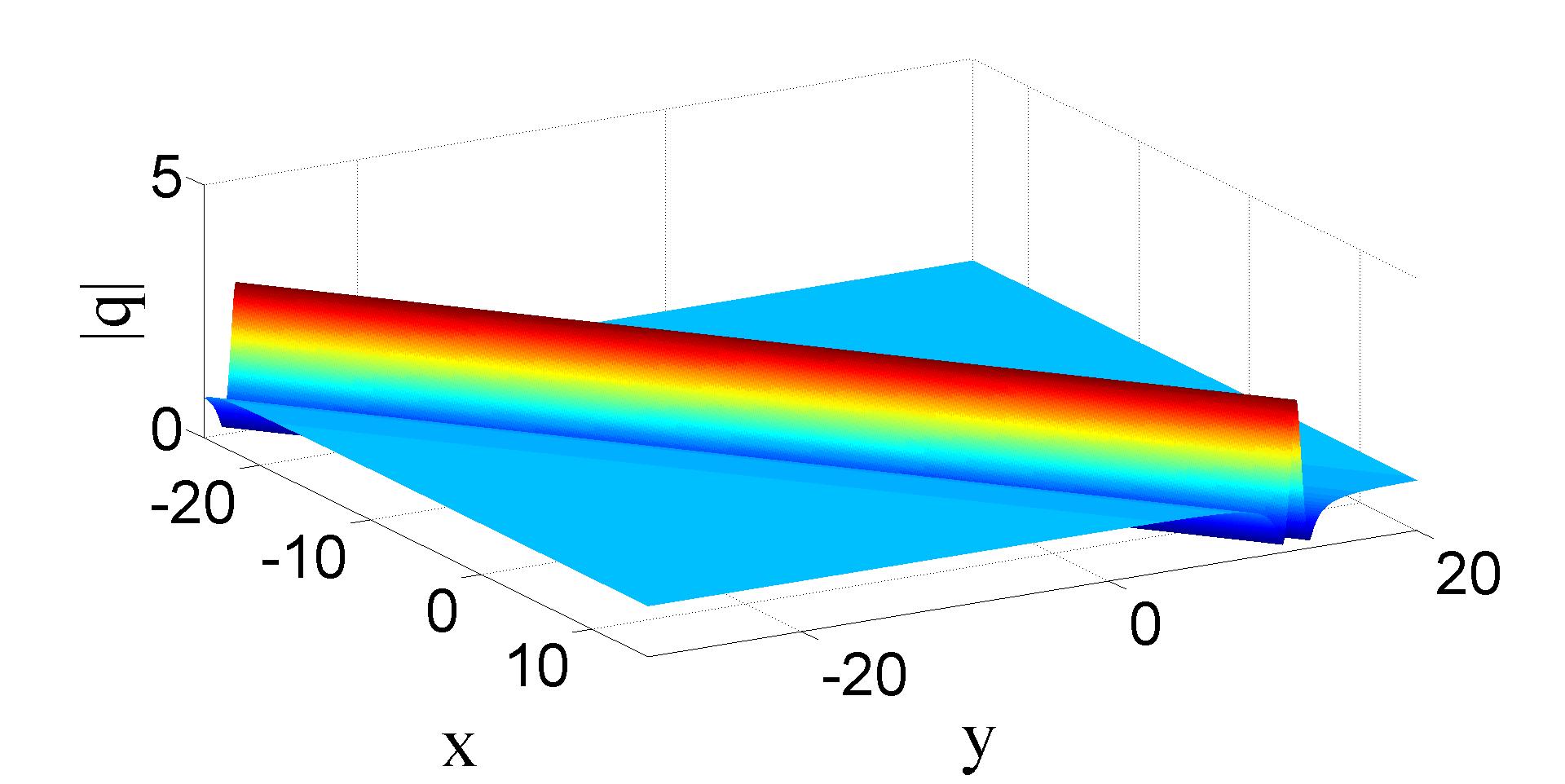}}
\caption{The time evolution of mutli-rational solutions  consisting of a fundamental rogue wave and a $W$-shaped soliton of the DSIII equation given by equation \eqref{multi-2} with parameters $p_{1}=1\,,p_{2}=2\,,\gamma_{11}=0\,,\gamma_{12}=0\,,\gamma_{21}=0\,,\gamma_{22}=0\,,c_{11}=0\,,c_{21}=0$.~}\label{fig8}
\end{figure}

Specially, when $p_{1}=1,p_{2}\neq1$, this solution becomes a hybrid of one rogue wave and one $W$-shaped soliton, which possess different dynamics comparing to the two-rogue wave shown in Fig.\ref{fig7},  see Fig.\ref{fig8}. It is seen that, when a line rogue wave arise from the constant background, the $W$-shaped soliton
is not a continuous line any more, and starts to disconnect into two part of lines gradually until they separate completely (see the $t=0$ panel).  In the intermediate time, this rational solution features the same wave pattern as multi-rogue waves in
DSI equation (the $t=0$ panel of Fig.2 in Ref.\cite{DSI}). At $t\rightarrow +\infty$, this solution turns out to be a $W$-shaped line soliton again.
%%%%%%%%%%%%%%%%%%%%%%%%%%%%%%%%%%%%%%%%%%%%%%%%%%%%%%%%%%%%%%%%%fig8

\begin{figure}[!htbp]
\centering
\subfigure[$t=-10$]{\includegraphics[height=6.0cm,width=7.5cm]{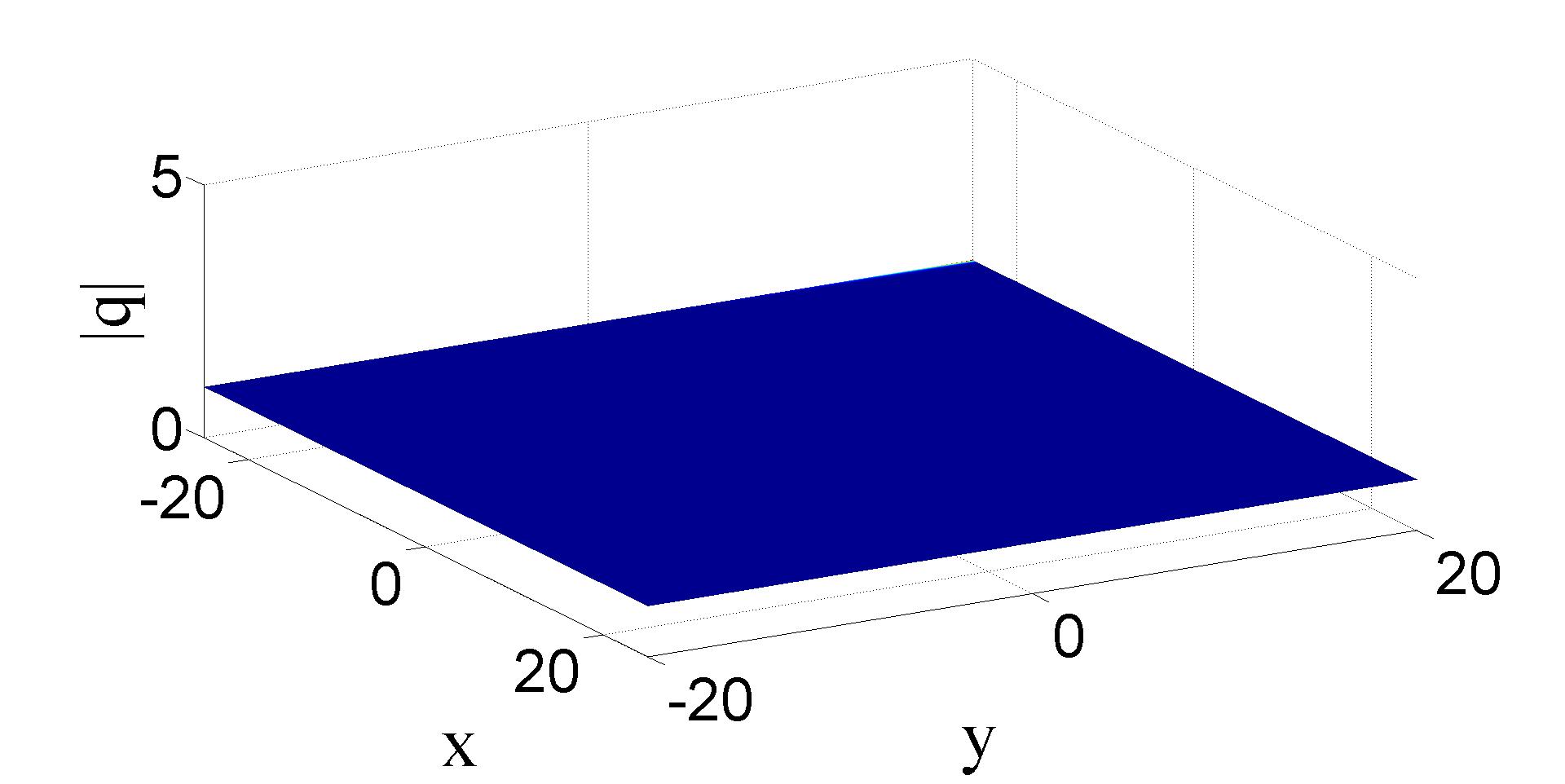}}\qquad
\subfigure[$t=-1$]{\includegraphics[height=6.0cm,width=7.5cm]{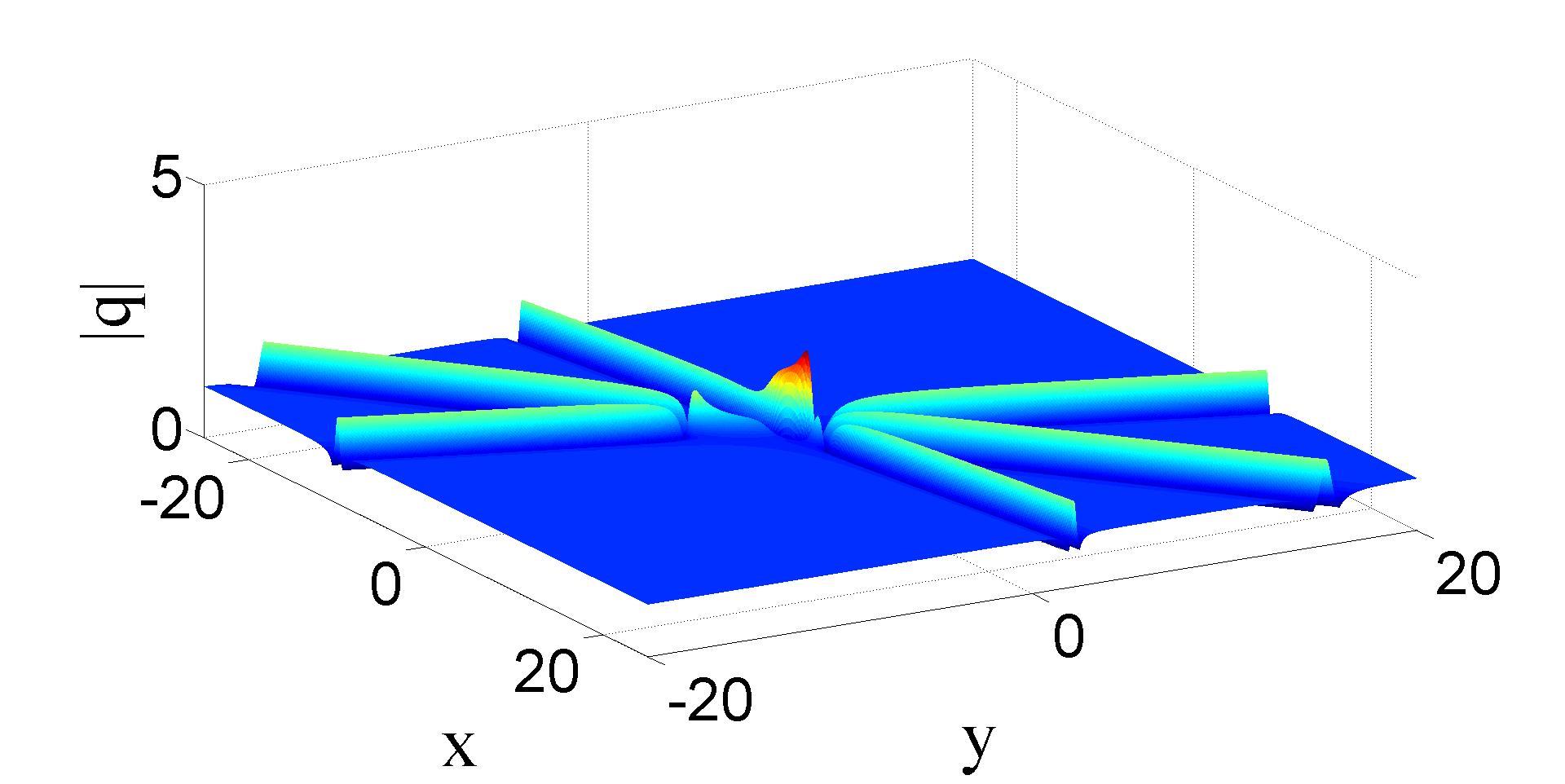}}
\subfigure[$t=0$]{\includegraphics[height=6.0cm,width=7.5cm]{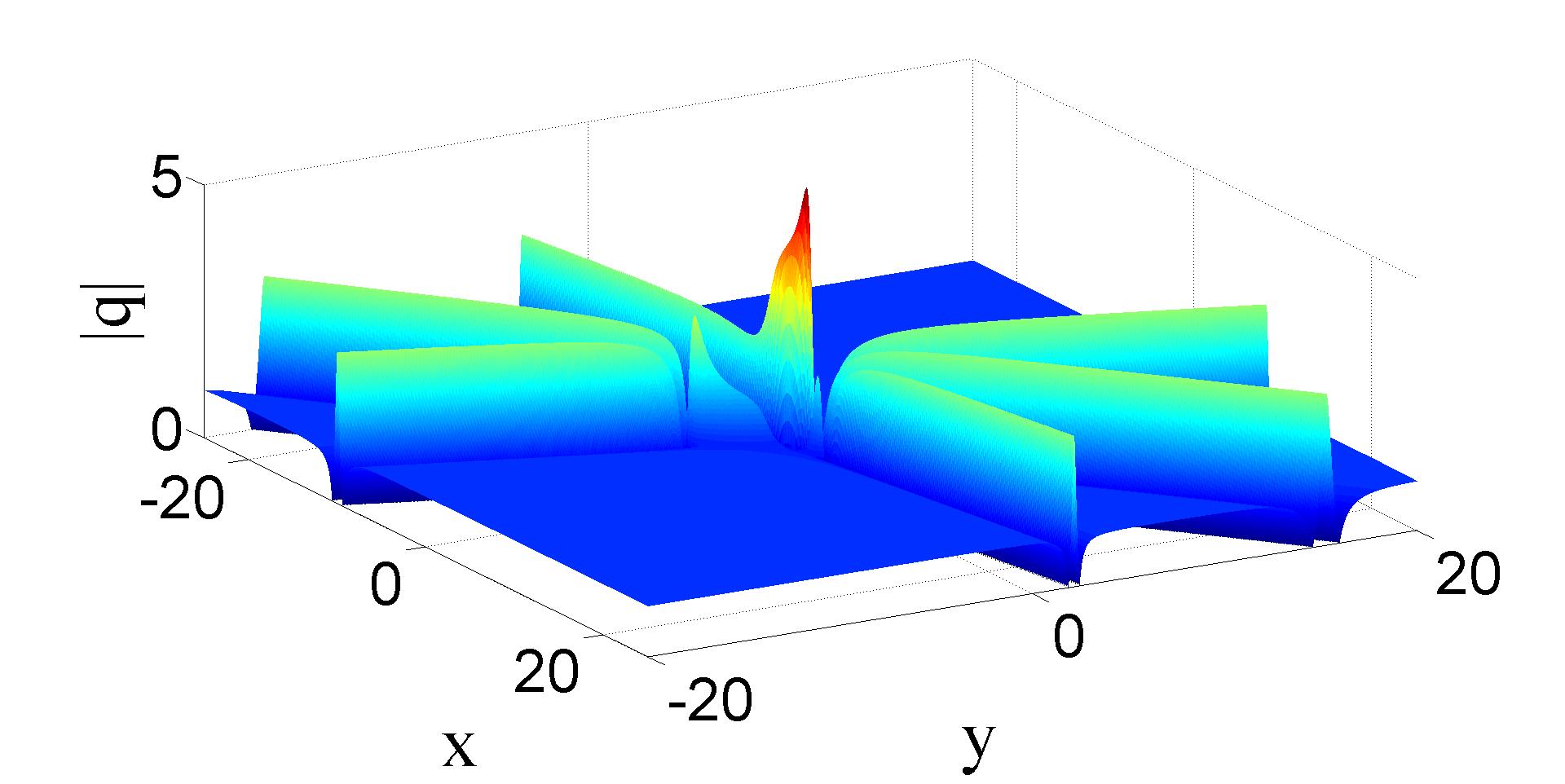}}\qquad
\subfigure[$t=10$]{\includegraphics[height=6.0cm,width=7.5cm]{3rw-1.jpg}}
\caption{The time evolution of a third-order rogue wave of the DSIII equation with parameters given by equation \eqref{pa-5}.~}\label{fig9}
\end{figure}

For larger $N$ and $\gamma_{ij}=0$ with real $p_{i}$,  these higher-order rogue waves behave qualitatively similar as the second-order rogue waves shown in Fig. \ref{fig7}, but more rogue waves interact with each
other, and the wave patterns become more complicated. Besides, they always hold the characteristics ``appear from nowhere and disappear without a trace"\cite{rw2}.  The Fig. \ref{fig9} shows a third-order rogue wave with parameters
\begin{equation}\label{pa-5}
\begin{aligned}
N=3\,,\lambda=1\,,p_{1}=\frac{1}{2}\,,p_{2}=\frac{2}{3}\,,p_{3}=\frac{3}{2}\,,\gamma_{ij}=0 \,,c_{i1}=0(\,1\leq i,j\leq3).
\end{aligned}
\end{equation}
Note that the maximum value of the solution $|q|$ can exceed 5 (i.e., five times the constant background) at some times, . Thus this interaction between these tree line rogue waves can generate higher peaks.
%%%%%%%%%%%%%%%%%%%%%%%%%%%%%%%%%%%%%%%%%%%%%%%%%%%%%%%%%%%%%%%%%
 %%%%%%%%%%%%%%%%%%%%%%%%%%%%%%%%%%%%%%%%%%%%%%%%%%%%%%%%%%%%%%%%%fig9

\begin{figure}[!htbp]
\centering
\subfigure[$t=-10$]{\includegraphics[height=6.0cm,width=7.5cm]{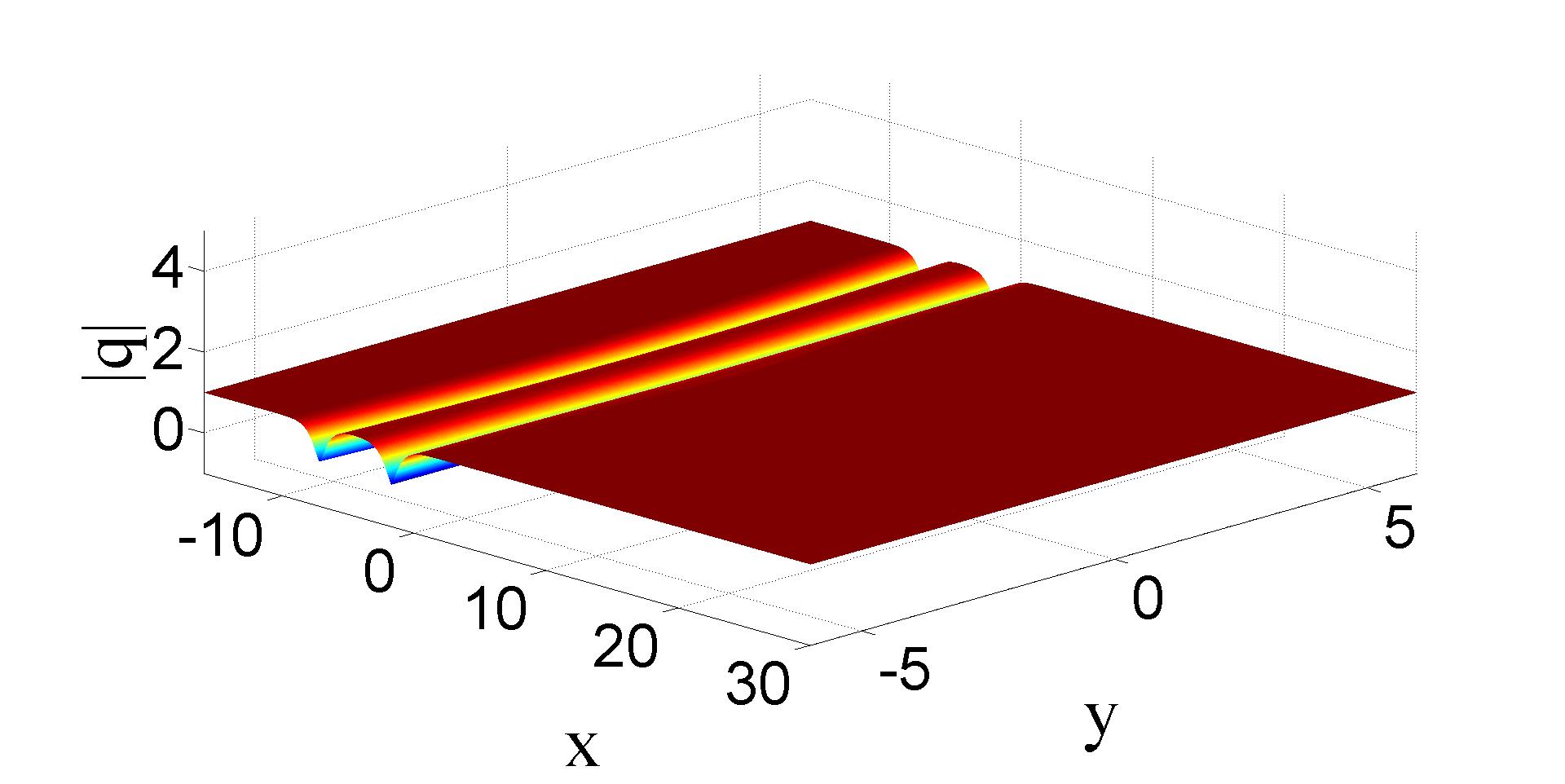}}\qquad
\subfigure[$t=0$]{\includegraphics[height=6.0cm,width=7.5cm]{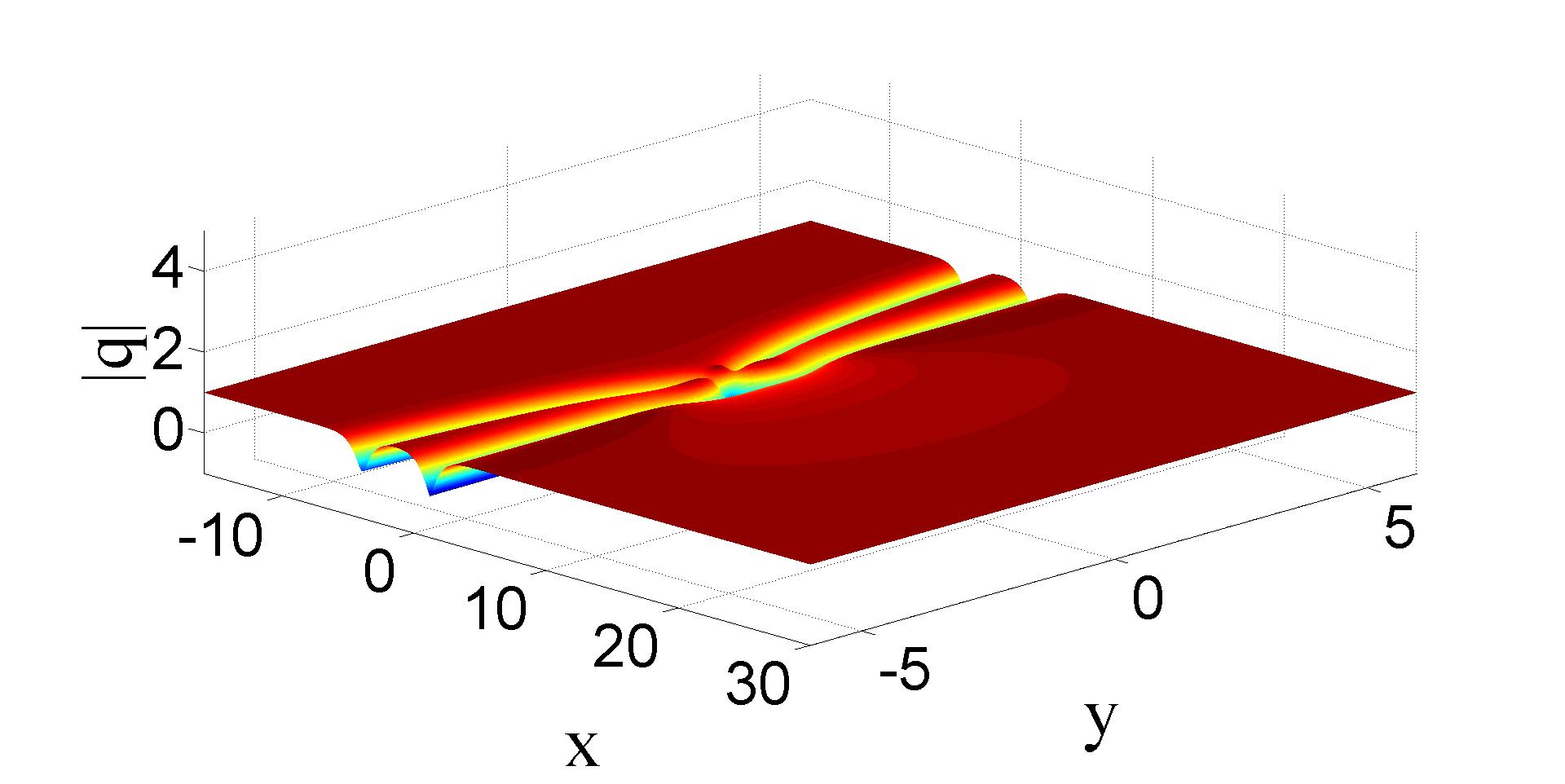}}
\subfigure[$t=1$]{\includegraphics[height=6.0cm,width=7.5cm]{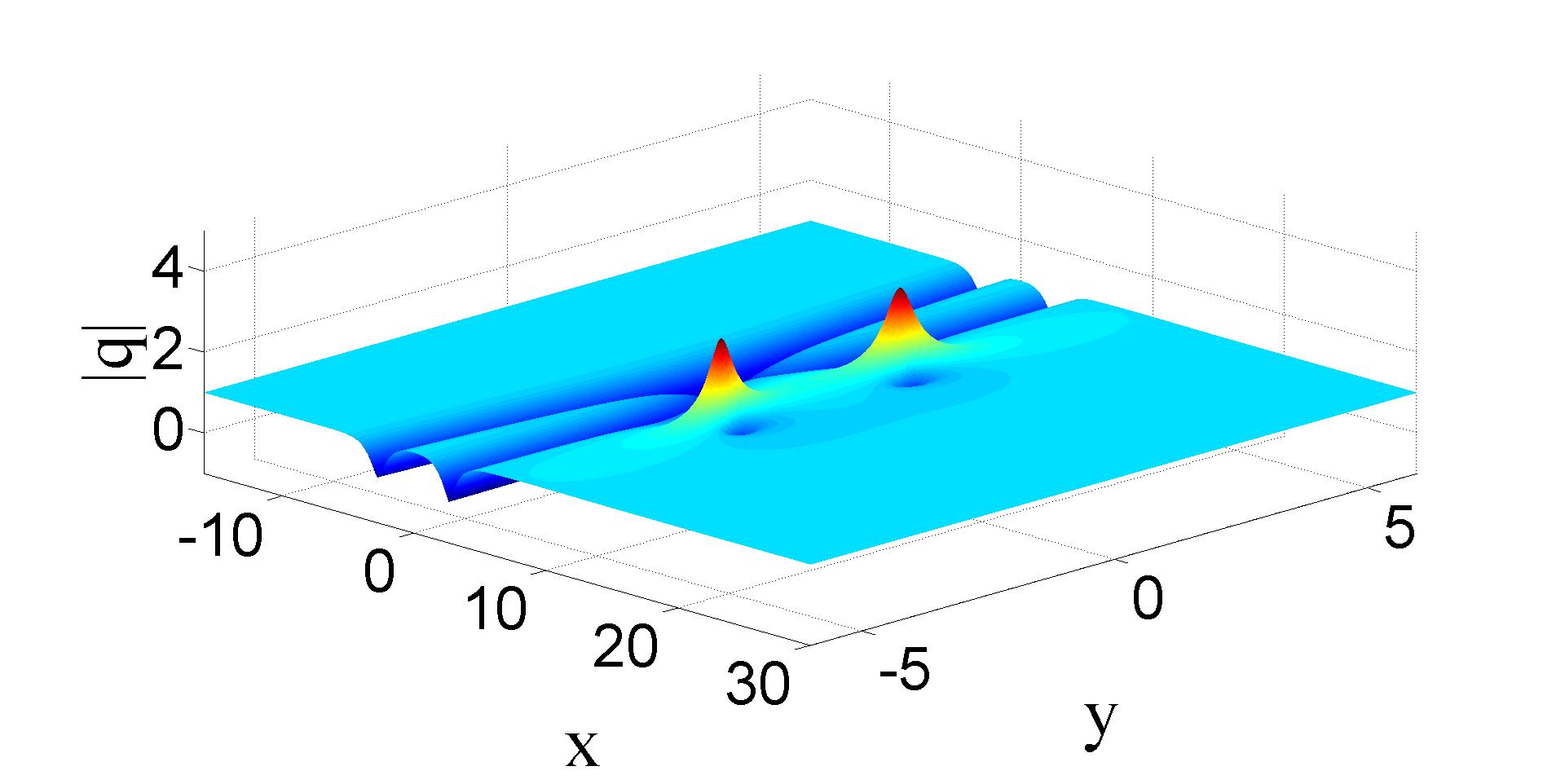}}\qquad
\subfigure[$t=10$]{\includegraphics[height=6.0cm,width=7.5cm]{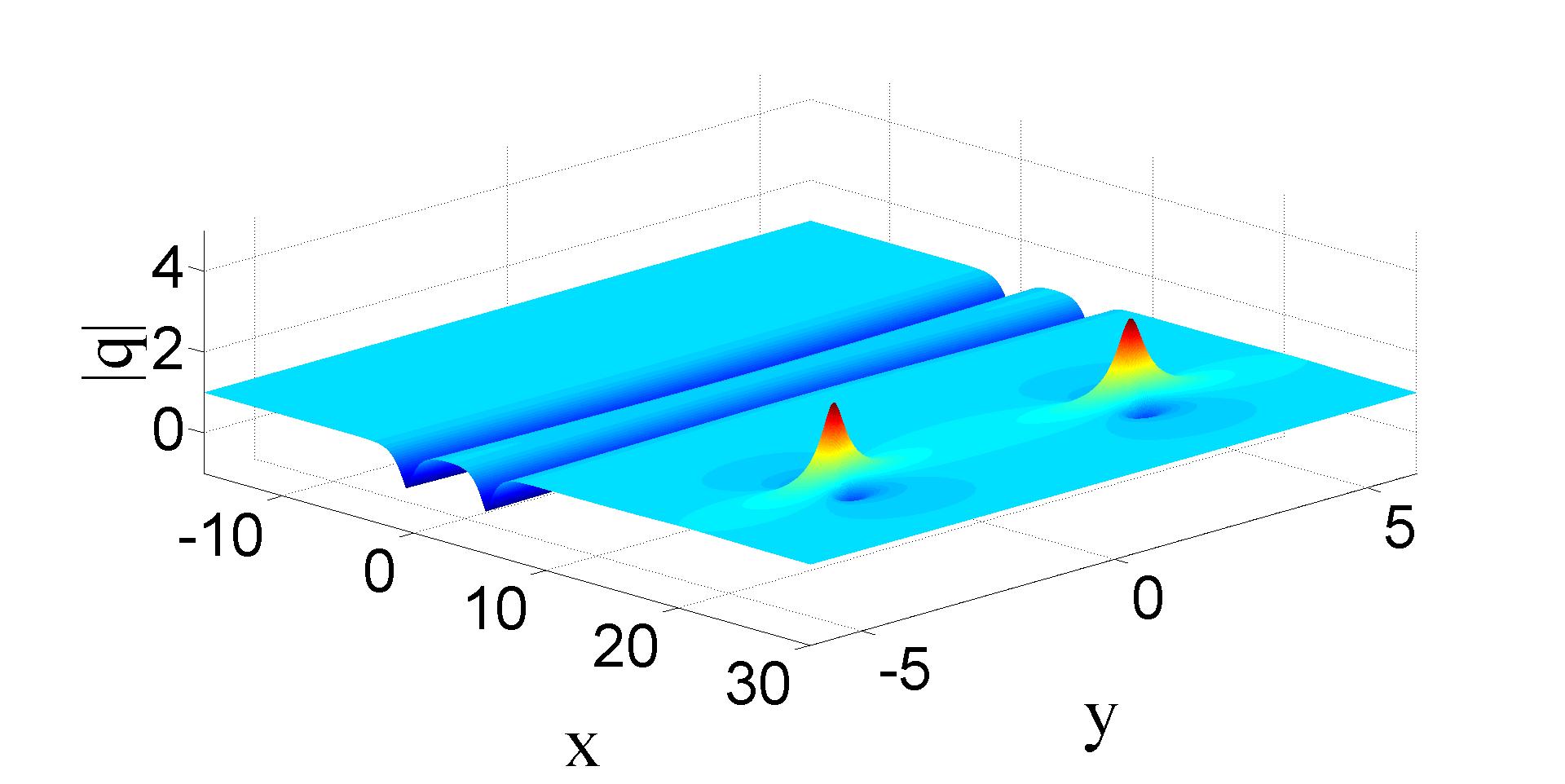}}
\caption{The time evolution of second-order lump on a second-order dark soliton background of the DSIII equation with parameters $\gamma_{11}=\gamma_{22}=1\,,\gamma_{12}=\gamma_{21}=0\,,\lambda=1\,,p_{1}=2+i\,,p_{2}=2+\frac{9}{10}i\,,c_{11}=10^{-4}\,,c_{21}=10^{-4}\,,\xi_{10}=\xi_{20}=0$.~}\label{fig10}
\end{figure}

(ii)  When one takes $\gamma_{ii}=1\,,\gamma_{ij}=0\,(\,i\neq j\,)$ and real parameters $p_{i}$ ( $p_{i}\neq1$ ) in equation (\ref{multi-2}), this solution demonstrates the second-order rogue waves on a two-dark-soliton background. However, when one takes $\gamma_{ii}=1\,,\gamma_{ij}=0\,(\,i\neq j\,)$ and complex parameters $p_{i}$ in equation (\ref{multi-2}), this solution describes two-dark soliton when $t\rightarrow-\infty$, see Fig.\ref{fig10}. In the intermediate time, two lumps form on the two-dark soliton
gradually (see the panels at $t=0,1$). When $t\rightarrow+\infty$, the original two-dark soliton develops into two-dark soliton and two lumps (see the panel at $t=10$).
%%%%%%%%%%%%%%%%%%%%%%%%%%%%%%%fig10

\begin{figure}[!htbp]
\centering
\subfigure[$t=-10$]{\includegraphics[height=6.0cm,width=7.5cm]{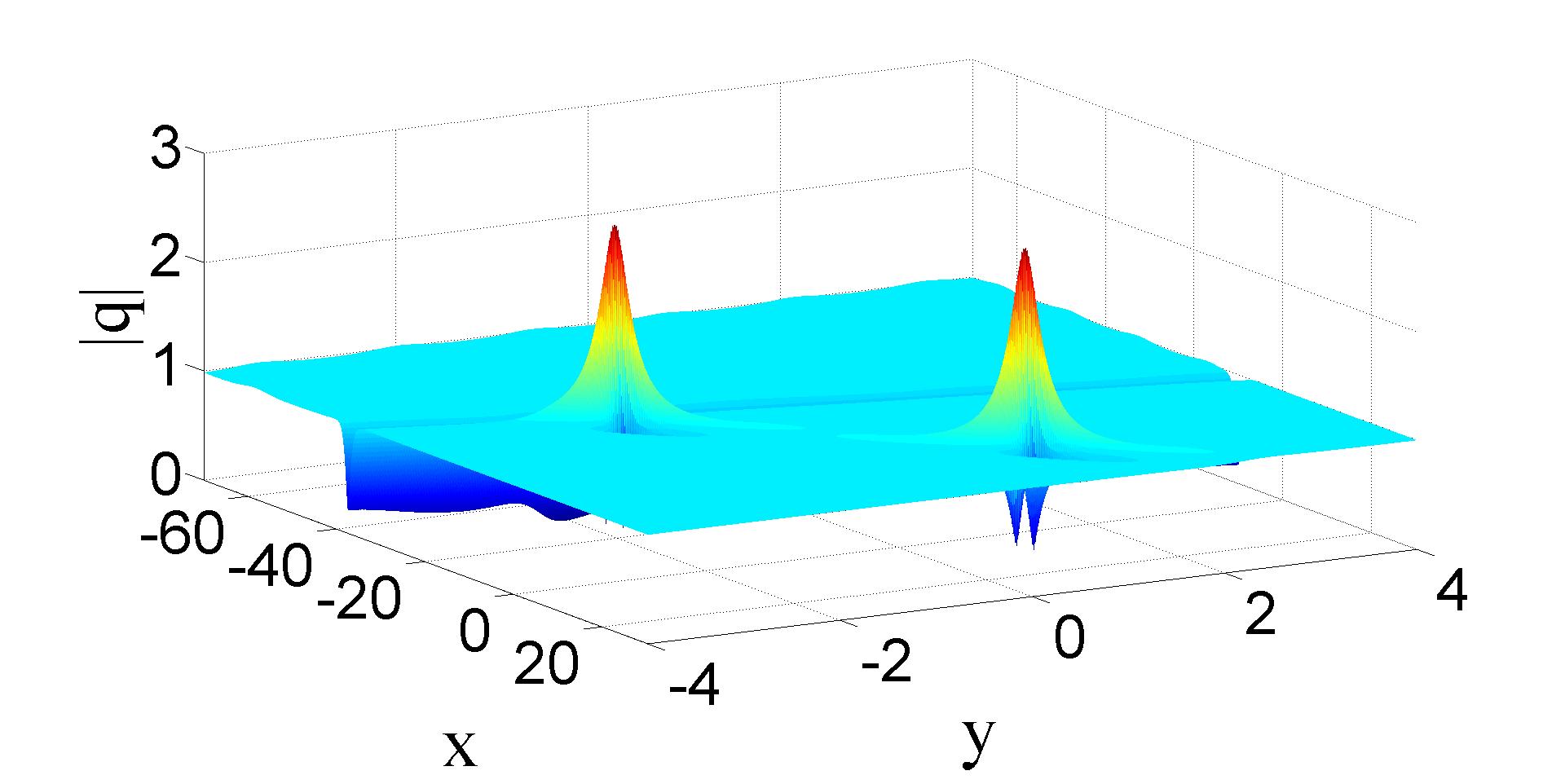}}\qquad
\subfigure[$t=-1$]{\includegraphics[height=6.0cm,width=7.5cm]{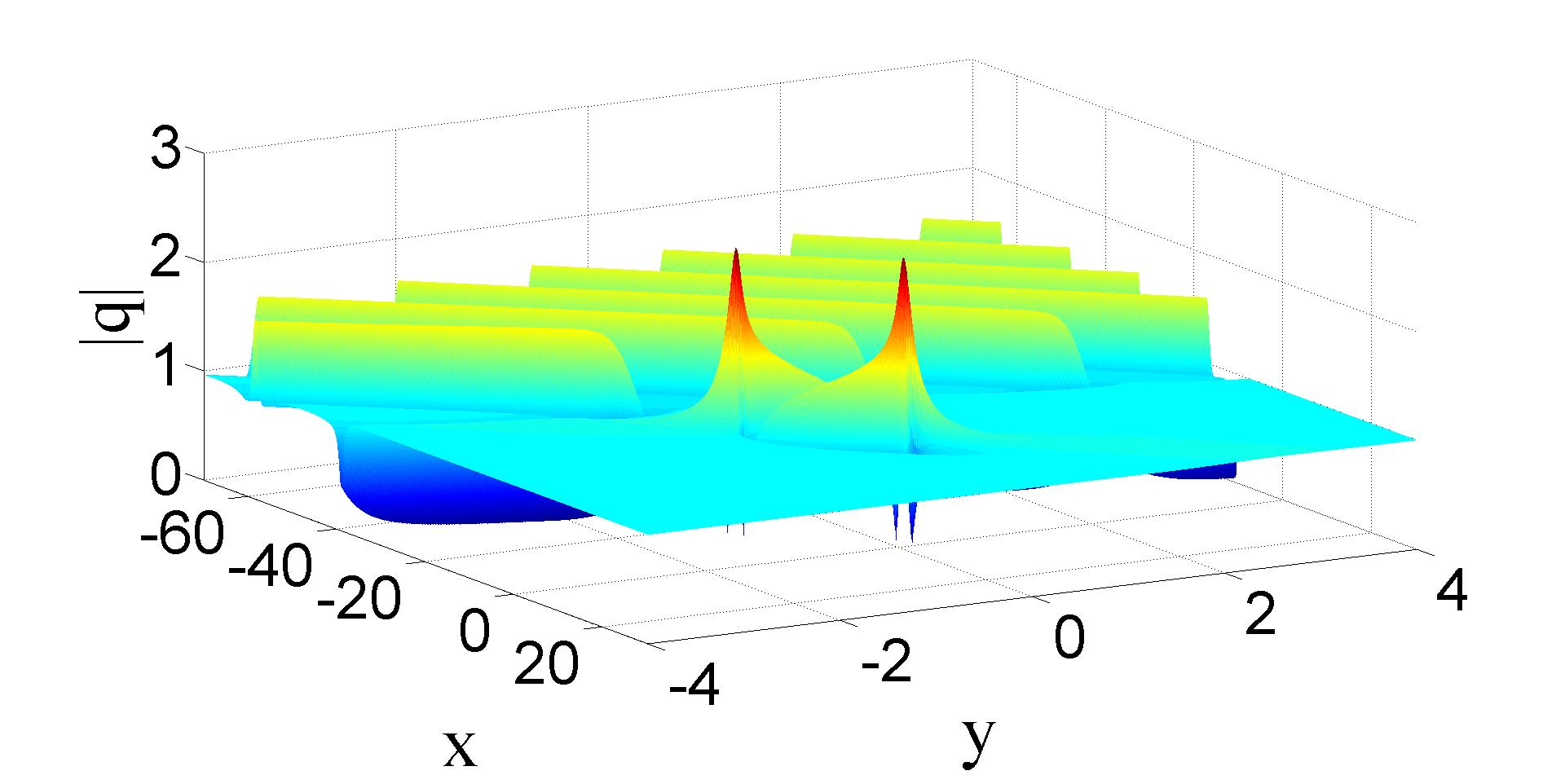}}
\subfigure[$t=0$]{\includegraphics[height=6.0cm,width=7.5cm]{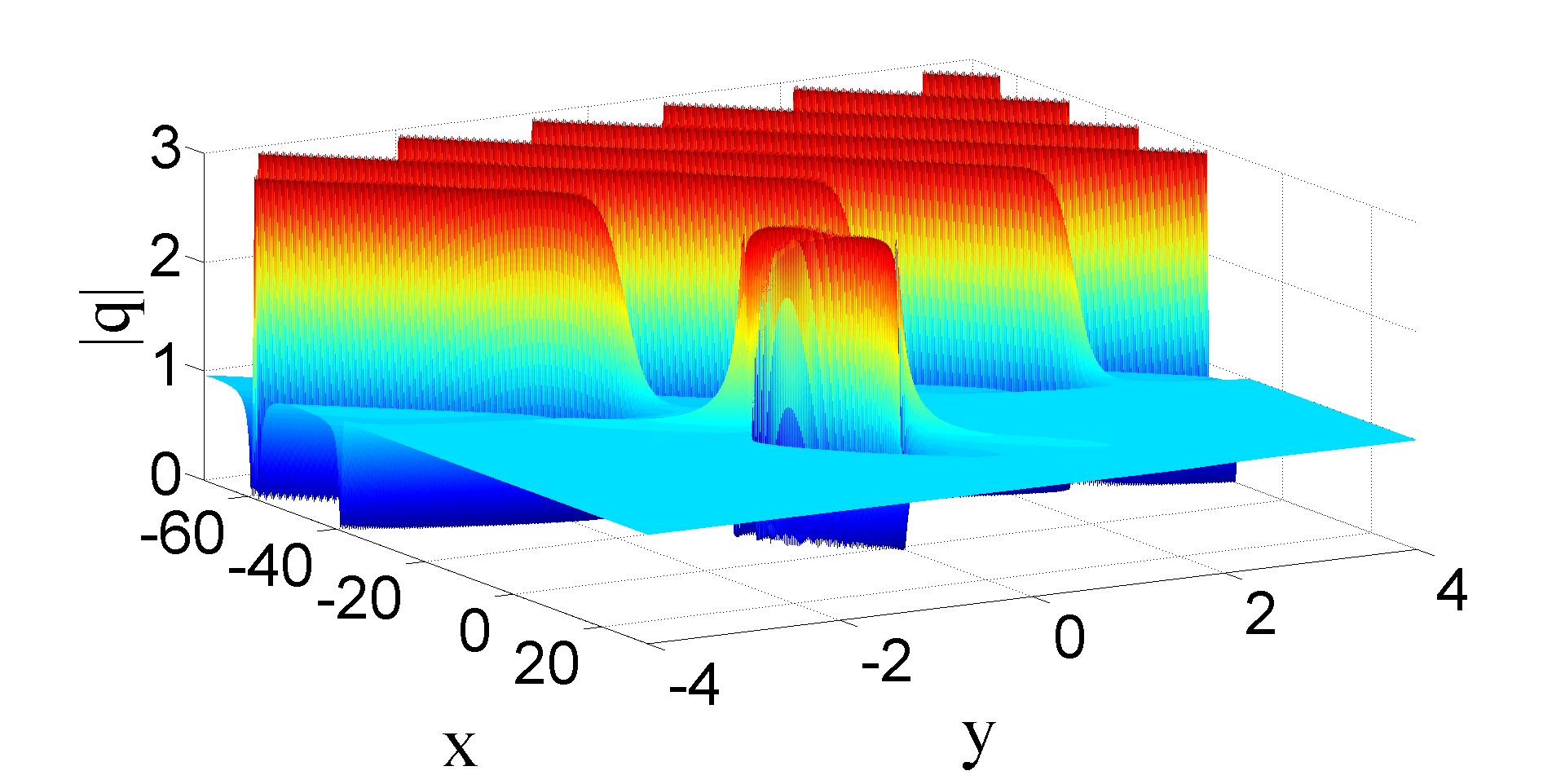}}\qquad
\subfigure[$t=10$]{\includegraphics[height=6.0cm,width=7.5cm]{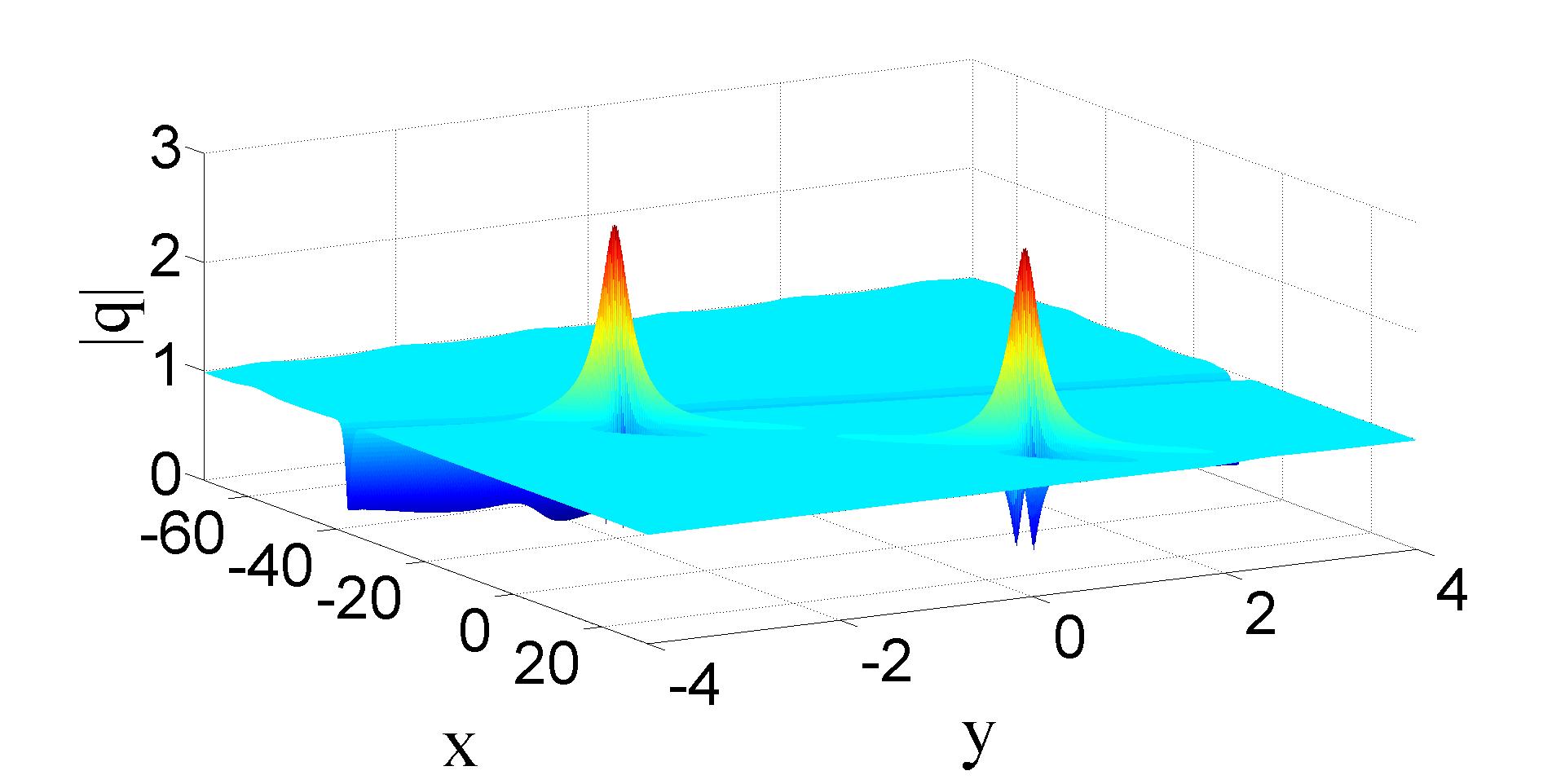}}
\caption{The time evolution of mutli-rational solutions \eqref{multi-2} consisting of two lumps, a dark soliton and a line breather of the DSIII equation with parameters \eqref{pa-6}.~}\label{fig11}
\end{figure}

(iii)  When one takes $\gamma_{ij}=1\,(\,1\leq i, j\leq2\,)$ and complex parameters $p_{i}$ satisfies $p_{1}=p_{2}^{*}$, this solution is a hybrid of two lumps, one dark soliton and a line breather. This solution with parameters
\begin{equation}\label{pa-6}
\begin{aligned}
N=2\,,\lambda=1\,,p_{1}=1+\frac{1}{2}i\,,p_{2}=1-\frac{1}{2}i\,,\gamma_{ij}=1 \,,c_{i1}=0(\,1\leq i,j\leq2),
\end{aligned}
\end{equation}
 is shown in Fig.\ref{fig11}. As can be seen, when $t\rightarrow \pm\infty$, this solution is only composed of two lumps and a dark soliton  ( see the panels at $t=\pm 10$ ).  Differently, these two lumps exist on the background for all times, which form on the dark soliton as shown in Fig.\ref{fig10}. In the intermediate time, the line breather arises from the constant background , and retreat back to it at larger time (see the panels at $t=-1,0$ ). During this short period, this solution is made up of two lumps, a dark soliton and one line breather.  To the best of our knowledge, this kind of mixed solution has not been reported before.
 %%%%%%%%%%%%%%%%%%%%%%%%%%%%%%%%fig10
%%%%%%%%%%%%%%%%%%%%%%%%%%%%%%%%%%%%%%%%%%%%%%%%%%%%%%%%%%%%%%%%%%%%%%%%%%%%%%%%%%%%%%%%%%%%%%%%%%%%%%%%%%%%%%%%%%%%%%%%%%%%fig12

\subsection{Higher-order semi-rational solutions}$\\$

Another subclass of non-fundamental semi-rational solutions is the higher-order semi-rational solutions, which can be derived by taking $N=1\,,n_{i}>1$ in equation (\ref{rational}). This subclass of semi-rational solutions can be classified into different patterns:

(a) Higher order rogue waves. When one takes $\gamma_{11}=0$ and real parameter $p_{1}$ except $\pm1$ in equation (\ref{rational}), the corresponding semi-rational solutions become rational solutions, which is the higher order rogue waves in the DSIII equation.

(b) Higher order rogue waves on a dark soliton background. When one takes $\gamma_{11}=1$ and real parameter $p_{1}$ except $\pm1$, the semi-rational solutions are mixed solutions consisting of higher order rogue waves and a dark soliton. In this case, these rogue waves are also localized both in space and time in two dimensions.

(c) Higher order lumps on a dark soliton background. When one takes $\gamma_{11}=0$ and complex parameter $p_{1}$, these solutions describe $n_{i}$ lumps form on the dark soliton background gradually. When $t\rightarrow -\infty$, these semirational solutions are a dark soliton, while they are a hybrid of $n_{i}$ lumps and a dark soliton when $t\rightarrow +\infty$.

\begin{figure}[!htbp]
\centering
\subfigure[$t=-10$]{\includegraphics[height=6.0cm,width=7.5cm]{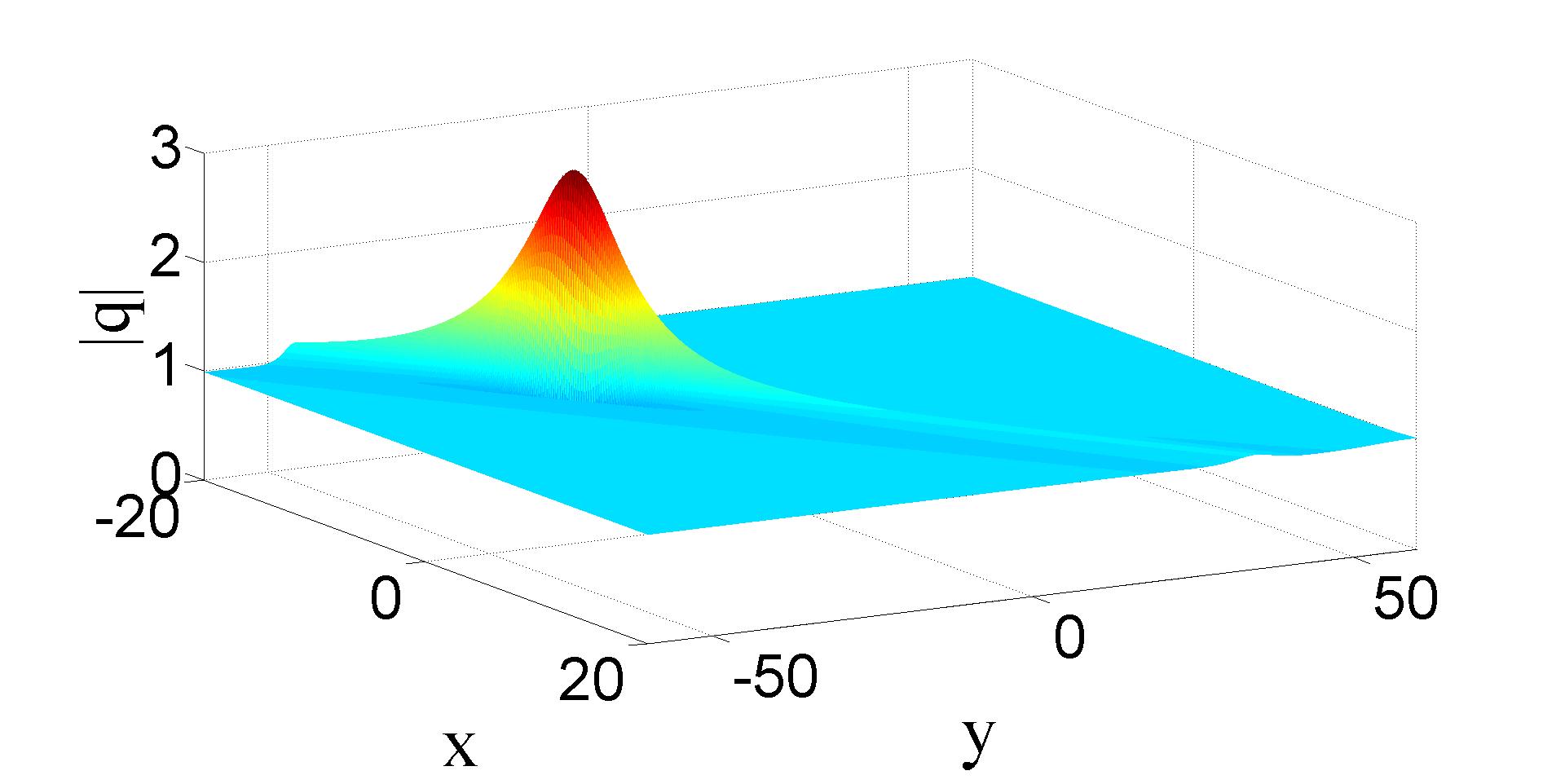}}\qquad
\subfigure[$t=-1$]{\includegraphics[height=6.0cm,width=7.5cm]{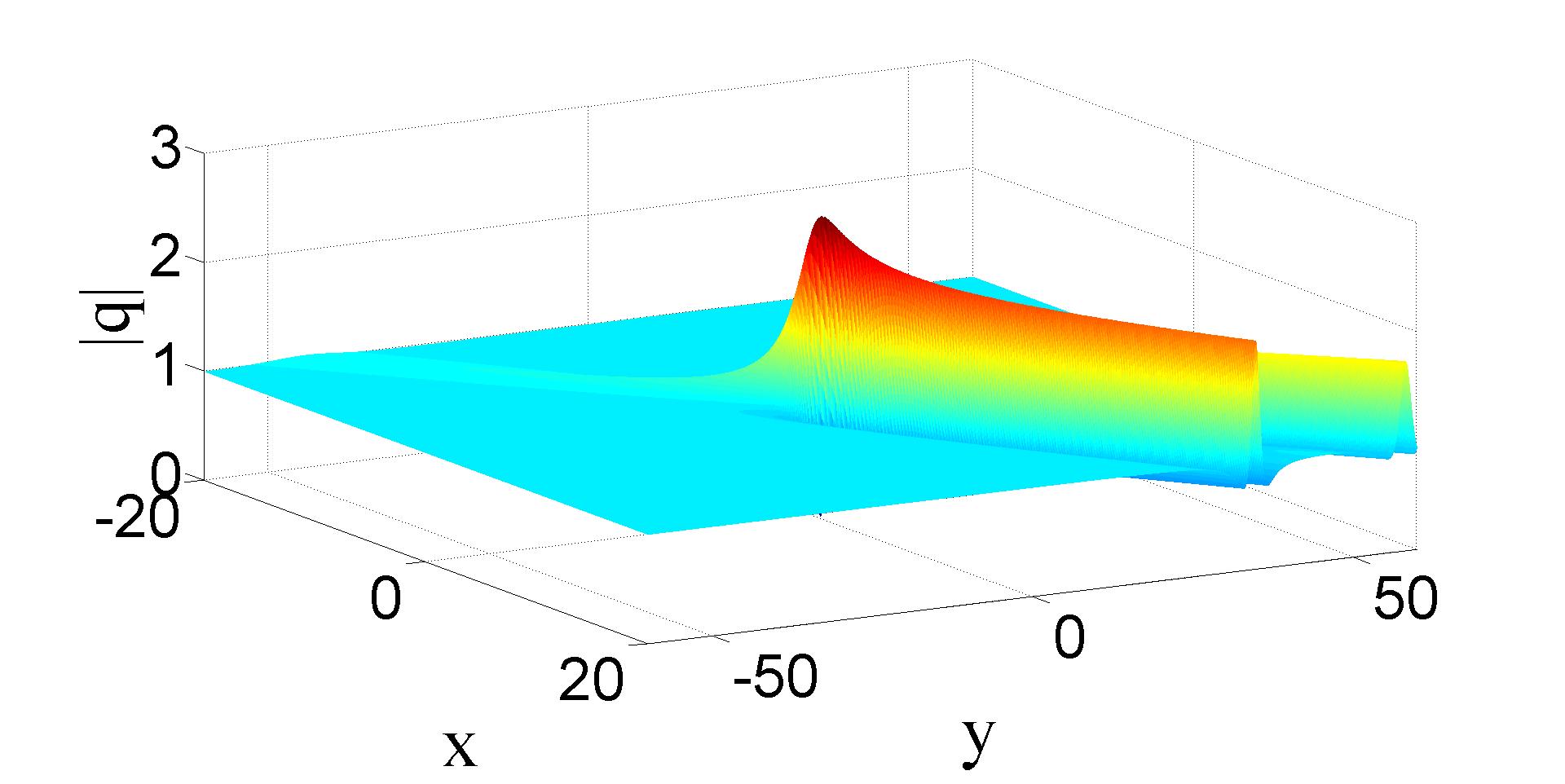}}
\subfigure[$t=0$]{\includegraphics[height=6.0cm,width=7.5cm]{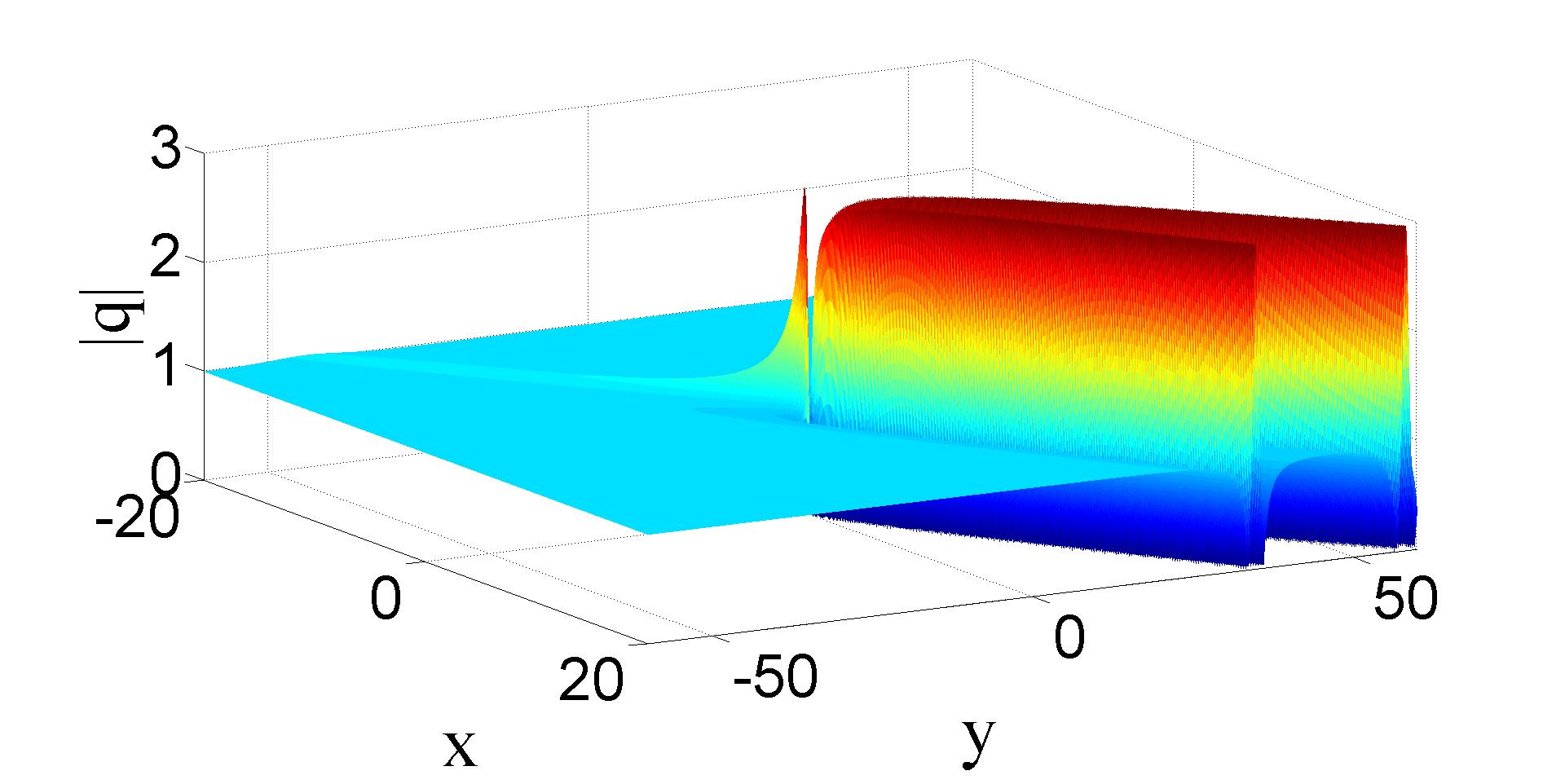}}\qquad
\subfigure[$t=10$]{\includegraphics[height=6.0cm,width=7.5cm]{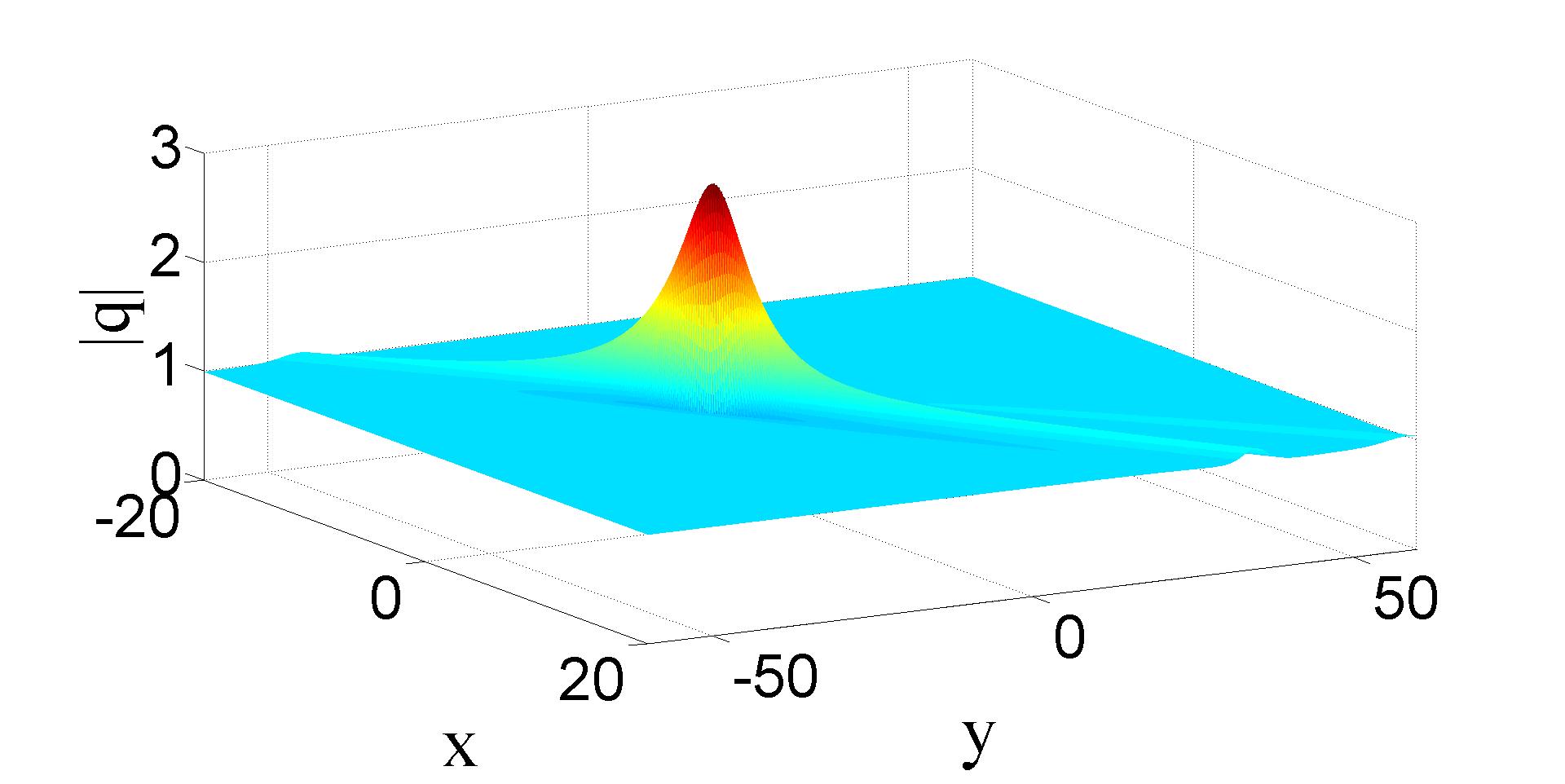}}
\caption{The time evolution of second-order rogue waves  of the DSIII equation defined in equation \eqref{higher-2} with parameters $\gamma_{11}=0\,,p_{1}=2\,,c_{11}=0\,,c_{12}=0\,,\xi_{10}=0$.~}\label{fig12}
\end{figure}

To demonstrate this subclass of non-fundamental semi-rational solutions in details, we take the case of $n_{i}=2$ as an example. In this case, the explicit forms of the second-order semi-rational solutions of the DSIII equation are given by the following formulae :
\begin{equation}\label{higher-2}
\begin{aligned}
q=\frac{g}{f}\,,
V={\rm -\lambda\,log} (f)_{yy}\,,
U={\rm -\lambda\,log} (f)_{xx},
\end{aligned}
\end{equation}
with
\begin{equation}\label{hfg}
\begin{aligned}
f=&e^{\xi_{1}+\xi^{*}_{1}}[(p_{1}\partial_{p_{1}}+\xi_{1}^{'})^{2}+c_{12}][(p^{*}_{1}\partial_{p^{*}_{1}}+\xi_{1}^{'*})^{2}+c_{12}^{*}]\frac{1}{p_{1}+p_{1}^{*}}+\gamma_{11}c_{12}c^{*}_{12}\,,\\
g=&e^{\xi_{1}+\xi^{*}_{1}}[(p_{1}\partial_{p_{1}}+\xi_{1}^{'}+1)^{2}+c_{12}][(p^{*}_{1}\partial_{p^{*}_{1}}+\xi_{1}^{'*}-1)^{2}+c_{12}^{*}]\frac{1}{p_{1}+p_{1}^{*}}+\gamma_{11}c_{12}c^{*}_{12}\,,
\end{aligned}
\end{equation}
and $\xi_{1}\,,\xi_{1}^{'}$ are defined in Theorem 2. By different parameters $p_{1}$ and $\gamma_{11}$, these solutions can be classified into following patterns:

(1) When taking $\gamma_{11}=0$ and real parameter $p_{1}$, the corresponding solution is a second-order rogue wave, see Fig.\ref{fig12}.  Different from the second-order rogue wave shown in Fig.\ref{fig7}, when $t\rightarrow\pm\infty$, this rogue wave features a localized lump traveling with the constant background (see the panel $t=\pm10$ ). In the intermediate time, the  lump disappears and the rogue wave features a parabola-shaped line waves(see the panel at $t=0$ ) .
%%%%%%%%%%%%%%%%%%%%%%%%%%%%%%%%%%fig12

\begin{figure}[!htbp]
\centering
\subfigure[$t=-10$]{\includegraphics[height=5.0cm,width=5.0cm]{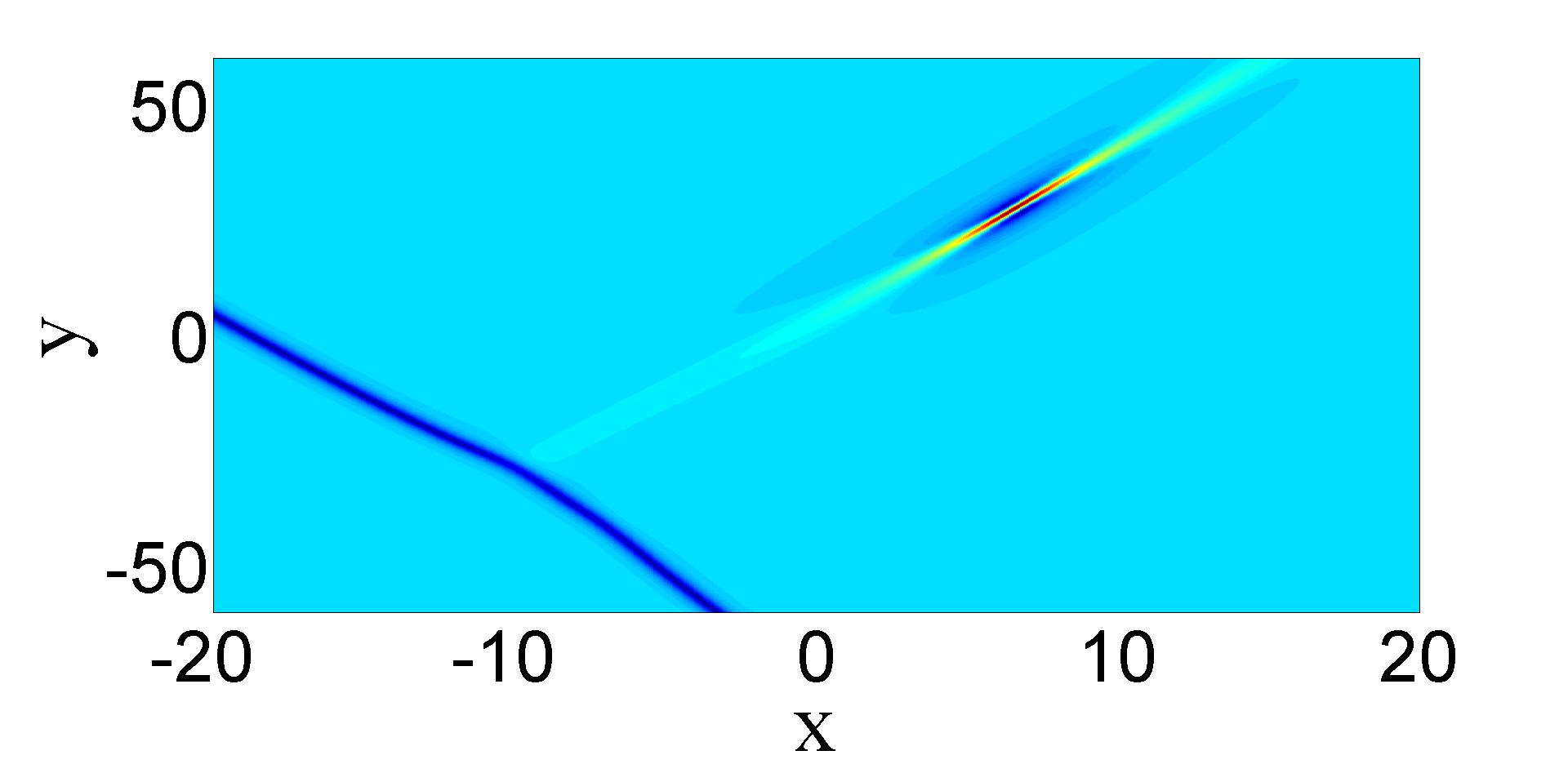}}\quad
\subfigure[$t=-1$]{\includegraphics[height=5.0cm,width=5.0cm]{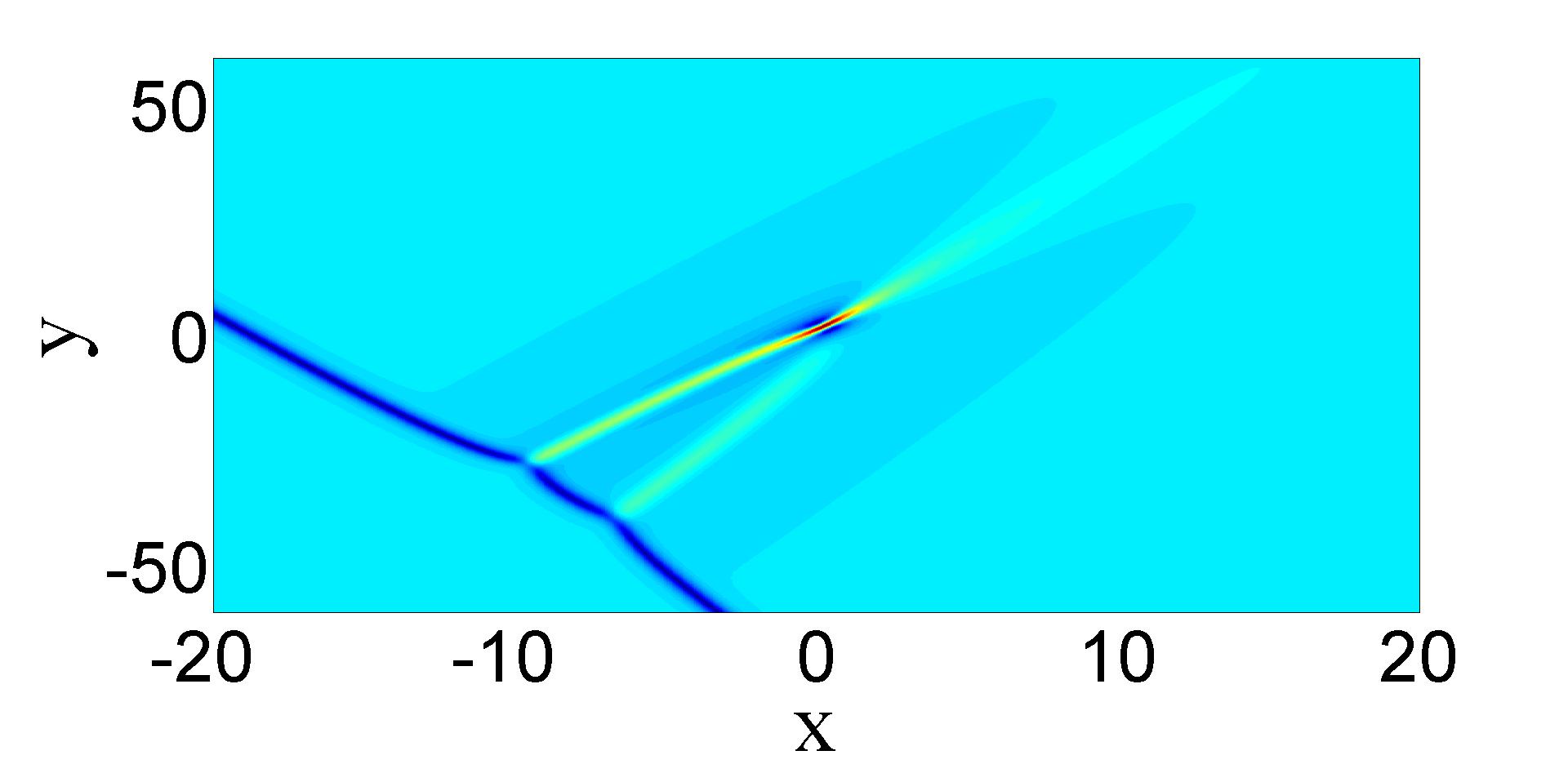}}\quad
\subfigure[$t=0$]{\includegraphics[height=5.0cm,width=5.0cm]{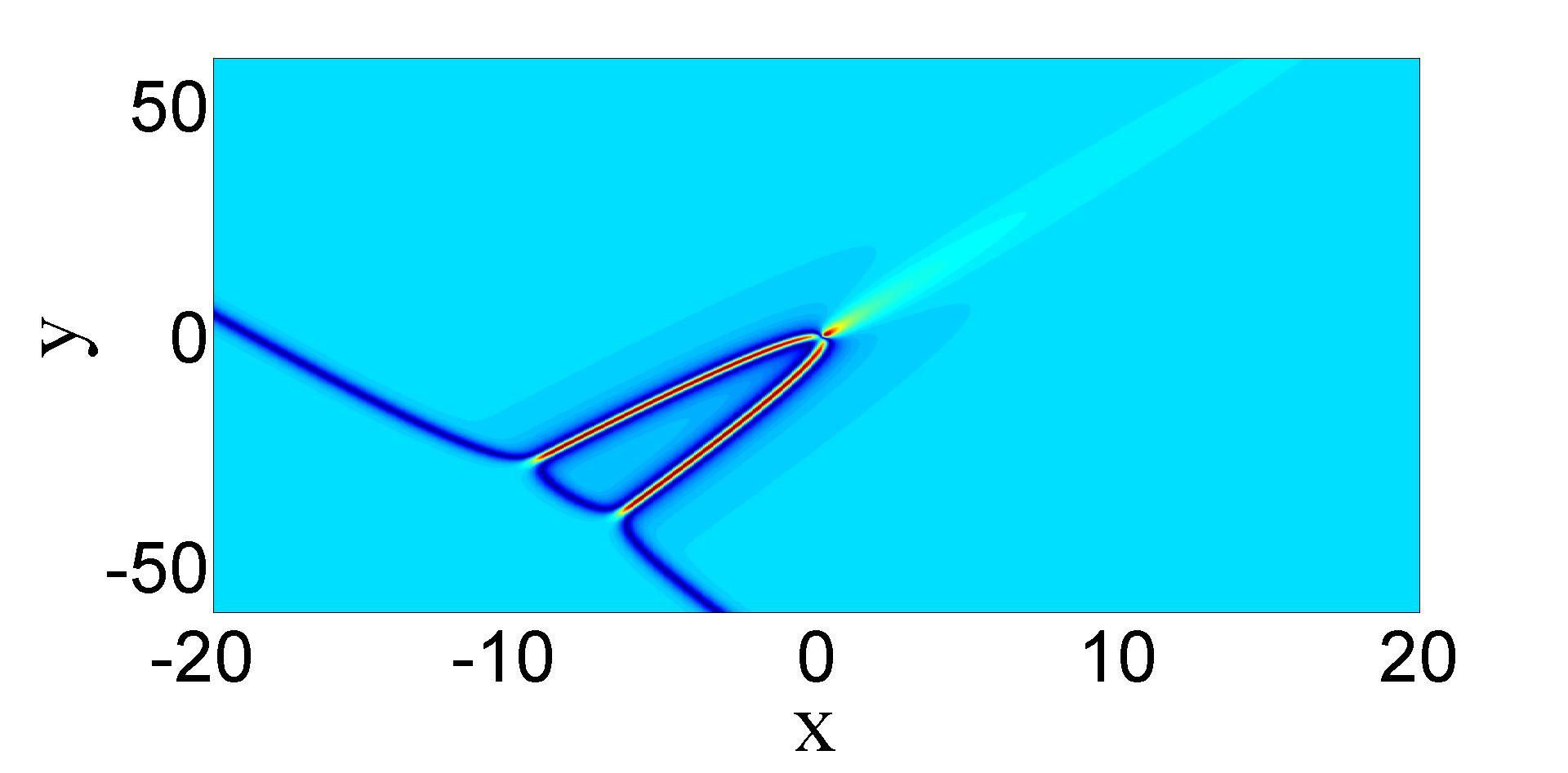}}
\subfigure[$t=1$]{\includegraphics[height=5.0cm,width=5.0cm]{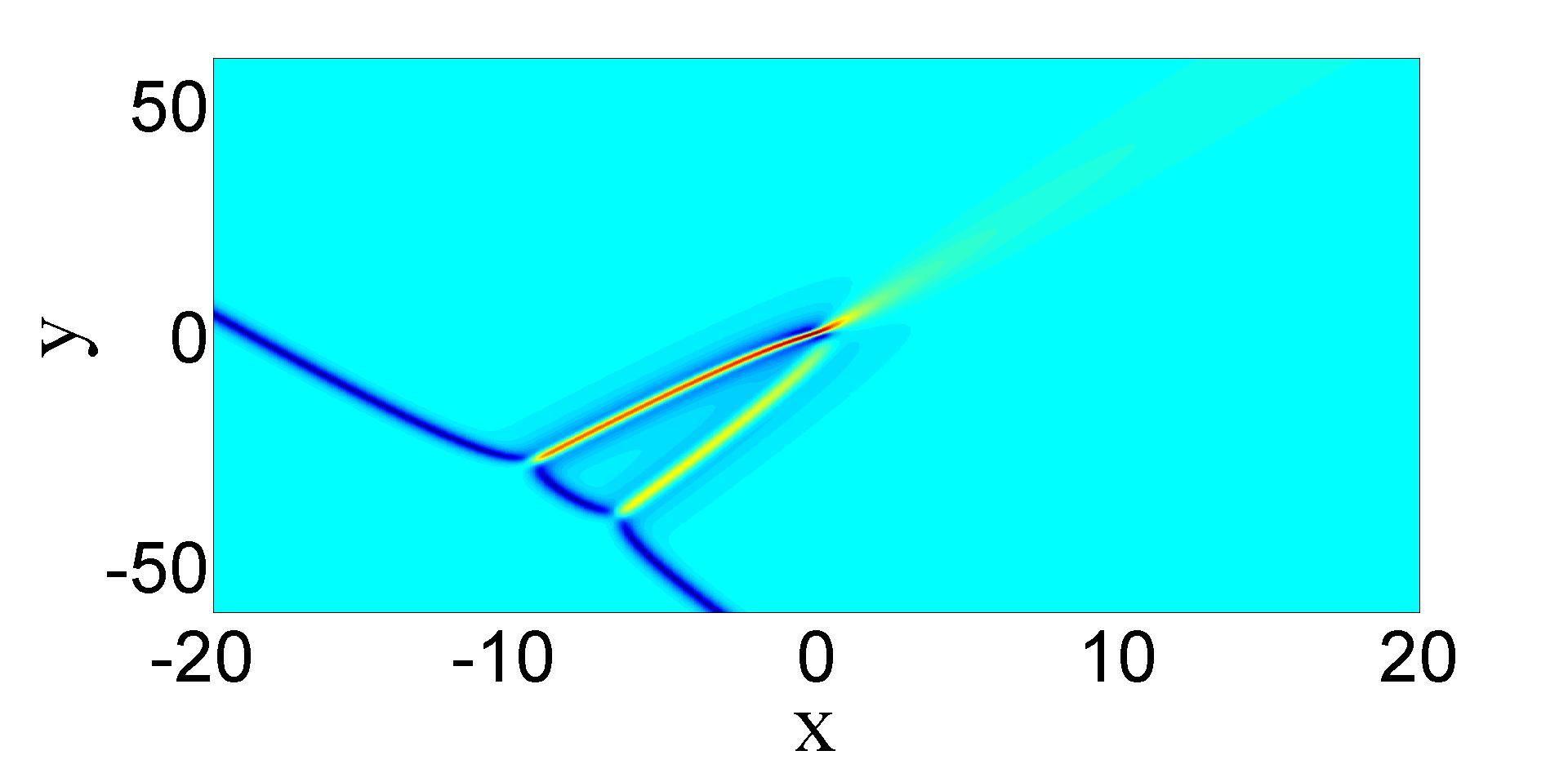}}\quad
\subfigure[$t=5$]{\includegraphics[height=5.0cm,width=5.0cm]{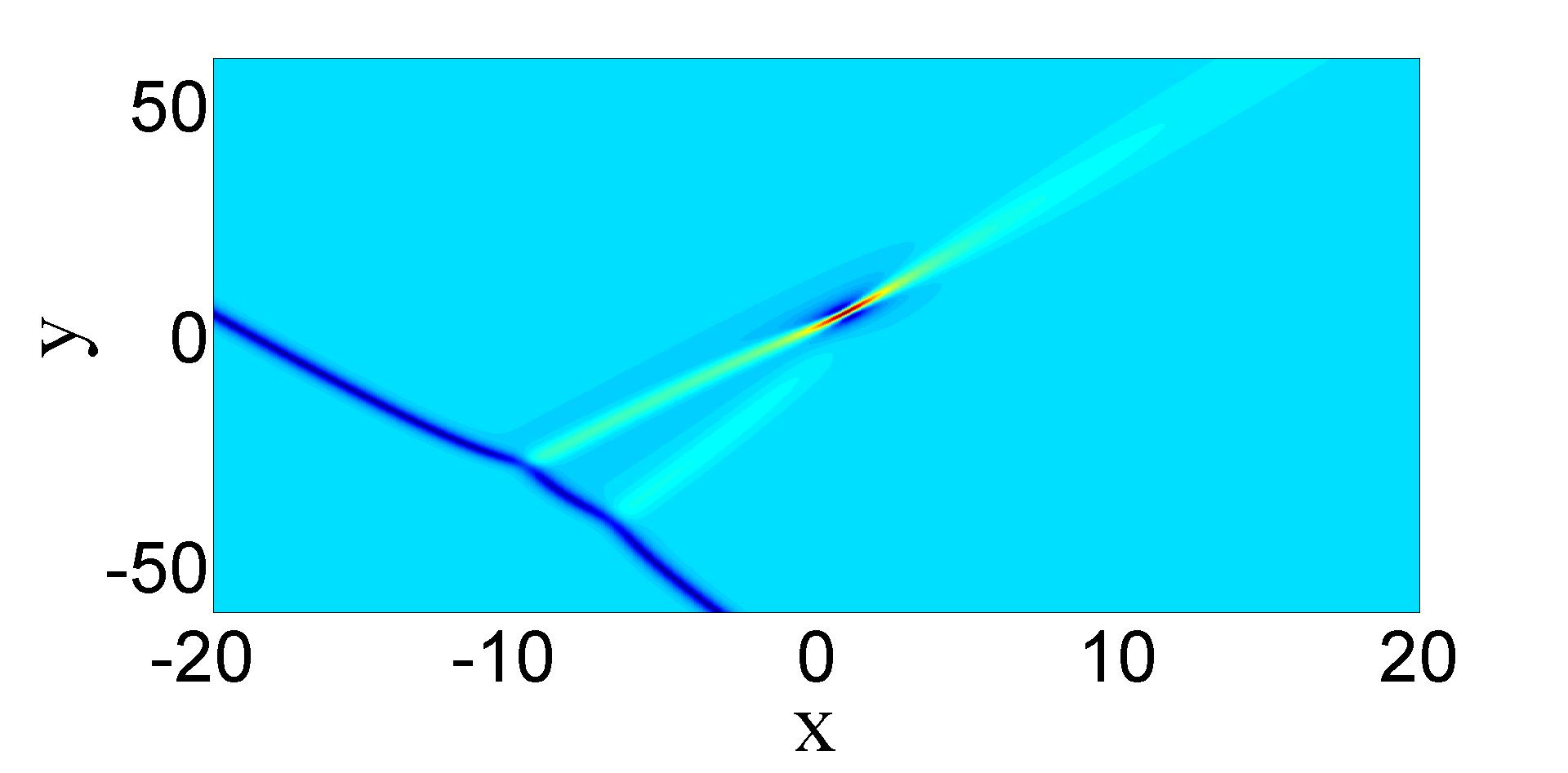}}\quad
\subfigure[$t=10$]{\includegraphics[height=5.0cm,width=5.0cm]{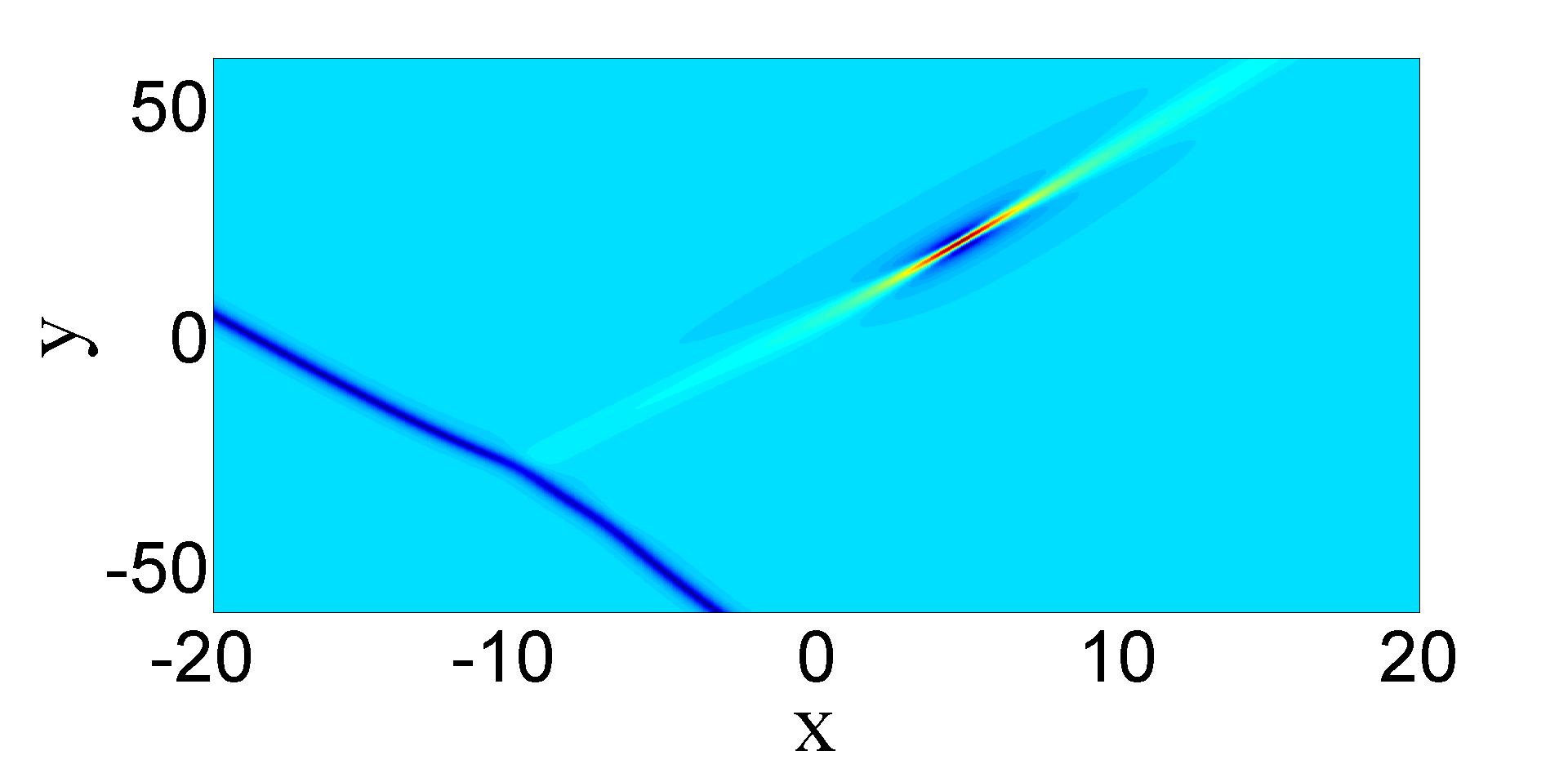}}
\caption{The time evolution of second-order rogue waves \eqref{higher-2} on a dark soliton background in the DSIII equation with parameters $\gamma_{11}=1\,,p_{1}=2\,,c_{11}=0\,,c_{12}=1\,,\xi_{10}=0$. The plotted is $|q|$ field. The constant background value is $1$.~}\label{fig13}
\end{figure}

(2) When taking $\gamma_{11}=1$ and real parameter $p_{1}$, these semi-rational solutions consisting of a second-order rogue wave and a dark soliton  possess a new phenomenon:
 line rogue wave localized both in space and time. As shown in Fig. \ref{fig13}, after the lump immerses into the line rogue wave,  this second-order line rogue wave approaches maximum amplitudes. It is to be noted that this higher order rogue wave just exist in a localized space (see the panel at $t=0$ ). At larger time, it disappears into the constant background and retreat back to a lump.  To the best of our knowledge, this type of line rogue waves which are localized in both space and time have never been reported before.
 %%%%%%%%%%%%%%%%%%%%%%%%%%%%%%%%%%%%%%%%%%%%%%fig113
 %%%%%%%%%%%%%%%%%%%%%%%%%%%%%%%%%%%%%%%%%%%%%%%%%%%%%%%%%%%%%%%%%fig13
 %%%%%%%%%%%%%%%%%%%%%%%%%%%%%%%%%%%%%%%%%%%%%%%%%%%%%%%%%%%%%%%%%

\begin{figure}[!htbp]
\centering
\subfigure[$t=-10$]{\includegraphics[height=6.0cm,width=7.5cm]{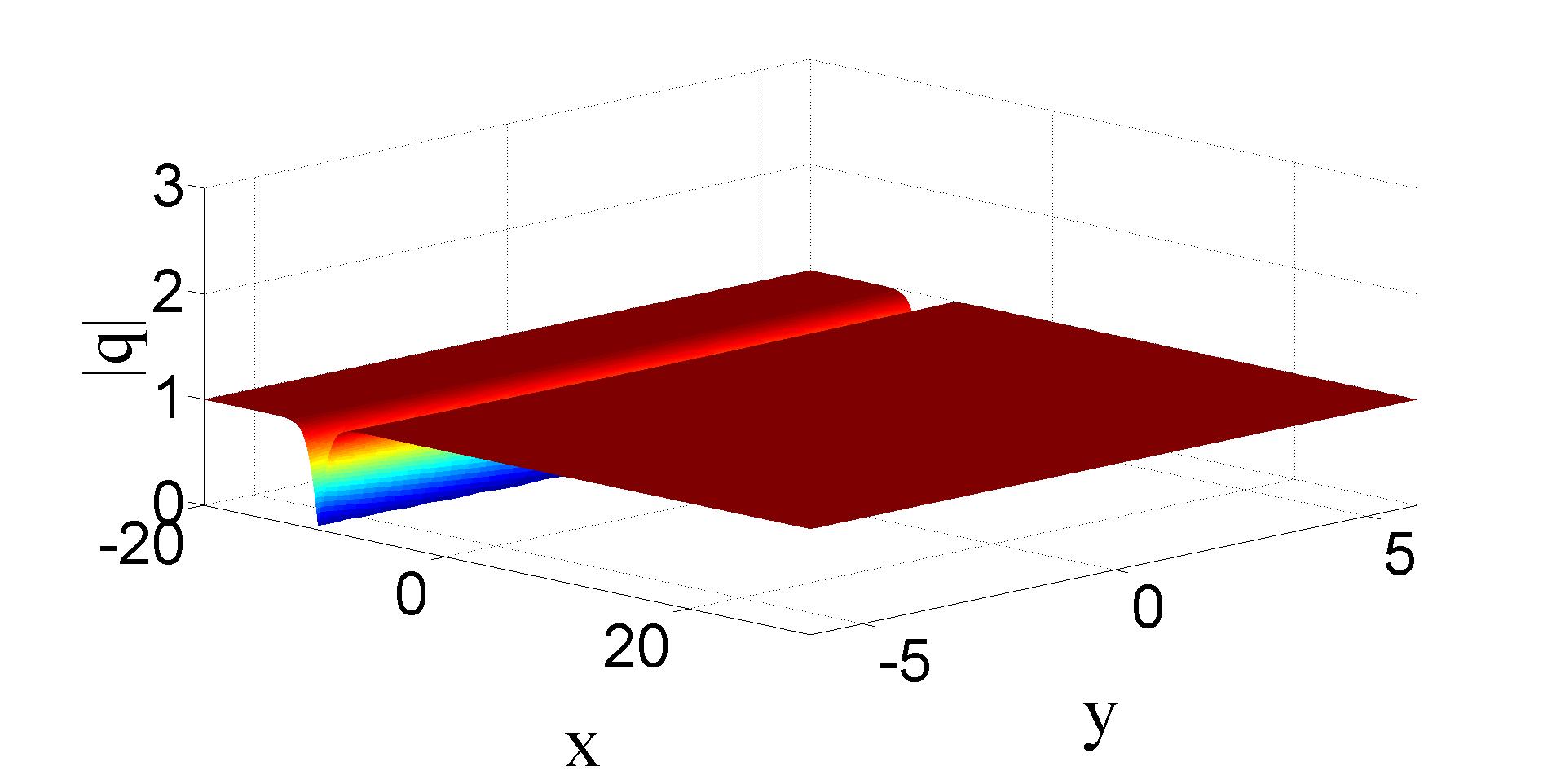}}\qquad
\subfigure[$t=-1$]{\includegraphics[height=6.0cm,width=7.5cm]{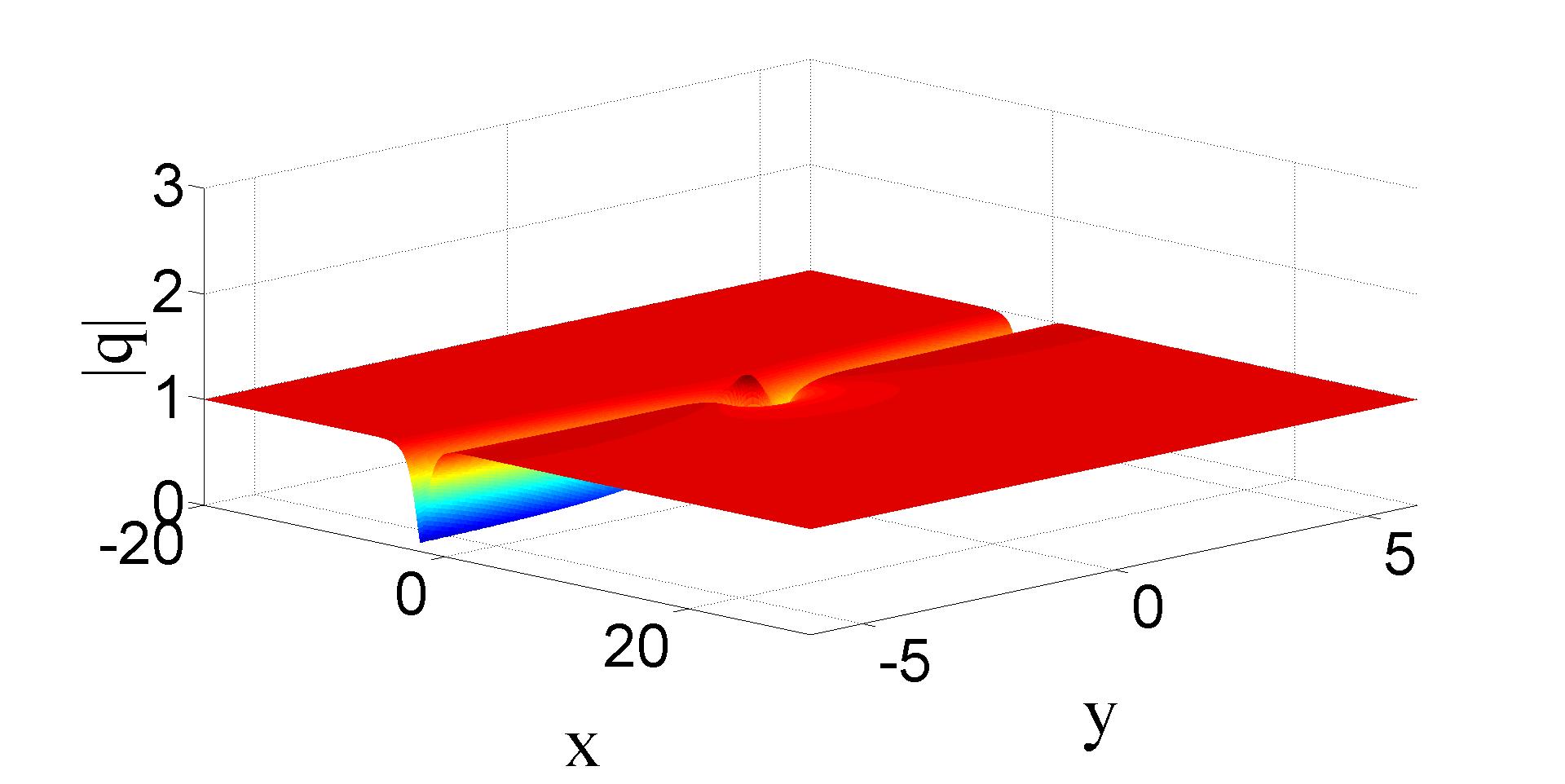}}
\subfigure[$t=0$]{\includegraphics[height=6.0cm,width=7.5cm]{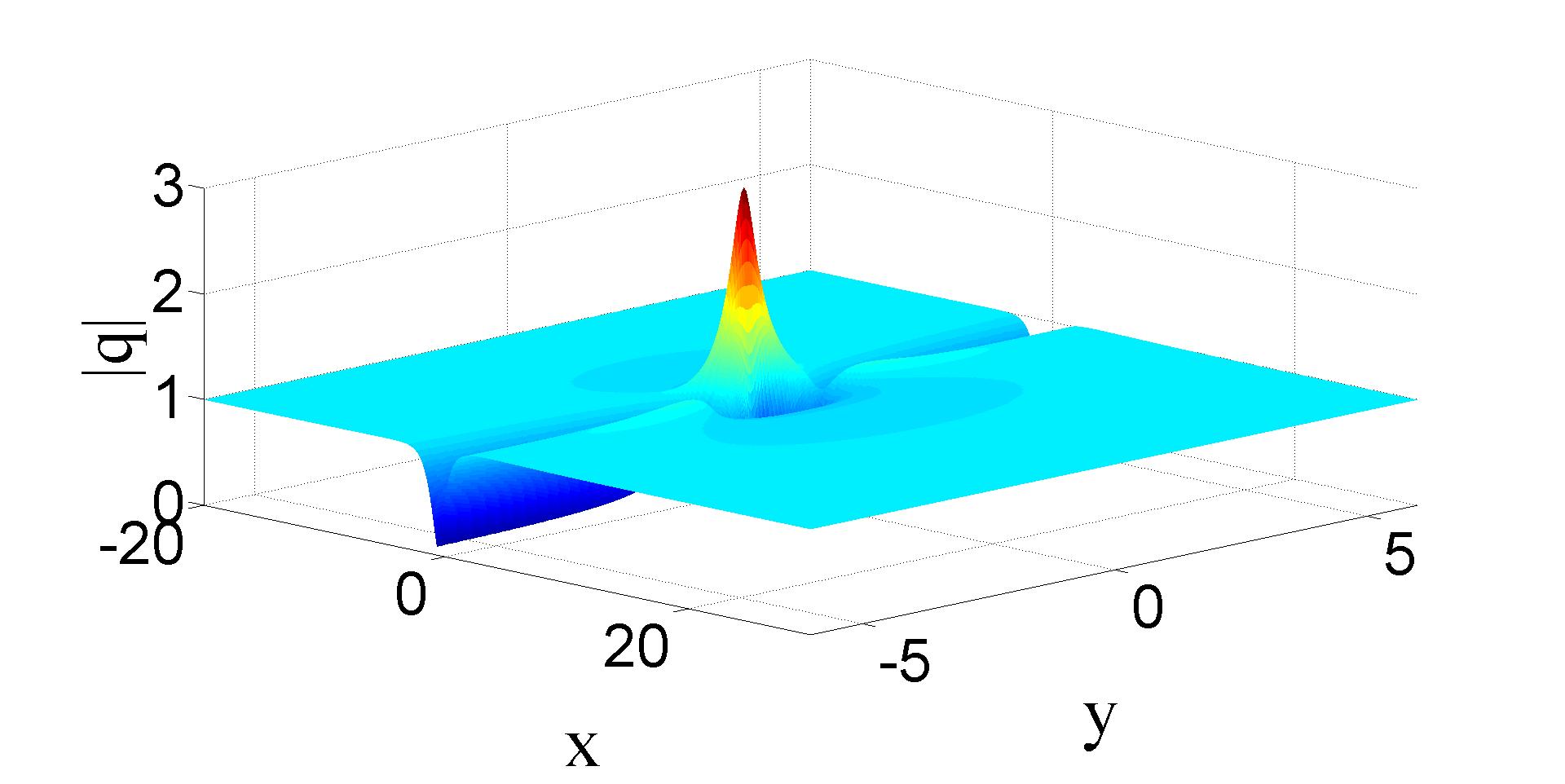}}\qquad
\subfigure[$t=10$]{\includegraphics[height=6.0cm,width=7.5cm]{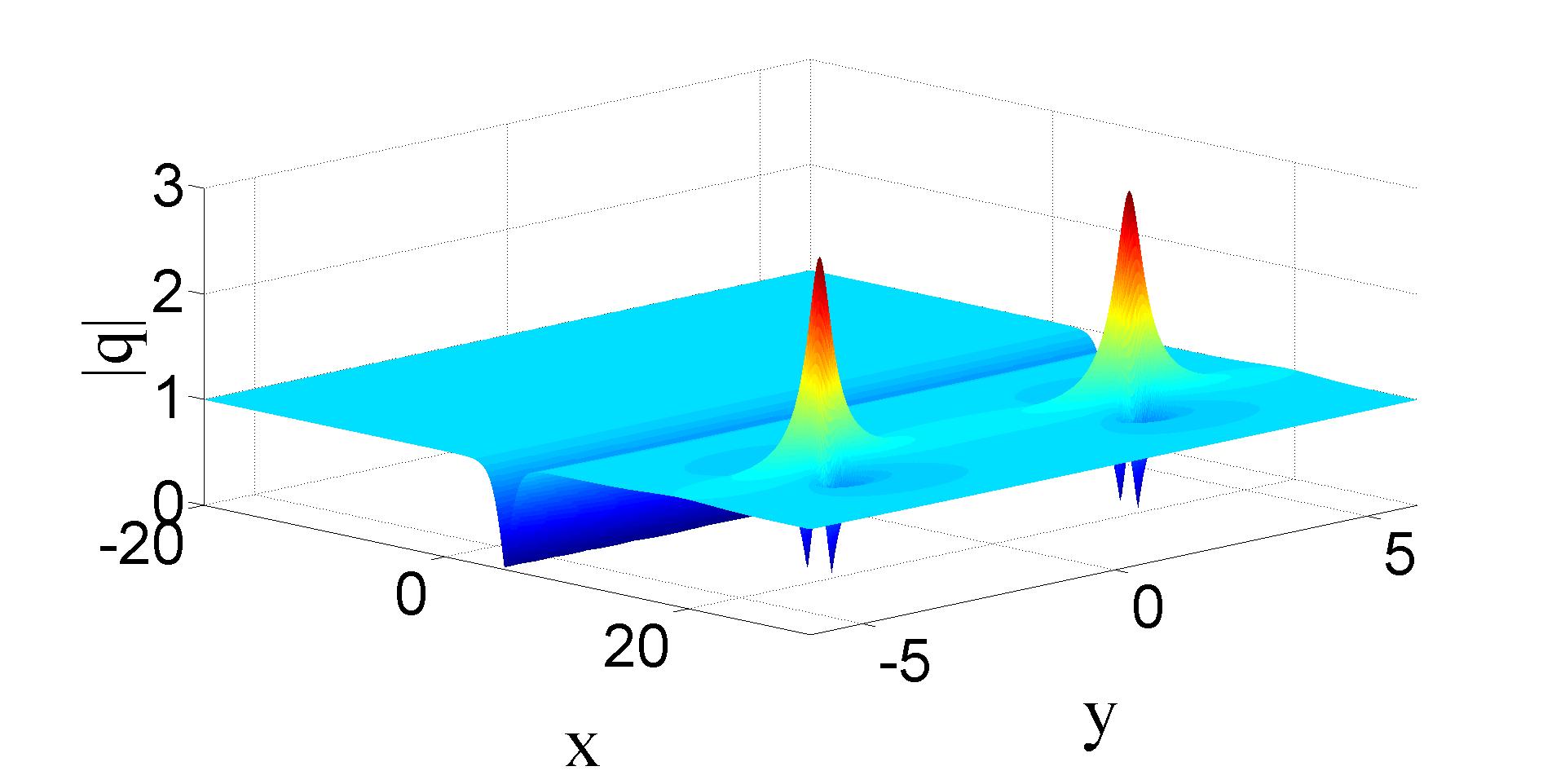}}
\caption{The time evolution of high-order semi-rational solutions consisting of two lumps and a dark soliton of the DSIII equation given by equation \eqref{higher-2} with parameters $\gamma_{11}=1\,,p_{1}=2+i\,,c_{11}=0\,,c_{12}=1\,,\xi_{10}=0$.~}\label{fig14}
\end{figure}

(3) When taking $\gamma_{11}=1$ and complex parameter $p_{1}$, in this situation, the corresponding solution describes a kind of unique dynamics, see Fig. \ref{fig14}.  As $t\rightarrow-\infty$, this solution is a dark soliton (see the panel at $t=-10$ ). However,
in the intermediate time, we observe gradual separation of  two lumps separated from the dark soliton, and the original dark soliton splits into one dark soliton and two lumps as $t\rightarrow+\infty$ ( see the panel at $t=10$ ). This unique dynamical behaviour corresponds to the numerical study of
blow up solutions in KP equation \cite{XX-1,XX-2}.

%%%%%%%%%%%%%%%%%%%%%%%%%%%%%%%%%%%%%%%%%%%%%%%%%%%%%%%%%%%%%%%%%%%%%%%%%%%%%%%%%%%%%%%%%%%%%%%%%%%%%%%%%%%%%%%%%%%%%%fig14

 \section{Summary and discussion}\label{5}
In this paper, we derived general $N$th-order dark solitons of the DSIII equation by reducing the Gram-type solutions of the KP hierarchy. Under suitable constraint in terms of parameters, a new kind of mixed solutions consisting of
 usual breathers, line breathers and dark solitons are given through Gram determinants, see Fig.\ref{fig2}. By introducing differential operators $A_{i}\,,B_{j}$, general semi-rational solutions containing rogue waves, breathers and dark solitons are generated. We showed that the fundamental rogue waves are line rogue waves, which arise from the constant background and retreat back to the constant background again. Except fundamental rogue waves interacting with  dark solitons, these fundamental semi-rational solutions also demonstrate another unique phenomenon: a lump form on the dark soliton and separate from the dark soliton (see Fig. \ref{fig6}).  We also showed that multi-rogue waves demonstrate the interaction of several fundamental line rogue waves, and interesting curvy wave patterns appear due to the interaction (see Fig.\ref{fig7} and Fig. \ref{fig9}).
 The multi-semi-rational solutions possess various dynamics, such as more lumps form on  multi-dark solitons (see Fig.\ref{fig10}), such as interaction of lumps, line breathers and dark solitons (see Fig. \ref{fig11}).
 Differently, the higher order rogue waves start from a localized lump and retreat back to it (see Fig.\ref{fig12}). A new kind of line rogue waves which are localized both in space and time are  illustrated, see Fig. \ref{fig13}. The higher order semi-rational solutions illustrating the dynamics that more lumps separated from the one soliton gradually have also been shown (see Fig. \ref{fig14}), these dynamics correspond to the numerical study of
blow up solutions in KP equation \cite{XX-1,XX-2}.

Thus, our reported  results in this paper  are more richer in nature when  compared with the earlier reported results on the DSIII equation.  Another advantage of this paper is that the expressions of obtained solutions are very simple, which are given in terms of determinants.

{\bf Acknowledgments}  {\noindent \small
This work is supported by the NSF of China under Grant No. 11671219,  and the K.C. Wong Magna Fund in Ningbo University.  We thank other members in our group at Ningbo University for their many discussions and suggestions on the paper.  K.P. also acknowledges support from the National Board for Higher Mathematics (NBHM), Department of Science
and Technology-Science and Engineering Research Board (DST-SERB), Indo-French Centre for the Promotion of Advanced Research IFCPAR (5104-2) and Council of Scientific and Industrial
Research (CSIR), Government of India.
}

\clearpage

 \end{document}